\RequirePackage{fix-cm}
\documentclass{svjour3}                     
\smartqed  
\usepackage{graphicx}
\graphicspath{{images/}}

\usepackage{verbatim,xcolor,soul,relsize,hyperref}
\usepackage{amsmath,lineno,amssymb,mathtools,url}
\usepackage{algpseudocode}
\usepackage{algorithm}
\usepackage{multirow}
\usepackage{mhchem}
\usepackage{makecell}

\newtheorem{dfn}{Definition}
\newtheorem{pro}[dfn]{Problem}
\newtheorem{thm}[dfn]{Theorem}
\newtheorem{exa}[dfn]{Example}

\newtheorem{cor}[dfn]{Corollary}
\newtheorem{con}[dfn]{Conjecture}
\newtheorem{lem}[dfn]{Lemma}

\counterwithin{dfn}{section}

\newcommand{\C}{\mathbb C}
\newcommand{\R}{\mathbb R}
\newcommand{\Z}{\mathbb Z}

\newcommand{\ga}{\gamma}

\newcommand{\De}{\Delta}
\newcommand{\la}{\lambda}
\newcommand{\La}{\Lambda}

\newcommand{\ti}{\tilde}
\newcommand{\ep}{\varepsilon}

\newcommand{\mut}{\mu^{(t)}}
\newcommand{\muo}{\mu^{(1)}}
\newcommand{\angstrom}{\textup{\AA}}

\newcommand{\Eu}{\mathrm{E}}

\newcommand{\CIS}{\mathrm{CIS}}

\newcommand{\PSD}{\mathrm{PSD}}

\newcommand{\RMS}{\mathrm{RMS}}

\newcommand{\PDF}{\mathrm{PDF}}

\newcommand{\PDD}{\mathrm{PDD}}

\newcommand{\PDDt}{\mathrm{PDD}^{\{2\}}}

\newcommand{\PDDall}{\mathrm{PDD}^{(h)}}
\newcommand{\PDDh}{\mathrm{PDD}^{\{h\}}}
\newcommand{\ADAh}{\mathrm{ADA}^{\{h\}}}
\newcommand{\PDAh}{\mathrm{PDA}^{\{h\}}}

\newcommand{\LND}{\mathrm{LND}}
\newcommand{\ADA}{\mathrm{ADA}}
\newcommand{\PDA}{\mathrm{PDA}}

\newcommand{\AMD}{\mathrm{AMD}}
\newcommand{\PPC}{\mathrm{PPC}}

\newcommand{\EMD}{\mathrm{EMD}}
\newcommand{\EMDh}{\mathrm{EMD}^{\{h\}}}
\newcommand{\EMDm}{\mathrm{EMD}^{(h)}}
\newcommand{\vol}{\mathrm{vol}}

\newcommand{\bs}{} 

\newcommand{\lra}{\leftrightarrow}

\newcommand{\vect}[2]{ \left( \begin{array}{c} 
 #1 \\ #2 \end{array} \right)}

\begin{document}


\title{Higher-order, generically complete, continuous, and polynomial-time isometry invariants of periodic sets} 
\subtitle{}

\titlerunning{Higher-order isometry invariants of periodic sets}        

\author{Daniel Widdowson %
\and        Vitaliy Kurlin 
}


\institute{D.Widdowson and V.Kurlin \at 
Computer Science, University of Liverpool, UK \email{vitaliy.kurlin@liverpool.ac.uk}
}

\date{Received: date / Accepted: date}
\maketitle

\begin{abstract}
Periodic point sets model all solid crystalline materials (crystals) whose atoms can be considered zero-sized points with or without atomic types.
This paper addresses the fundamental problem of checking whether claimed crystals are novel, not noisy perturbations of known materials obtained by unrealistic atomic replacements.
Such near-duplicates have skewed ground-truth because past comparisons relied on unstable cells and symmetries.   
The proposed Lipschitz continuity under noise is a new essential requirement for machine learning on any data objects that have ambiguous representations and live in continuous spaces. 
For periodic point sets under isometry (any distance-preserving transformation), we designed invariants that distinguish all known counter-examples to the completeness of past descriptors and detect thousands of (near-)duplicates in large high-profile databases of crystals within two days on a modest desktop computer.


\noindent 
Communicated by

\end{abstract}


\begin{acknowledgements}
This research was supported by the EPSRC New Horizons grant `Inverse design of periodic crystals' (2022-2025) and the Royal Society APEX fellowship `New geometric methods for mapping the space of periodic crystals' (2023-2025). 
\end{acknowledgements}

\section{The key questions of mathematical data science for real applications}
\label{sec:intro}

Many real data objects have infinitely many different representations.
For example, any rigid object such as a solid crystalline material can be given by atomic coordinates that strongly depend on a chosen basis in Euclidean space $\R^3$.
Hence the \textbf{first question} that mathematical data science should ask about any objects is \emph{Same or different?} \cite{sacchi2020same}.
To make this question meaningful, we should rigorously define what objects can be called \emph{the same} (or \emph{equivalent}) as formalized below.
\smallskip

An \emph{equivalence} is a binary relation (denoted by $S\sim Q$) satisfying three axioms: 
(1) \emph{reflexivity:} any object $S\sim S$;
(2) \emph{symmetry:} if $S\sim Q$ then $Q\sim S$;
(3) \emph{transitivity:} if $S\sim Q$ and $Q\sim T$ then $S\sim T$.
Any classification needs an equivalence satisfying these axioms to split all objects into disjoint classes: the \emph{equivalence class} $[S]$ of an object $S$ consists of all $Q$ equivalent to $S$.
If two classes $[S]$ and $[T]$ share a common object $Q$, then $[S]=[T]$ by
the transitivity axiom.
\smallskip

For any collection of objects, one can consider many different equivalences.
For instance, any finite or periodic configurations of atoms (molecules or crystals) can be called equivalent if they have the same chemical composition.
However, we know many \emph{polymorphic} materials (such as diamond and graphite) that have the same composition but differ by other properties.  
In this case, we need a stronger equivalence that would split all atomic configurations into as many different classes as practically necessary to uniquely identify all physical and chemical properties.
\smallskip

The following equivalence is crucial for many real objects, including molecules and materials whose structures are determined in a rigid form \cite{anosova2024importance}: a \emph{rigid motion} is a composition of translation and rotations in $\R^n$, which preserves all object properties under the same ambient conditions such as temperature and pressure.
Indeed, there is no sense in distinguishing atomic configurations that can be exactly matched by rigid motion, but it is important to see differences in \emph{rigid shapes} (equivalence classes under rigid motion) that can affect their properties.
\smallskip

If we consider compositions of a rigid motion with mirror reflections, we get a slightly weaker equivalence: an \emph{isometry} (denoted by $S\simeq Q$) is any distance-preserving transformation.  
Since mirror images can be distinguished by a sign of orientation, we focus on isometries, which form the full Euclidean group $\Eu(n)$.
\smallskip

After an equivalence (isometry in our case) is fixed, 
objects can be distinguished by an isometry \emph{invariant} $I$ that is a function mapping a given object $S$ to a numerical value (vector or a matrix) $I(S)$ preserved under any isometry, i.e. if $S\simeq Q$, then $I(S)=I(Q)$.
An example invariant of a finite set $S$ is its size (the number of points).
Any non-constant invariant $I$ can distinguish some (not necessarily) all non-isometric sets, i.e. if $I(S)\neq I(Q)$ then $S\not\simeq Q$ by definition.
\smallskip

The invariance is stronger than the \emph{equivariance} requiring that any isometry $f$ maps $I(S)$ to $T_f(I(S))$, where a transformation $T_f$ depends on $f$. 
For example, any linear combination $e(S)$ of coordinates of a finite set $S\subset\R^n$ 
is equivariant, not invariant, and hence allows a \emph{false negative} that is a pair of objects $S\simeq Q$ with $e(S)\neq e(Q)$.
The invariance is much stronger 
by requiring that $T_f$ is the identity.
Then $I(S)\neq I(Q)$ always guarantees that $S\not\simeq Q$ are not isometric.
\smallskip
 
A full answer to the question `\emph{Same or different?}' requires a \emph{complete} invariant $I$ satisfying the much harder inverse implication: if $I(S)=I(Q)$ then $S\simeq Q$.
In other words, $I$ has no \emph{false positives} that are pairs $S\not\simeq Q$ with $I(S)=I(Q)$.   
All triangles $S$ (sets of three points) have a complete invariant $I(S)$ of three inter-point distances due to the side-side-side (SSS) theorem.
Any complete invariant is similar to a DNA-style code that uniquely identifies any object under isometry.
\smallskip

A simple input of real objects is a discrete set of points, which can represent corners, edge pixels, or atomic centers in a molecule or a material.
In the finite case, if given points $p_1,\dots,p_m\in\R^n$ are ordered, they are uniquely determined under isometry \cite{schoenberg1935remarks,kruskal1978multidimensional} by the matrix of pairwise Euclidean distances $|p_i-p_j|$ or the Gram matrix of scalar products $p_i\cdot p_j$, see \cite[chapter 2.9]{weyl1946classical} and \cite{villar2021scalars}.
\smallskip

However, most points in real objects are unordered, e.g. many materials consist of indistinguishable atoms.
A brute-force extension of distance matrices to $m$ unordered points is impractical due to the exponential cost of $m!$ permutations.
In this unordered case, \cite{boutin2004reconstructing} proved that the vector of sorted pairwise distances is \emph{generically complete} meaning that this invariant distinguishes all non-isometric finite sets in $\R^n$ outside some measure 0 subspace of singular sets of points.
\smallskip

After the case of 3 points was settled by the SSS theorem 2000+ years ago, even $m=4$ unordered points in $\R^2$ did not have a better than a brute-force complete isometry invariant based on $4!=24$ permutations, partially due to infinitely many pairs of non-isometric 4-point clouds with the same 6 pairwise distances \cite{caelli2006generating}.  
The finite case was solved in 2023 \cite{widdowson2023recognizing} for any number $m$ of unordered points under rigid motion in $\R^n$, see \cite[Theorem~5.3]{widdowson2021pointwise} for a simpler complete invariant for 4 points under isometry in $\R^n$. 
We now focus on the much harder periodic case.

\begin{dfn}[lattice, motif, $l$-periodic set] 
\label{dfn:periodic}
Vectors $v_1,\dots,v_n\in\R^n$ form a \emph{basis} if any vector in $\R^n$ can be written as $v=\sum\limits_{i=1}^n t_i v_i$ for unique $t_1,\dots,t_n\in\R$. 
For $1\leq l\leq n$, the first $l$ vectors define the \emph{lattice} $\La=\{\sum\limits_{i=1}^l c_i v_i \mid c_1,\dots,c_l\in\Z\}$ and the \emph{unit cell} $U=\{\sum\limits_{i=1}^n x_i v_i \mid x_1,\dots,x_l\in[0,1), x_{l+1},\dots,x_n\in\R\}\subset\R^n$.
If $l=n$, then $U$ is an $n$-dimensional parallelepiped.
If $l<n$, then $U$ is an infinite slab over an $l$-dimensional parallelepiped on $v_1,\dots,v_l$.   
For any finite \emph{motif} of points $M\subset U$, the sum $S=M+\La=\{p+v \mid p\in M, v\in\La\}$ is an \emph{$l$-periodic point set}.
\end{dfn}

\begin{figure}[h]
\includegraphics[height=19mm]{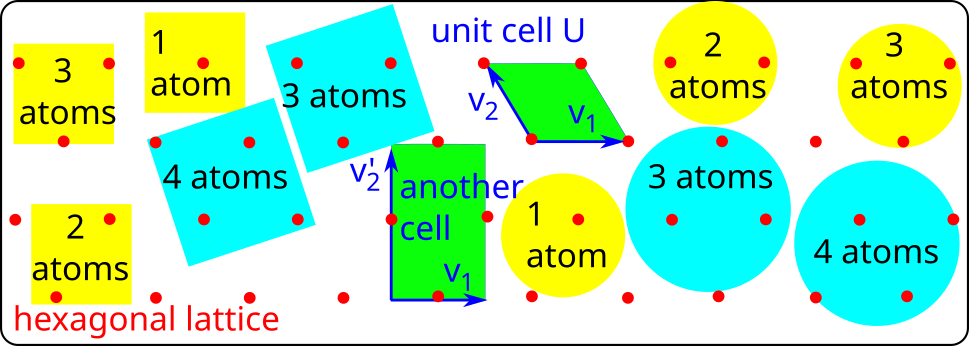}
\hspace*{0.5mm}
\includegraphics[height=19mm]{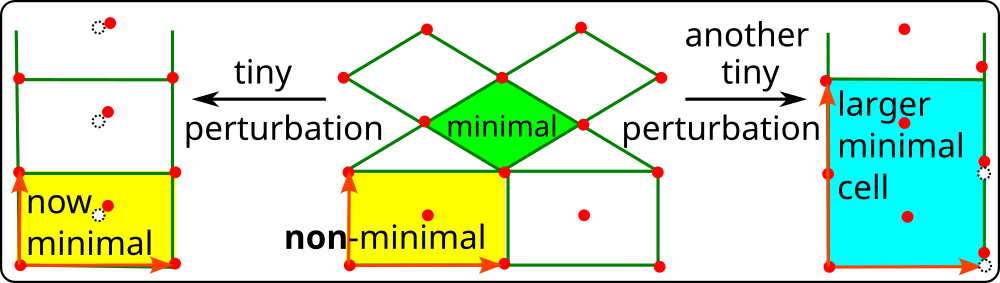}
\vspace*{-2mm}
\caption{\textbf{Left}: any periodic point set can be given by many pairs (cell, motif), see Definition~\ref{dfn:periodic}.
Any periodic set has vastly different finite subsets within boxes or balls of the same cut-off size.
\textbf{Right}: almost any perturbation can arbitrarily scale up a unit cell and break the symmetry.}
\label{fig:hexagonal_lattice} 
\end{figure}

A classification of periodic point sets under isometry cannot be easily reduced to the finite case.
Indeed, the hexagonal lattice of red points in Fig.~\ref{fig:hexagonal_lattice}~(left)  has many non-isometric finite subsets of points within differently positioned boxes or balls of the same cut-off radii.
A motif of points within a unit cell is also ambiguous because any lattice can be generated by infinitely many different bases, which span \emph{primitive} (minimal by volume) unit cells of various shapes.
Crystallographers developed a unique Niggli cell \cite{niggli1928krystallographische} but any such cell discontinuously scales by an arbitrary factor \cite[Theorem~15]{widdowson2022average} under almost all perturbations because of experimental noise \cite{lawton1965reduced} and atomic vibrations, see Fig.~\ref{fig:hexagonal_lattice}~(right). 
\smallskip
 
Even if a complete invariant distinguishes all different objects, the space of equivalence classes is often continuous in the sense that a small perturbation produces a near-duplicate of a slightly different class.
One past approach was to ignore all perturbations up to a small threshold $\ep>0$.
Then the transitivity axiom can make all sets of the same size equivalent through a long enough chain of perturbations $S_1\sim\dots\sim S_k$, each time shifting points up to a fixed Euclidean distance $\ep>0$.
Similarly, adding a single outlier should make finite sets non-equivalent, otherwise all sets of different sizes become equivalent by the transitivity axiom.
\smallskip

This \emph{sorites} paradox \cite{hyde2011sorites} has been discussed from ancient times: while removing grains from a heap of sand one by one, when will a heap of sand suddenly stop being a heap?
The discontinuity problem remained unresolved for materials \cite{zwart2008surprises} because isometry classes of periodic crystals still have no well-defined continuous metric.
The challenges of continuous measurements are important for real objects under many equivalences and motivate the \textbf{second question} `\emph{If different, by how much?}' in Geometric Data Science, which we formalize below for periodic sets.

\begin{pro}
\label{pro:invariants}
For all periodic sets $S\subset\R^n$ with up to $m$ of points in a unit cell, find an invariant $I$ with values in a metric space satisfying the conditions below.
\smallskip

\noindent
(a) \textbf{Completeness} (injectivity): 
any periodic point sets $S,Q\subset\R^n$ are isometric if and only $I(S)=I(Q)$, i.e. $I$ has \emph{no false negatives} and \emph{no false positives}.
\smallskip

\noindent
(b) \textbf{Invertibility} (reconstruction):
any periodic point set $S\subset\R^n$ can be reconstructed from its invariant $I(S)$, uniquely under isometry of $\R^n$. 
\smallskip

\noindent
(c) \textbf{Lipschitz continuity}:
there is a distance metric $d$ on invariant values satisfying all metric axioms
(1) $d(a,b)=0$ if and only if $a=b$,
(2) $d(a,b)=d(b,a)$,
(3) \emph{triangle inequality} 
$d(a,b)+d(b,c)\geq d(a,c)$ for all $a,b,c$;
 and a constant $\la$ such that, for any $\ep>0$, if a periodic point set $Q$ is obtained by perturbing every point of a periodic point set $S$ up to Euclidean distance $\ep$, then $d(I(S),I(Q))\leq \la\ep$.
\smallskip

\noindent
(d) \textbf{Computability}:
for a fixed dimension $n$, the invariant $I(S)$, the metric $d$ and the reconstruction of $S\subset\R^n$ can be obtained in polynomial time of the motif size.
\end{pro}

The reconstruction in condition~\ref{pro:invariants}(b) is stronger than the completeness in \ref{pro:invariants}(a) because a complete invariant can be too complicated with no explicit inversion to an original object.
For example, a DNA code is practically used for identifying humans, but cannot (yet) grow a genetic replica of a living person.
\smallskip

Conditions~\ref{pro:invariants}(a,b) become practically meaningful only with a Lipschitz continuous metric in condition~\ref{pro:invariants}(c) because any noise makes all real objects at least slightly different as in Fig.~\ref{fig:hexagonal_lattice}~(right).
This discontinuity allowed anyone to \emph{claim known materials as new} \cite{chawla2024crystallography} by perturbing atomic positions, scaling up a minimal cell, and changing atomic types to make comparisons by symmetries, unit cells, and chemical compositions unreliable.
As a result, many simulated crystals can be artificially generated, e.g. the report of ``2.2 million new crystals – equivalent to nearly 800 years’ worth of knowledge" from \cite{google2023millions} was rebutted by experts \cite{cheetham2024artificial,widdowson2025geographic}. 
\smallskip

The metric axioms are essential for recognizing isometric sets $S\simeq Q$ by checking if a complete invariant $I$ satisfies $d(I(S),I(Q))=0$. 
If the triangle inequality in \ref{pro:invariants}(c) fails with any positive error, outputs of $k$-means and DBSCAN clustering may be pre-determined for a non-metric and hence are not trustworthy \cite{rass2024metricizing}.
Polynomial-time condition \ref{pro:invariants}(d) makes Problem~\ref{pro:invariants} notoriously hard, else one can design a complete infinite-size invariant by taking all isometric images of $S$. 
\smallskip 

An invariant $I$ satisfying all the conditions above is similar to geographic coordinates that continuously parametrize the surface of Earth.
Hence, Problem~\ref{pro:invariants} is interpreted as geographic-style mapping of the \emph{Crystal Isometry Space} $\CIS(\R^n;m)$ defined as the \emph{moduli} space of all periodic sets with up to $m$ points in a unit cell under isometry in $\R^n$.
An invariant $I$ can be considered a function on the union $\bigcup\limits_{m\geq 1}\CIS(\R^n;m)$ with values in a metric space, where all computations 
should be faster than in $\CIS(\R^n;m)$, i.e. in polynomial time in $m$ for a fixed dimension $n$.
\smallskip 

\textbf{Contributions}.
We extend the Pointwise Distance Distribution (PDD) \cite{widdowson2022resolving} to stronger (also generically complete) invariants $\PDDall$ for higher orders $h>1$ by keeping the Lipschitz continuity under bounded noise and polynomial-time computability for fixed $n,h$.
The invariants $\PDDt$ distinguish all known examples 
$S\not\simeq Q$ with $\PDD(S)=\PDD(Q)$ in $\R^3$ and experimentally confirm thousands of near-duplicates in the world's largest databases of periodic materials in section~\ref{sec:experiments}.

\section{A review of open challenges in representations of periodic crystals}
\label{sec:past}

Problem~\ref{pro:invariants} makes sense for many real objects (finite point sets, embedded graphs, surfaces or complexes in $\R^n$) under other practical equivalences (affine and projective transformations). 
The graph isomorphism problem \cite{grohe2020graph} considers only conditions~\ref{pro:invariants}(a,d) without a continuous metric, which is needed for real lengths of edges.
Since pairwise distances \cite{boutin2004reconstructing} distinguish all generic sets of $m$ unordered points under isometry in $\R^n$ and the more recent complete invariants \cite{widdowson2023recognizing} continuously distinguish all finite sets under rigid motion in $\R^n$, we focus on periodic sets.
\smallskip

For $n=1$, Theorem~4 in \cite{grunbaum1995use} justified complete invariants 
for periodic sequences given by rational angles of the unit circle (in the complex plane $\C$) by using 6-factor products of complex numbers. 
Since the circle (a period) was fixed, these invariants are discontinuous under perturbations.
Indeed, the sequence $\Z$ of integers is infinitely close to $S=\{\ep,1,\dots,m\}+(m+1)\Z\subset\R$ for any small $\ep>0$, though their minimum periods $1$ and $m+1$ are arbitrarily different.
The much simpler complete invariant of a periodic sequence $S=\{p_1,\dots,p_m\}+L\Z\subset\R$ with a period $L$, where $0\leq p_1<\dots<p_m<L$, is the list of inter-point distances $p_{i+1}-p_i$ (under cyclic permutations) for $i=1,\dots,m$ and $p_{m+1}=p_1+L$.
\smallskip

A continuous metric $d(S,Q)$ on these cyclic classes of distance lists was introduced in \cite{kurlin2025complete} but such a metric requires an expansion to the least common multiple of the sizes $|S|,|Q|$ of motifs and doesn't come with a polynomial-time invariant.
The resulting brute force invariant for all periodic sequences $S$ with motifs up to $m$ points needs an expansion to at least $2^m$ points \cite[Theorem~5(1)]{farhi2007nontrivial}, which violates condition~\ref{pro:invariants}(d).
Problem~\ref{pro:invariants} remained open even in dimension $n=1$.
\smallskip

A finite approach to measuring the similarity between periodic point sets is to compare their finite subsets within a box or a ball of a large but fixed cut-off radius.
However, any periodic point set has many non-isometric finite subsets within differently positioned boxes or balls of the same size as in Fig.~\ref{fig:hexagonal_lattice}~(left). 
\smallskip

Local clusters centered at all points in a motif $M$ can be converted by Gaussian blurring into smooth functions \cite{batatia2022mace}, which can be decomposed in the infinite basis of spherical harmonics \cite{seeley1966spherical} and hence considered complete in the limit.
\cite{dusson2022atomic} discusses challenges of choosing several parameters (blurring, approximation, interaction order), including a cut-off radius that can discontinuously change these clusters due to new neighbors outside a smaller cut-off.
Even if this cut-off is smoothed out, a manually chosen value may not suffice or slow down computations \cite{parsaeifard2022manifolds,pozdnyakov2022comment}. 
\smallskip

Atomic vibrations are natural to measure by deviations of atoms from their initial positions, but a sum of small deviations over infinitely many points can be infinite and  also can give different values for different finite subsets.
However, a maximum deviation of atoms is well-defined as the bottleneck distance between any sets via bijections between atoms, which can be displaced but cannot vanish.

\begin{dfn}[bottleneck distance $d_B$]
\label{dfn:bottleneck}
The \emph{bottleneck distance} $d_B(S,Q)=\inf\limits_{g:S\to Q} \sup\limits_{p\in S}|p-g(p)|$ for any sets $S,Q\subset\R^n$ of the same cardinality is minimized for all bijections $g:S\to Q$ and maximized for all points $p\in S$.
\bs
\end{dfn}

Here $|p-q|$ denotes Euclidean distance between points $p,q\in\R^n$.
Though Definition~\ref{dfn:bottleneck} is impractical because of infinitely many bijections, $d_B$ can be efficiently computed \cite{efrat2001geometry} for finite sets if $|p-q|$ is replaced with $L_\infty(p,q)=\max\limits_{i=1,\dots,n}|p_i-q_i|$. 
\smallskip

If periodic point sets $S,Q$ have different densities (motif size $|S|$ divided by the cell volume), then $d_B(S,Q)$ is infinite \cite[Example~2.1]{widdowson2022resolving}.
Also, $d_B(S,Q)$ is discontinuous under perturbations of 2D lattices \cite{kurlin2024mathematics}whose \emph{primitive} cells have the same minimum volume \cite[Example~2.2]{widdowson2022resolving}.
Hence condition~\ref{pro:invariants}(c) of a Lipschitz continuous metric made Problem~\ref{pro:invariants} exceptionally hard in the periodic case.
\smallskip

\begin{dfn}[\emph{metrics} vs \emph{pseudo-metrics}]
\label{dfn:metric}
A distance $d$ between objects under an equivalence relation $\sim$ is a \emph{metric} if the following axioms hold:
\smallskip

\noindent
(1) \emph{coincidence:} $d(S,Q)=0$ if and only if $S\sim Q$;

\noindent
(2) \emph{symmetry:} $d(S,Q)=d(Q,S)$ for any objects $S,Q$; 

\noindent
(3) 
\emph{triangle inequality:} 
 $d(S,Q)+d(Q,T)\geq d(S,T)$ for any $S,Q,T$.
\smallskip
 
If the coincidence axiom (1) is replaced with 
$(1')$ $d(S,S)=0$ for any $S$, 
then non-equivalent $S\not\sim Q$ can have $d(S,Q)=0$, and $d$ is called a \emph{pseudo-metric}.
\bs
\end{dfn}

Definition~\ref{dfn:metric} guarantees positivity: 
$2d(S,Q)=d(S,Q)+d(Q,S)\geq d(S,S)=0$.
Many descriptors or invariants are compared by distances (such as Euclidean) that satisfy all metric axioms on descriptor values but define only pseudo-metrics on isometry classes due to the incompleteness of these invariants.
If $d(S,Q)>0$, then $S\not\sim Q$ by $(1')$, so a fast pseudo-metric can distinguish between some but not all objects.
Pseudo-metrics are weaker than metrics, e.g. the difference $||S|-|Q||$ of set sizes is a pseudo-metric not distinguishing any sets $S\not\simeq Q$ of the same size. 
\smallskip

Hence metrics satisfying all axioms (similar to complete invariants) are much more valuable than pseudo-metrics (similar to non-invariants or incomplete invariants). 
Any algorithm using an incomplete invariant $I$ cannot predict different properties of a \emph{false positive} pair of non-isometric sets $S\not\simeq Q$ with $I(S)=I(Q)$. 
\smallskip

That is why the \emph{discriminative} problem should be solved first (at least in general position) by designing complete and Lipschitz continuous invariants before \emph{generative} attempts can succeed.
Any non-complete invariant $I$ is not invertible 
 in the sense that different sets $S\not\simeq Q$ (false positives) can have $I(S)=I(Q)$. 
\smallskip

Now we review recent continuous invariants in the periodic case. 
Continuous metrics on lattices under rigid motion are known for dimension $n=2$ \cite{bright2023geographic,bright2023continuous}, not yet for $n=3$ \cite{kurlin2022complete}.
A generically complete and Lipschitz continuous invariant of periodic point sets $S\subset\R^3$ \cite{edelsbrunner2021density} is the sequence of \emph{density functions} $\psi_k(S;t)$ measuring the fractional volume of $k$-fold intersections $\bigcup\limits_{p_1,\dots,p_k\in S}(\bar B(p_1;t)\cap \dots\cap \bar B(p_k;t)\cap U)$ for any $k\geq 1$, where $U$ is a unit cell of $S$, and $\bar B(p;t)$ is the closed ball with a center $p\in S$ and a variable radius $t\geq 0$.
The infinite sequence $\{\psi_k\}_{k=1}^{+\infty}$ allows only an approximate distance and turned out to be incomplete \cite[Example~11]{anosova2022density}, but 
was analytically described for all periodic sequences of intervals in $\R$ \cite{anosova2023density}.
The \emph{periodic merge tree} \cite{edelsbrunner2024merge} is a continuous isometry invariant of periodic graphs with a slow interleaving pseudo-metric and a faster distance on simpler periodic 0-th barcodes.
The invariant below solved a weaker version of Problem~\ref{pro:invariants} for finite and periodic sets when completeness in 
(\ref{pro:invariants}a) is replaced with generic completeness.

\begin{dfn}[Pointwise Distance Distribution $\PDD$]
\label{dfn:PDD}
Let $S\subset\R^n$ be any $l$-periodic point set with a motif $M$ of $m$ points.
For any integer $k\geq 1$ and $p\in M$, let $d_1(p)\leq\dots\leq d_k(p)$ be the list of Euclidean distances from $p$ to its $k$ nearest neighbors within the whole set $S$.
These lists become rows of the $m\times k$ matrix $D(S,M;k)$.
Any $c>1$ identical rows are collapsed into a single row with the \emph{weight} $c/m$, which is written in the extra first column.
The resulting matrix $\PDD(S;k)$ of unordered rows with weights is called the \emph{Pointwise Distance Distribution} 
\cite{widdowson2022resolving}.
\bs
\end{dfn}

For finite sets, the PDD was studied under the name of a \emph{local distribution of distances} \cite{memoli2011gromov}.
The $\PDD$ can be considered a multiset of rows and a discrete probability \emph{distribution} with normalized weights interpreted as probabilities.
\smallskip

If a unit cell of $S$ is extended by a factor of $c $, then any point $p$ in the original motif has $c$ translationally equivalent copies in the extended motif $M_c$.
Then $D(S,M_c;k)$ has $c$ times more rows because each original row is expanded into $c$ identical rows but the invariant $\PDD(S;k)$ is the same weighted distribution of rows, independent of an initial cell of $S$.
The equality between weighted distributions is interpreted as a bijection between unordered sets respecting all weights.
This equality is best checked not by considering all bijections but by a metric that vanishes only on equal distributions due to the first metric axiom.
The PDD is Lipschitz continuous, computable (for a fixed dimension) in a near-linear time of $k,m$, and distinguishes all non-isometric sets in \emph{general position} (away from a measure 0 subspace), see \cite[Theorems 3.2, 4.3, 4.4, 5.1]{widdowson2022resolving} and proofs in \cite{widdowson2021pointwise}.

\begin{dfn}[homometric sets]
\label{dfn:homometric}
Finite or $l$-periodic sets $S,Q\subset\R^n$ are called \emph{homometric} \cite{patterson1939homometric} if they have the same \emph{Pair Distribution Function (PDF)}, which is a single distribution of all inter-point distances of $S$ (without considering their periodic copies), equivalent to a powder diffraction pattern 
without a cut-off radius.
\bs
\end{dfn}

\begin{figure}[h]
\includegraphics[width=\textwidth]{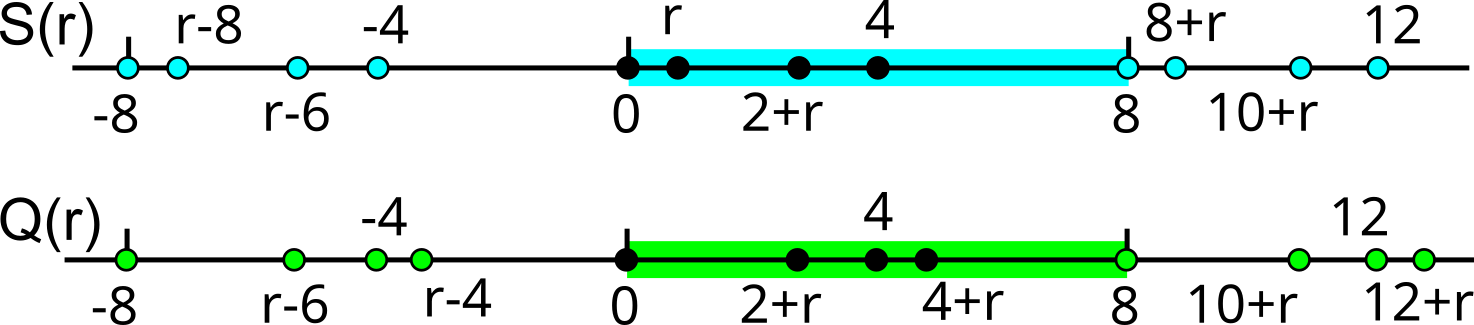}
\caption{For any $0<r\leq 1$, the homometric sets $S(r)=\{0,r,2+r,4\}+8\Z\not\cong Q(r)=\{0,r,2+r,4\}+8\Z$ have identical PDFs from Definition~\ref{dfn:homometric} but different $\PDD$s whose first columns we write as unordered sets: $\PDD(S(r);1)=\{r,r,2-r,2-r\}\neq \PDD(Q(r);1)=\{r,r,2-r,2+r\}$. 
}
\label{fig:1D_periodic_Sr+Qr} 
\end{figure}

The PDF is easily extractable from X-ray diffraction patterns and can be split into several distributions by fixing an atomic type (chemical element), say by listing average distances from all (say) carbon atoms to their neighbors in the full crystal.
The PDD does this splitting by geometry (all identical distances to neighbors) and is stronger than the PDF even for 1-dimensional periodic sequences in Fig.~\ref{fig:1D_periodic_Sr+Qr}. 
\smallskip

Almost any perturbation, as in Fig.~\ref{fig:hexagonal_lattice}~(right), can split every inter-point distance (say) $d$ into many $d_1,\dots,d_c$, which are all close to $d$ but are not copies of each other because the initial minimal cell was scaled by the factor $c$.
One attempt to resolve this discontinuity was to blur each distance by a Gaussian deviation and a smoothed PDF as a normalized sum of Gaussians around all 
distance values.
\smallskip

Discretizing the smoothed PDF for comparisons reduces its strength and creates the counter-intuitive pipeline: a discrete set $S$ $\to$ a smoothed $\PDF$ $\to$ a discrete sample of $\PDF(S)$.  
The discontinuity can be resolved by continuous metrics \cite{kantorovich1960mathematical,vaserstein1969markov} on PDDs interpreted as a probability distribution of rows of $k$ distances. 

\begin{figure}[h]
\includegraphics[width=\linewidth]{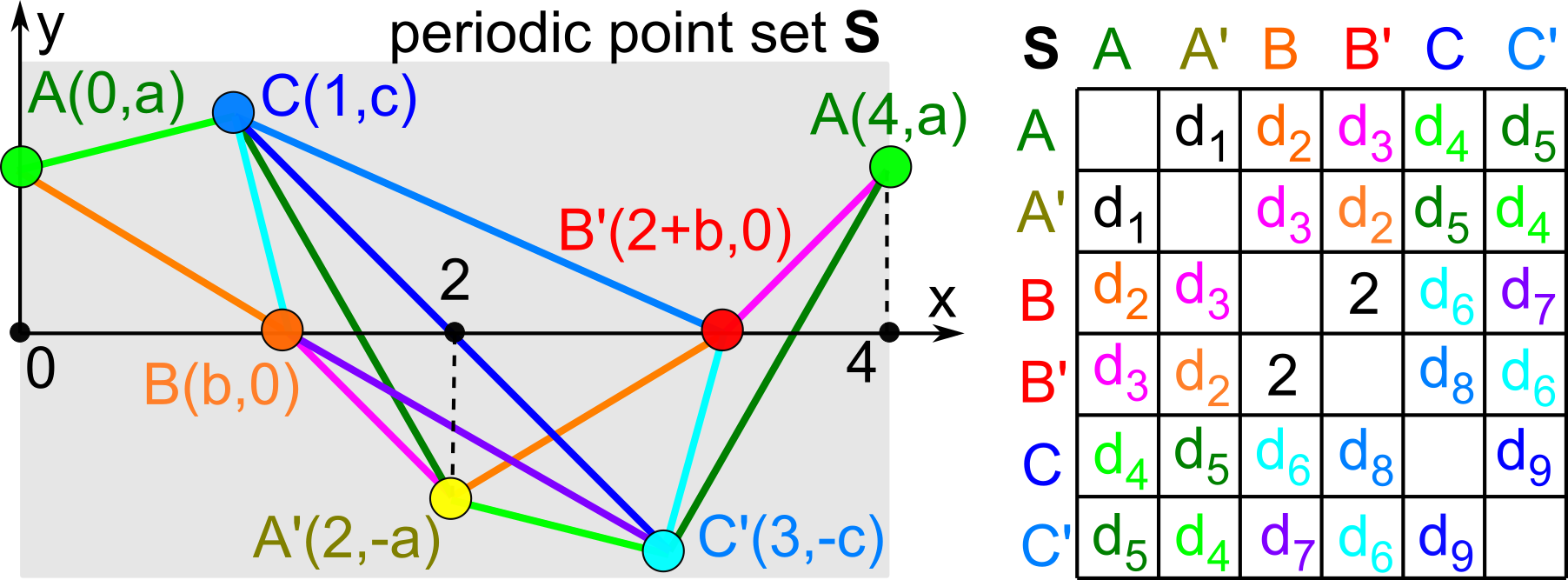}
\medskip

\includegraphics[width=\linewidth]{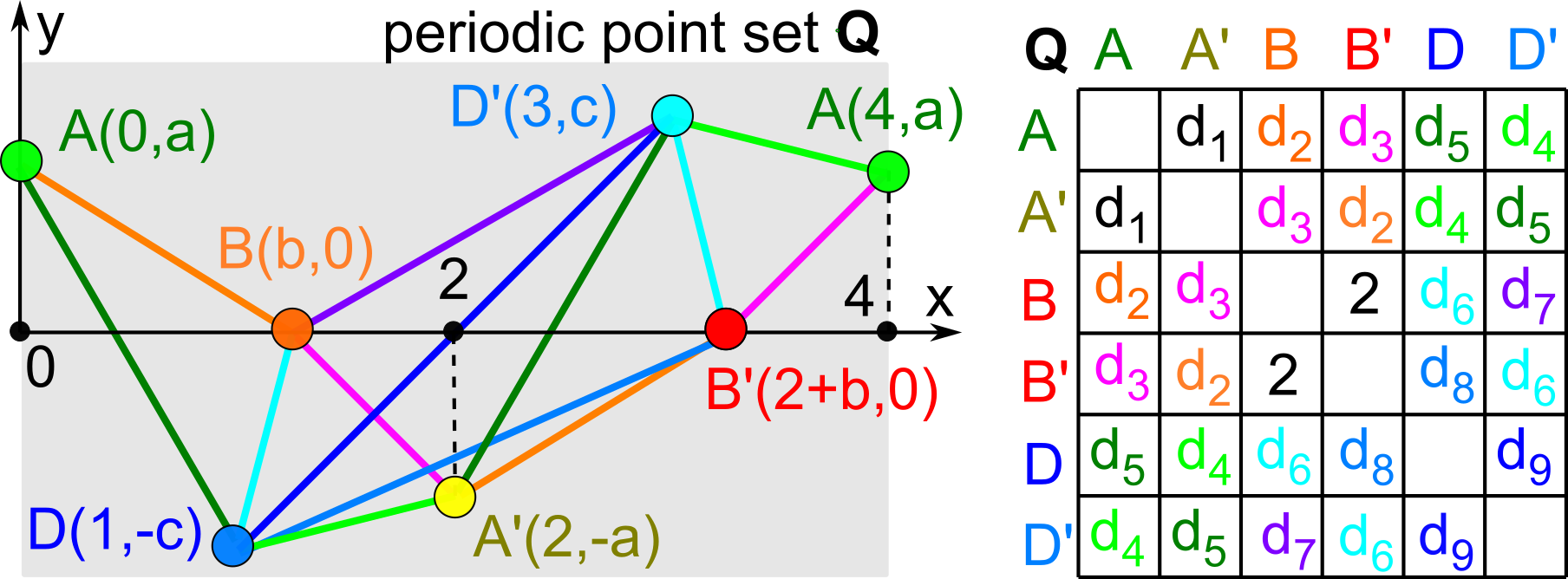}
\caption{The sets $S,Q$ are 1-periodic in the $x$-axis with period 4, e.g. $A$ denotes both $(0,a)$, $(4,a)$. 
\textbf{Right}: distances between closest points from classes modulo shifts by $4$ in $x$.
Then $\PDD(S;k)=\PDD(Q;k)$ by Example~\ref{exa:PDD} but $\PDDt(S;1)\neq\PDDt(Q;1)$ by Example~\ref{exa:6-point_pairs}.}
\label{fig:6-point_pairs} 
\end{figure}

\begin{exa}[sets with equal PDDs]
\label{exa:PDD}
The 1-periodic sets $S\not\simeq Q$ in \cite[Fig.~4]{pozdnyakov2022incompleteness} were designed to fail all iterations of the Weisfeiler-Leman test \cite{shervashidze2011weisfeiler}.
Fig.~\ref{fig:6-point_pairs} shows their 2D versions with period 4 in the $x$-axis and free parameters $a,b,c>0$.
\smallskip

The distances in Fig.~\ref{fig:6-point_pairs}~(right)
are for the closest representatives of 6 points. 
$\begin{array}{lll}
d_1=2\sqrt{a^2+1}, \quad & 
d_2=\sqrt{a^2+b^2}, \quad  &
d_3=\sqrt{a^2+(2-b)^2}, \\
d_4=\sqrt{1+(a-c)^2}, \quad &
d_5=\sqrt{1+(a+c)^2}, \quad &
d_6=\sqrt{(1-b)^2+c^2}, \\
d_7=\sqrt{(3-b)^2+c^2}, \quad  &
d_8=\sqrt{(1+b)^2+c^2}, \quad &
d_9=2\sqrt{c^2+1}.
\end{array}$
\smallskip

Then $\PDD(S;k)=\PDD(Q;k)$ 
because the equalities between distances (shown in the same color) in Fig.~\ref{fig:6-point_pairs}~(right) hold after adding any periodic translation, so if $d_1=d_2$ then $\sqrt{d_1^2+(4n)^2}=\sqrt{d_2^2+(4n)^2}$ for any $n\in\Z$.
\bs
\end{exa}

Simpler non-isometric finite sets in $\R^3$ with equal PDDs were distinguished by stronger invariants in \cite{widdowson2023recognizing}, which extended PDD by recording distances to subsets of more than one point.
In the periodic case, pairs of points behave discontinuously under cell extensions in Fig.~\ref{fig:hexagonal_lattice}.
Doubling a motif $M$ of $m$ points leads to $(2m)^2$ pairs including new distant neighbors from adjacent cells.
This obstacle motivated a `pointwise' approach to both finite and periodic sets in the next section.
\smallskip

Another `pointwise' \emph{isoset} \cite{anosova2021isometry} was proved to be complete for all periodic point sets in any $\R^n$.
A Lipschitz continuous metric on isosets was only approximated in polynomial time \cite{anosova2025recognition,mcmanus2025computing}, but condition~(\ref{pro:invariants}d) requires an exact computation.

\section{The new isometry invariants of finite and periodic sets of points}
\label{sec:invariants}

This section extends the PDD to higher order $h>1$ in Definition~\ref{dfn:PDDh} motivated by pairs of non-isometric sets $S\not\simeq Q$ with $\PDD(S;k)=\PDD(Q;k)$ in Example~\ref{exa:PDD}.
Definition~\ref{dfn:PDDh} makes sense for a finite set $S=M$ in any metric space.
\smallskip

\begin{dfn}[higher order $\PDDh(S;k)$] 
\label{dfn:PDDh}
Let $S\subset\R^n$ be any $l$-periodic point set with a motif $M$ of $m$ points.
Fix a point $p\in M$ and integers $h,k\geq 1$.
Consider any $h$ distinct points $p_1,\dots,p_h\in S\setminus\{p\}$ and the \emph{$h$-order average} $\dfrac{2}{h(h+1)}\sum\limits_{0\leq i<j\leq h}|p_i-p_j|$ of pairwise distances between the points $p=p_0,p_1,\dots,p_h$. 
Let $a(p;h,1)\leq\dots\leq a(p;h,k)$ be the list of $k$ smallest averages for the fixed point $p$ and variable points $p_1,\dots,p_h\in S\setminus\{p\}$.
These lists become rows of the $m\times k$ matrix $D(S,M;h,k)$, where we can collapse any $c>1$ equal rows to one row with the \emph{weight} $c/m$ written in the extra first column.
The final matrix of unordered rows with weights is the \emph{$h$-order Pointwise Distance Distribution} $\PDDh(S;k)$. 
\bs
\end{dfn}

Lemma~\ref{lem:PDDh_invariant} will prove that $\PDDh(S;k)$ is independent of a motif $M$ to justify the notation without $M$. 
In Definition~\ref{dfn:PDDh}, we can 
keep $m$ rows of average sums with equal weights $1/m$.
The matrices $\PDD^{\{1\}},\dots\PDDh$ can  be concatenated into a single matrix $\PDDall(S;k)$ of $m$ unordered rows and $kh$ ordered columns.

\begin{exa}[$\PDD^{(2)}$ for the sequences in {Fig.~\ref{fig:1D_periodic_Sr+Qr}}]
The sum $\sum\limits_{0\leq i<j\leq 2}|p_i-p_j|$ is the perimeter of the triangle on the points $p_0\in M$ and $p_1,p_2\in S$.
The row of a point $p\in M$ in 
$\PDD^{(2)}(S;k)$ consists of the $k$ shortest distances followed by $k$ smallest perimeters (divided by $3$) of triangles at $p$.  
In Fig.~\ref{fig:1D_periodic_Sr+Qr}, the point $p_0=0$ in the motif of $S(r)=\{0,r,2+r,4\}+8\Z$ has the nearest neighbors $p_1=r$, $p_2=2+r$ at the distances $r,2+r$, and two smallest averaged perimeters $2(2+r)/3,8/3$.
The point $p_0=0$ in $Q(r)=\{0,r,2+r,4\}+8\Z$ has the nearest neighbors at the distances $2+r,4-r$, and two smallest averaged perimeters $\frac{8}{3},\frac{8}{3}$.
Then 
$\PDD^{(2)}(S(r);2)=\left( \begin{array}{cc|cc}
 r & 2+r & \frac{2(2+r)}{3} & \frac{8}{3} \\
 r & 2 & \frac{2(2+r)}{3} & \frac{2(4-r)}{3} \\
 2-r & 2 & \frac{2(2+r)}{3} & \frac{2(4-r)}{3}  \\
 2-r & 4-r & \frac{2(4-r)}{3} & \frac{8}{3} 
 \end{array}\right)\neq
\PDD^{(2)}(Q(r);2)=\left( \begin{array}{cc|cc}
 2+r & 4-r & \frac{8}{3} & \frac{8}{3} \\
 2-r & 2+r & \frac{4}{3} & \frac{8}{3} \\
 r & 2-r & \frac{4}{3} & \frac{8}{3}  \\
 r & 2 & \frac{4}{3} & \frac{8}{3} 
 \end{array}\right)$, where
 all rows have equal weights $\frac{1}{4}$, so we have skipped these weights for brevity.
 \bs
\end{exa}

The factor $\dfrac{2}{h(h+1)}$ was chosen to guarantee the Lipschitz continuity with $\la=2$ in (\ref{pro:invariants}c).  
Examples~\ref{exa:6-point_pairs},~\ref{exa:Pauling_homometric} show that $\PDDt$ distinguishes all known homometric sets for $n=2,3$, which have identical PDDs.
Any increase in $k$ adds extra columns with larger values to $\PDDh(S;k)$ without changing any previous values.
So the number $k$ 
is considered a degree of approximation, not a parameter like a cut-off radius whose changes substantially affect local atomic clouds.
\smallskip

Lemma~\ref{lem:PDDh_invariant} proves the invariance of $\PDDall$ under isometry in $\R^n$ and under changes of a cell. 
If $k$ is greater than the number $\binom{m-1}{h-1}$ of $h$-tuples with a fixed $p\in S$, we set all non-existing sums in Definition~\ref{dfn:PDDh} to the largest existing value. 

\begin{lem}[invariance of $\PDDh(S;k)$]
\label{lem:PDDh_invariant}
For any integers $h,k\geq 1\leq l\leq n$ and any finite unordered set $S$ in a metric space or any $l$-periodic point set $S\subset\R^n$, the higher-order 
$\PDDh(S;k)$ from Definition~\ref{dfn:PDDh} is an isometry invariant of $S$.
\end{lem}
\begin{proof}
First, for any $l$-periodic point set $S\subset\R^n$, we show that scaling up a unit cell $U$ to a non-primitive cell keeps $\PDD$ invariant.
It suffices to scale up $U$ by a factor $c$, say along the first basis vector $v_1$ of $U$, then the number $m$ of motif points of $S$ is multiplied by $c$.
Then the matrix $D(S, S\cap (cU);h,k)$ consisting of $k$ smallest average sums of pairwise distances between $h+1$ points in Definition~\ref{dfn:PDDh} has the larger size $cm\times k$ in comparison with the original $m\times k$ matrix $D(S,S\cap U;h,k)$ but each row is repeated $c$ times for the shifted points $p+i v_1$, where $p$ is any point from the original motif $M=S\cap U$ of the $l$-periodic set $S$, for $i=0,\dots,c-1$.
\smallskip

Second, we show that the matrix $D(S,S\cap U;h,k)$ and hence $\PDDh(S;k)$ is independent of a primitive cell $U$. 
Let $U,V$ be primitive cells of any $l$-periodic set $S\subset\R^n$ with a lattice $\La$.
Any point $q\in S\cap V$ can be translated by a vector of $\La$ to a point $p\in S\cap U$ and vice versa.
These translations preserve distances and establish a bijection between the motifs $S\cap U\lra S\cap V$, and a bijection between all rows of the matrices $D(S,S\cap U;h,k)\lra D(S,S\cap V;h,k)$.
\smallskip

Third, we prove that $\PDDh(S;k)$ is preserved under any isometry $f:S\to Q$ of $l$-periodic point sets.
Any primitive cell $U$ of $S$ is bijectively mapped by $f$ to the unit cell $f(U)$ of $Q$, which should be also primitive.
Indeed, if $Q$ is preserved by a translation along a vector $v$ that doesn't have all integer coefficients in the basis of $f(U)$, then $S=f^{-1}(Q)$ is preserved by the translation along $f^{-1}(v)$, which doesn't have all integer coefficients in the basis of $U$, so $U$ was non-primitive.
Since $U$ and $f(U)$ have the same number of points from $S$ and $Q=f(S)$, the isometry $f$ gives a bijection between the motifs $S\cap U\lra Q\cap f(U)$.
\smallskip

For any discrete sets $S,Q$, 
the $k$ smallest average sums of all distances between any point $p\in S\cap U$ and $p_1,\dots,p_h\in S$, equal the same sums for $f(p)\in Q\cap f(U)$ and $f(p_1),\dots,f(p_h)\in Q$, respectively.
These coincidences imply that $\PDD(S;k_1,\dots,k_h)=\PDDh(Q;k_1,\dots,k_h)$ up to a permutation of rows.  
\qed
\end{proof}

\begin{exa}[{$\PDDt$} distinguishes $S,Q$ in {Example~\ref{exa:PDD}}]
\label{exa:6-point_pairs} 
We start with singular cases when $S,Q$ are identical.
If $c=0$, then $C=D$, $C'=D'$, so $S,Q$ are identical in Fig.~\ref{fig:6-point_pairs}.
If $b\in\{0,1,2\}$, then the periodic shifts of $B\cup B'$ (hence $S,Q$) become mirror images with respect to the vertical line $x=2$.
We now assume that $1<b<2$.
Then $d_2>d_3$, $d_5>\max\{d_4,d_6\}$, and $\min\{d_7,d_8,d_9\}>d_6$.
\smallskip

The set $S$ in Fig.~\ref{fig:6-point_pairs} has a motif of 6 points, which generate isometric triangles $\triangle ABC\simeq\triangle A'B'C'$ with the perimeter $d_2+d_4+d_6$, see details in Example~\ref{exa:PDD}.
The other potentially smaller perimeters of triangles on points of $S$ are
$d_3+d_5+d_6$, $d_3+d_4+d_7$.
The smallest perimeter for $S$ is the minimum of these sums.
The smallest perimeter for $Q$ is $\min\{d_2+d_4+d_5, \; d_2+d_5+d_6,\; d_3+d_4+d_6\}$.
\smallskip

\noindent
If $t=d_2+d_4+d_6$ equals one of the last sums, one of the following cases holds:
if $d_2=d_3$ then $b=1$,
if $d_4=d_5$ then $c=0$,
if $d_6=d_7$ then $b=2$ or $0$, so $S\simeq Q$. 
\smallskip

\noindent
If $t=d_3+d_5+d_6$ is a minimal perimeter for $S$, then $t$ cannot equal any of the three sums for $Q$.
Indeed, if $t=d_2+d_5+d_6$ then $d_2=d_3$.
If $t=d_3+d_4+d_6$ then $d_4=d_5$.
The minimality of the sum $t$ for the set $S$ means that $d_3+d_6<d_2+d_4$, so $t=d_3+d_5+d_6$ cannot equal $d_2+d_4+d_5$ for $Q$.
\smallskip

\noindent
If $t=d_3+d_4+d_7$ is a minimal perimeter for $S$, then $t$ cannot equal any of the three sums for $Q$.
Indeed, if $t=d_3+d_4+d_6$ then $d_6=d_7$.
The minimality of $t$ for $S$ means that $d_3+d_7<d_2+d_6<d_2+d_5$, so 
$t=d_3+d_4+d_7<d_2+d_4+d_5$ for $Q$.
Similarly, if $d_4+d_7<d_5+d_6$ then $t=d_3+d_4+d_7<d_3+d_5+d_6<d_2+d_5+d_6$.
\smallskip

In all these cases, $S,Q$ become isometric.
Hence the smallest perimeters 
in $\PDDt$ for $k=1$ distinguish 
all pairs of the homometric sets $S,Q$. 
The same conclusion holds for more general sets obtained from $S,Q$ by periodic translations in other directions (along the $y$-axis or even in any $\R^n$), see  \cite[Fig.~10]{pozdnyakov2022incompleteness}, when extra periods are large and don't affect any triangles with the smallest perimeters.
\bs
\end{exa}

The rows of $\PDDh(S;k)$ are unordered to guarantee the continuity under perturbations, though we can lexicographically order the rows for convenience.  
Recall that $u=(u_1,\dots,u_n)$ is \emph{lexicographically} smaller than $v=(v_1,\dots,v_n)$ in $\R^n$ (written $u<v$) if $u_i=v_i$ for $i=1,\dots,k$ and $u_{k+1}<v_{k+1}$ for some $k<n$.  
\smallskip

We can convert any $\PDDh$ into a fixed-size matrix, which can be flattened into a vector for easy comparisons, while keeping the continuity and almost all invariant data. 
Any distribution of $m$ unordered values can be reconstructed from its $m$ moments defined below.
When all weights $w_i$ are rational as in our case, the distribution can be expanded to equal-weighted values $a_1,\dots,a_m$.
The $m$ moments can recover all $a_1,\dots,a_m$ as roots of a polynomial of degree $m$ whose coefficients are expressed via the $m$ moments \cite{macdonald1998symmetric}. 
For example, any reals $a,b$ are the roots of the quadratic polynomial $x^2-(a+b)x+ab$, where $ab=\frac{1}{2}((a+b)^2-(a^2+b^2))$.
\smallskip

Let $A$ be any unordered set of real numbers $a_1,\dots,a_m$ with weights $w_1,\dots,w_m$, respectively, such that $\sum\limits_{i=1}^m w_i=1$.
For any integer $t\geq 1$ , the $t$-th \emph{moment} \cite[section~2.7]{keeping1995introduction} 
is $\mu_t(A)=\sqrt[t]{m^{1-t}\sum\limits_{i=1}^m w_i a_i^t}$, so
 $\mu_1(A)=\sum\limits_{i=1}^m w_i a_i$ is the usual average.
For $t\geq 2$, we normalize the sum (before taking the $t$-th root) by the factor $m^{(1/t)-1}$ to prove continuity of all moments with the Lipschitz constant $\la=2$. 
\smallskip

\begin{dfn}[the $t$-moments matrix {$\mu^{(t)}[\PDDh]$}]
\label{dfn:moments}
Fix any integers $h,k,t\geq 1\leq l\leq n$, and a finite or $l$-periodic point set $S\subset\R^n$.
For every column $A$ of the matrix $\PDDh(S;k)$ from Definition~\ref{dfn:PDDh}, which consists of unordered numbers $a_1,\dots,a_m$ with weights, write the new column $(\mu_1(A),\dots,\mu_t(A))$.
The resulting \emph{$t$-moments} matrix of sizes $t\times k$ is denoted by $\mu^{(t)}[\PDDh(S;k)]$. 
For $t=h=1$, the $1\times k$ matrix $\mu^{(1)}[\PDD(S;k)]$ was called the vector of \emph{Average Minimum Distances} \cite{widdowson2022average} and was also denoted by $\AMD(S;k)=(\AMD_1,\dots,\AMD_k)$.
\bs
\end{dfn}

The matrix  $\mu^{(t)}[\PDDh(S;k)]$ has $t$ ordered rows and $k$ ordered columns but is a bit weaker than the original distribution $\PDDh(S;k)$ with the same parameters $h,k$, because each column is reconstructable from its moments for $t\geq m$ only up to a permutation.
However,  to faster filter distant crystals, we can flatten any matrix $\mu^{(t)}[\PDD(S;k)]$ with indexed entries to a vector of $tk$ coordinates.
\smallskip

For a finite set $S\subset\R$, a simple complete invariant under translations is the ordered sequence of inter-point distances.
However, a naive extension to periodic sets is discontinuous, e.g. $\Z$ is $\ep$-close to $\{0,1+\ep\}+2\Z$ but their periods $1$ and $2$ are not close.  
Definition~\ref{dfn:PSD} introduces a distribution whose  completeness and Lipschitz continuity for $n=1$ will be proved by Theorem~\ref{thm:PSD} and Lemma~\ref{lem:PSD_continuous}.
 
\begin{dfn}[\emph{Pointwise Shift Distribution} $\PSD$]
\label{dfn:PSD}
For any periodic point set (sequence) $S\subset\R$ with a motif $M$ of $m$ points, write down distances from each $p\in M$ to its $k$ nearest neighbors $q>p$ in increasing order in a row of an $m\times k$-matrix.
Collapse any $c>1$ equal rows to one row with the weight $c/m$ in an extra first column.
The resulting matrix $\PSD(S;k)$ is the \emph{Pointwise Shift Distribution} and makes sense for any finite set $S=M\subset\R$ of $m\geq k+1$ unordered points.
\bs
\end{dfn}

$\PSD(S;k)$ differs from $\PDD(S;k)$ because we consider only neighbors $q$ to the right of a point $p$ in the line $\R$, so $\PSD$ consists of shifts (distances to the right).  

\begin{thm}[completeness for $n=1$]
\label{thm:PSD}
\textbf{(a)}
A finite set $M\subset\R$ of $m$ unordered points is reconstructable from $\PDD(M;m-1)$ uniquely under isometry.

\noindent
\textbf{(b)}
For all periodic sets $S\subset\R$ with $m$ points in a motif, $\PSD(S;m)$ is a complete invariant under rigid motion and can be computed in time $O(m^2)$. 
\end{thm}
\begin{proof}
\textbf{(a)}
For a finite set $S\subset\R$ of $m$ unordered points, we prove that $S$ can be reconstructed from $\PSD(S;m-1)$ uniquely under isometry.  
Indeed, the number $m$ can be assumed to be known as one plus the number of columns in $\PSD(S;m-1)$.
Find a row $R$ whose last distance $d$ is maximal in $\PSD(S;m-1)$.
This maximal distance is achieved exactly for two most distant points of $S$, else $\PSD(S;m-1)$ is unrealizable by $m$ distinct points.
These two most distant points can be fixed at the positions $0$ and $d$ up to isometry of $\R$. 
All other $m-2$ points of $S$ are uniquely determined by the first $m-2$ distances in the row $R$, which should be distinct.
\medskip

\noindent
\textbf{(b)}
The time to compute $\PSD(S;k)$ is linear in the size $m$ of a motif and in the number $k$ of neighbors.
Let $S$ have a motif $M$ of $m$ points $0=p_0<p_1<\dots<p_{m-1}<p_m$ and period $L=p_m-p_0$.
For any point $p_i\in M$, the distance to its $k$-th neighbor is $p_{i+k-mN}-p_i+LN$, where 
$N=[k/m]$ is the integer part and $p_j=p_{j-m}+L$ for $m\leq j<2m$.
So all $k$ neighbors of $p_i$ are computed in linear time in both $k,m$, hence the total time over $m$ points of $M$ is quadratic in $m$. 
\smallskip

Now we prove that any periodic point set $S\subset\R$ can be reconstructed (uniquely under translation) from any row $a_1<\dots<a_{m-1}<a_m$ of $\PSD(S;m)$ by writing the points of a motif as $p_k=a_{k+1}-a_1$ for $k=0,\dots,m-1$, where $p_0=0$, and setting the period of $S$ to $d_m$. 
The number $m$ is given as the number of columns of $\PSD(S;m)$.
The completeness can be stated as follows: any periodic sequences $S,Q\subset\R$ whose motifs have at most $m$ points are related by translation if and only if $\PSD(S;m)=\PSD(Q;m)$ as weighted distributions of unordered rows.
\qed
\end{proof}

The invariant $\PSD(S;k)$ can be enhanced to a complete invariant under isometry (including reflections) in $\R$ as follows.  
Let $\bar S$ be the mirror image of $S$ under reflection $x\mapsto -x$. 
In any row $a_1<\dots<a_k$ of $\PSD(S;k)$ for $k\geq m$, we can use the $m$-th distance $a_m$ equal to the period $L$ to write the corresponding row 
$$L-a_{m-1}<\dots<L-a_{1}<2L-a_{m-1}$$ 
in the matrix $\PSD(\bar S;k)$.
Then any periodic sequences $S,Q$ are related by isometry in $\R$ if and only if $\PSD(S;m)=\PSD(Q;m)$ or $\PSD(\bar S;m)=\PSD(Q;m)$.
\smallskip

Theorem~\ref{thm:PSD} with the Lipschitz continuity in Lemma~\ref{lem:PSD_continuous} will show that the $\PSD$ solves Problem~\ref{pro:invariants} for all periodic sets (under rigid motion) for $n=1$.
\smallskip

The generic completeness of $\PDD(S;k)$ (with a motif size $|S|$ and a lattice of $S$) in \cite[Theorem~4.4]{widdowson2022resolving} and Examples~\ref{exa:6-point_pairs},~\ref{exa:Pauling_homometric} motivate the following conjecture.

\begin{con}[completeness of $\PDDall$ under isometry in $\R^h$]
\label{con:PDDh_complete}
For $h\geq 1$, any periodic point set $S\subset\R^h$ can be reconstructed (uniquely under isometry) from the invariant $\PDDall(S;k)$ for a sufficiently large $k$ in Definition~\ref{dfn:PDDh}. 
\end{con}

\section{Lipschitz continuous metrics on higher-order invariants}
\label{sec:continuity}
 
This section introduces metrics on 
$\PDDh$ invariants and proves their Lipschitz continuity. 
Any vectors $u,v\in\R^m$ of distances or their average sums can be compared by the \emph{Minkowski} metric $L_q(u,v)=(\sum\limits_{i=1}^m|u_i-v_i|^q)^{1/q}$ for $q\in[1,+\infty)$ and $L_\infty(u,v)=\max\limits_{i=1,\dots,m}|u_i-v_i|$ in the limit case $q=+\infty$.
The \emph{Root Mean Square} metric $\RMS(u,v)=\dfrac{L_2(u,v)}{\sqrt{m}}$ is the Euclidean metric normalized by the (square root of the) number $m$ of coordinates.
These metrics $L_q$ and $\RMS$ controllably change under perturbations of distances and will play the role of a `ground' metric $d$ to compare unordered distributions $\PDDh$ by the EMD metric below.

\begin{dfn}[Earth Mover's Distance $\EMD$ {\cite{rubner2000earth}}]
\label{dfn:EMD}
\textbf{(a)}
Let $X$ be a space with a ground metric $d$.
Any unordered set $\{(R_i,w_i)\}_{i=1}^m$ of objects $R_i\in X$ with weights $w_i>0$ such that $\sum\limits_{i=1}^m w_i=1$ is called a (normalized) \emph{weighted} distribution. 
For any such weighted distributions $A=\{(R_i(A),w_i(A))\}_{i=1}^{m(A)}$ and $B=\{(R_j(B),w_j(B))\}_{j=1}^{m(B)}$, the \emph{Earth Mover's Distance} is defined as
$$\EMD(A,B)=\min\limits_{f_{ij}\in[0,1]} \sum\limits_{i=1}^{m(A)}\sum\limits_{j=1}^{m(B)} f_{ij} d(R_i(A),R_j(B))$$ 
subject to 
$\sum\limits_{i=1}^{m(A)}f_{ij}\leq w_j(B)$, $\sum\limits_{j=1}^{m(B)}f_{ij}\leq w_i(A)$, and $\sum\limits_{i=1}^{m(A)}\sum\limits_{j=1}^{m(B)} f_{ij}=1$.
\smallskip

\noindent
\textbf{(b)}
For any real $q\in[1,+\infty]$, any integers $n,h,k\geq 1$, and any periodic point sets $S,Q\subset\R^n$, the distance $\EMDh_q[k](S,Q)$ is the $\EMD$ from part (a) between the distributions $\PDDh(S;k)$ and $\PDDh(Q;k)$ with the ground metric $L_q$.
Define the distance 
$\EMDm_q[k](S,Q)=\max\limits_{i=1,\dots,h}\{\EMD^{\{i\}}_q[k](S,Q)\}$. 
The notation $\EMD$ without a subscript $q$ is used for the (default) ground metric $\RMS$ instead of $L_q$.
\bs
\end{dfn}

Our experiments use $\RMS$ or the Minkowski metric $L_\infty$, because these ground metrics will give the Lipschitz constant 2 for the $\EMD$ on $\PDDh$ in Theorem~\ref{thm:PDDh_continuous}.
Definition~\ref{dfn:EMD}(b) introduced the distance $\EMDm_q[k](S,Q)$ as the maximum of $\EMD$s over orders $1,\dots,h$ to keep the Lipschitz constant small.
This maximum can be replaced with a sum or another metric transform, see \cite[section~4.1]{deza2009encyclopedia}.

\begin{figure}[h!]
\includegraphics[height=40mm]{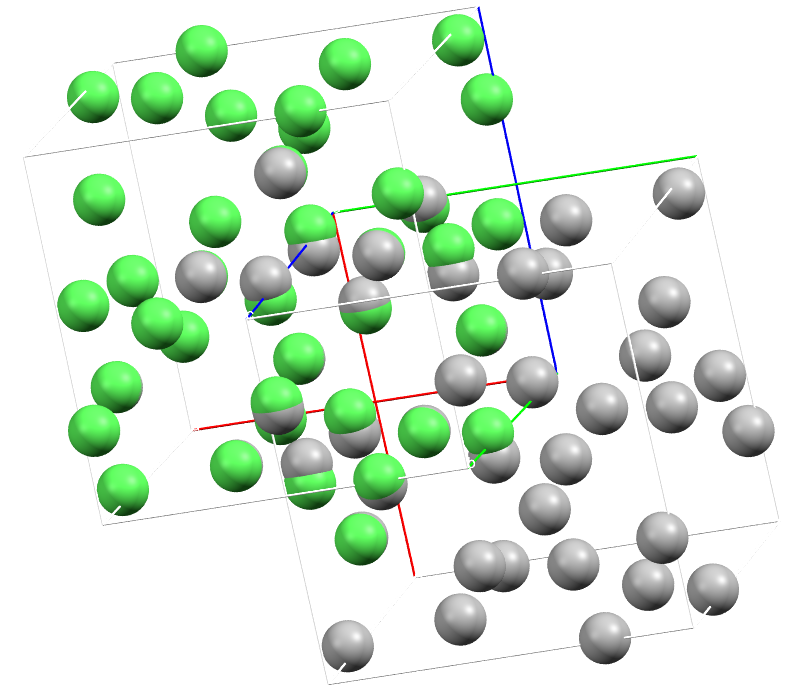}
\hspace*{2mm}
\includegraphics[height=40mm]{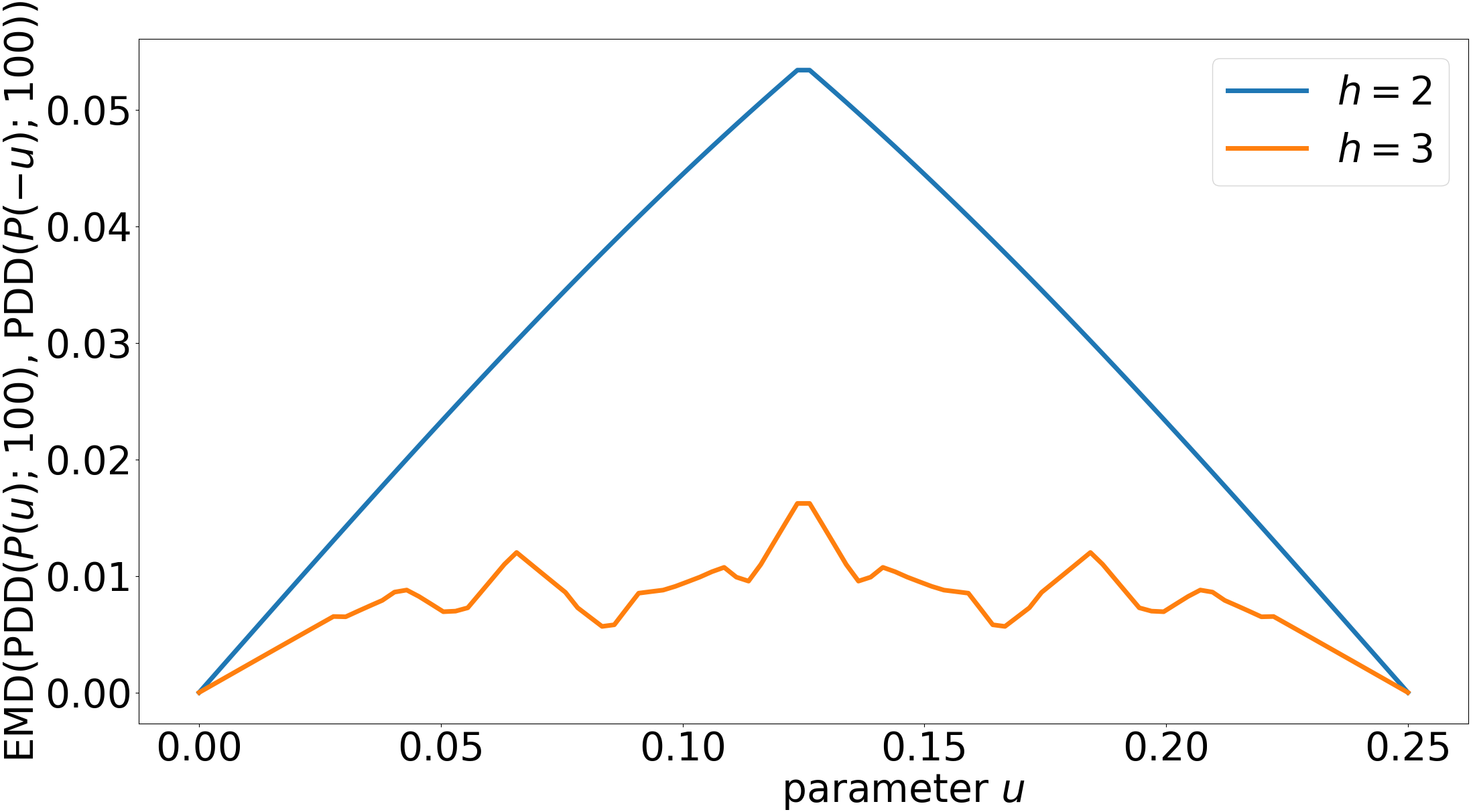}
\caption{\textbf{Left}: a comparison of Pauling's crystals $P(\pm u)$ for $u=0.03$ \cite{pauling1930crystal}, by COMPACK \cite{chisholm2005compack}, which aligns subsets of 15 atoms. 
The atoms from different $P(\pm 0.03)$ are shown in green and gray.
\textbf{Right}: $\EMD_\infty$ from Definition~\ref{dfn:EMD}(b) is between $\PDDh$ for $k=100$ and Pauling's crystals $P(\pm u)$, which depend on $u\in[0,0.25]$ and are identical at the boundary values.
}
\label{fig:Pauling} 
\end{figure}

\begin{exa}[$\PDDt$ distinguishes Pauling's crystals]
\label{exa:Pauling_homometric}
Fig.~\ref{fig:Pauling}~(left) shows a pair of overlaid Pauling's crystals $P(\pm 0.03)$ with 24 atoms in a cubic cell \cite{pauling1930crystal}. 
The importance of $\PDDt$ in comparison with $\PDD$ is demonstrated by the infinite series of periodic sets $P(\pm u)\subset\R^3$, which have the same $\PDD(P(u);k)=\PDD(P(-u);k)$ for all parameters $u\in(0,0.25)$ and $k\geq 1$ but are distinguished by $\PDDh(S;100)$ due to $\EMD_\infty^{\{h\}}[100]>0$ for $h=2,3$ in Fig.~\ref{fig:Pauling}~(right).
\bs
\end{exa}
 
For any discrete set $S\subset\R^n$, the \emph{packing radius} $r(S)$ is the minimum half-distance between any points of $S$.
Recall the brief notation from Definition~\ref{dfn:EMD}(b): 
$$\begin{array}{l}
\EMDh_q[k](S,Q) = \EMD_q\big(\, \PDDh(S;k), \, \PDDh(Q;k) \, \big).
\end{array}$$

\begin{thm}[Lipschitz continuity of $\PDDh$]
\label{thm:PDDh_continuous}
Fix integers $h,k\geq 1\leq l\leq n$.
Let $Q$ be a finite or an $l$-periodic point set obtained from a finite or an $l$-periodic point set $S\subset\R^n$, respectively, by perturbing every point of $S$ up to a Euclidean distance $\ep\in[0,r(S))$.
Then 
$\EMDh_q[k](S,Q)\leq 2\ep\sqrt[q]{k}$, where $\sqrt[q]{k}=1$ for $q=+\infty$, and
$\EMD\big(\, \PDDh(S;k), \, \PDDh(Q;k) \, \big)\leq 2\ep$, where the ground metric is $\RMS$. 
\end{thm}

\begin{figure}[h]
\centering
\includegraphics[width=0.48\linewidth]{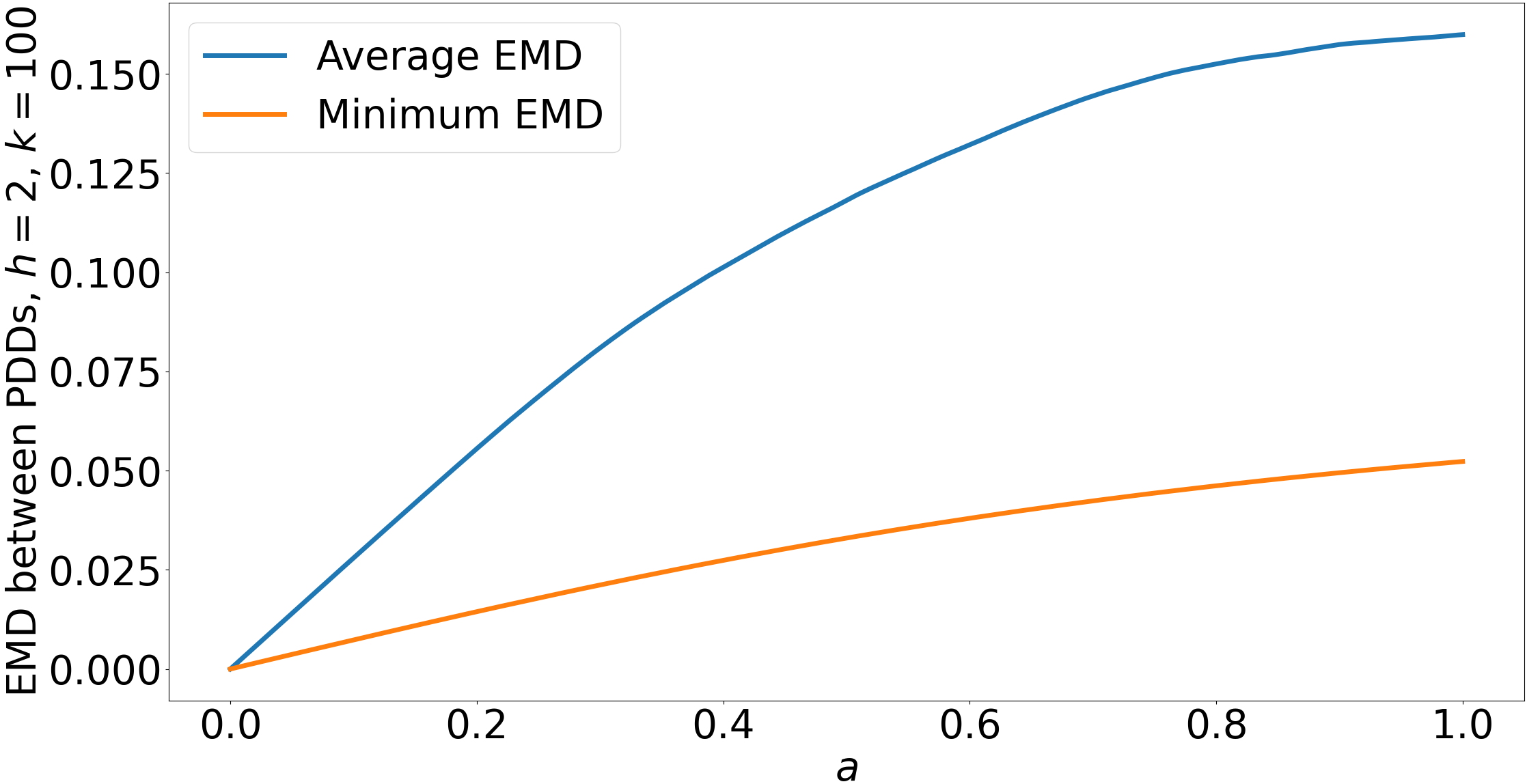}
\hspace*{1mm}
\includegraphics[width=0.48\linewidth]{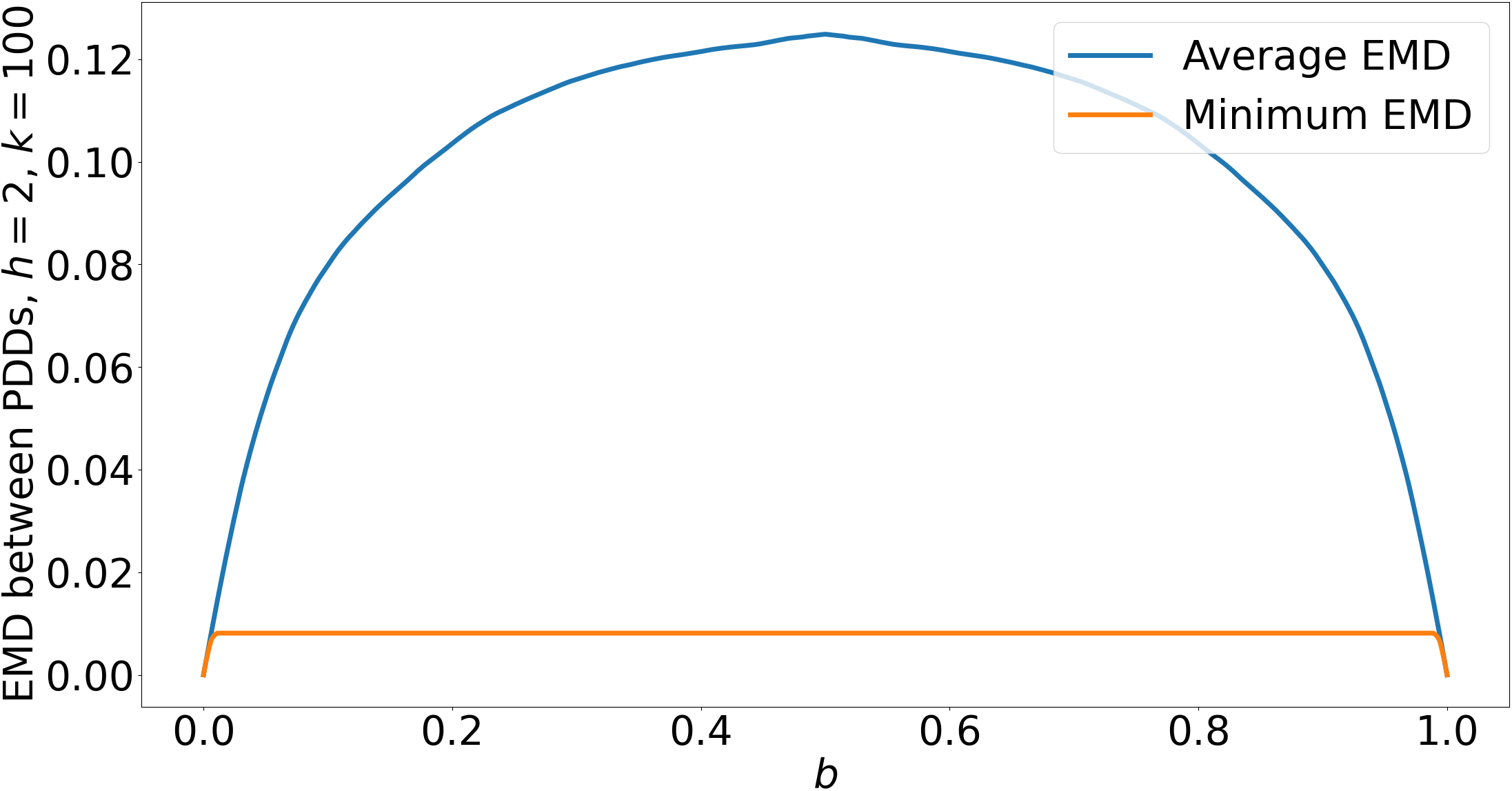}
\caption{The distance $\EMD_\infty^{\{2\}}[100]$ between the 1-periodic sets $S,Q$ in Fig.~\ref{fig:6-point_pairs}, which have identical $\PDD$s.
The average and minimum of $\EMD_\infty^{\{2\}}[100]$ were computed for uniformly sampled parameters $a,b,c$ from Example~\ref{exa:6-point_pairs}.
These sets $S,Q$ are isometric for $b\in\{0,1\}$ but $\EMD_\infty^{\{2\}}[100]>0$ for $0<b<1$ experimentally confirms that $S\not\simeq Q$, see Example~\ref{exa:6-point_pairs}.
}
\label{fig:6-point_pairs_EMD} 
\end{figure}

Fig.~\ref{fig:6-point_pairs_EMD} shows how $\EMD_\infty^{\{2\}}[100]$  continuously changes under perturbations.  

\begin{lem}[perturbation of an ordered vector]
\label{lem:perturb_list}
Let $v_1\leq\dots\leq v_k$ be a vector $v$ of ordered real numbers.
For some $\ep\geq 0$, let a map $g$ perturb each $a_i$ to $\ti v_i=g(v_i)$ so that $|v_i-\ti v_i|\leq\ep$ for $i=1,\dots,k$.
Let $\ti v$ be the vector obtained by putting $\ti v_1,\dots,\ti v_k$ in increasing order.
Then $L_\infty(v,\ti v)\leq\ep$.
\end{lem}
\begin{proof}
It suffices to prove that the $i$-th number $u_i$ in the ordered vector $\ti v$ is $\ep$-close to the $i$-th number $v_i$ in the original vector $v$, so $v_i-\ep\leq u_i\leq v_i+\ep$ for $i=1,\dots,k$.
Assume by contradiction that $u_i<v_i-\ep$.
Since every component of $v$ was perturbed by at most $\ep$, the $i$ numbers $u_1\leq\dots\leq u_i<v_i-\ep$ can be obtained only as perturbations of components from $v$ that are strictly less than $v_i$.
However, the ordered vector $A$ has at most $i-1$ numbers that are less $v_i$.
This contradiction proves that  $u_i\geq v_i-\ep$.
A similar argument proves that $u_i\leq v_i+\ep$.
\qed
\end{proof}

\begin{lem}[upper bound of $\EMD$]
\label{lem:EMD_upper_bound}
Consider any weighted distributions $A=\{(R_i(A),w_i)\}_{i=1}^m$ and $B=\{(R_i(B),w_i)\}_{i=1}^m$ of matched objects with equal weights and ground distances $d(R_i(A),R_i(B))\leq\ep$ for $i=1,\dots,m$.
Then $\EMD(A,B)\leq\ep$.
\end{lem}
\begin{proof}
Define the flows $f_{ij}$ from the $m$ objects of $A$ to the corresponding $m$ objects of $B$ by setting $f_{ii}=\dfrac{1}{m}$ and
$f_{ij}=0$ for $i\neq j$, $i,j=1,\dots,m$.
Then
$$\EMD(A,B)\leq\sum\limits_{i,j=1}^m f_{ij} d(R_i(S),R_j(Q))
=\dfrac{1}{m}\sum\limits_{i=1}^m d(R_i(S),R_i(Q))
\leq \dfrac{1}{m}\sum\limits_{i=1}^m \ep = \ep$$
since $\EMD(A,B)$ is the minimum over all $f_{ij}\in [0,1]$, see  Definition~\ref{dfn:EMD}(a).
\qed
\end{proof}

\textbf{Proof of Theorem}~\ref{thm:PDDh_continuous}. 
In the periodic case, if the perturbation satisfies $\ep<r(S)$, \cite[Lemma~4.1]{edelsbrunner2021density} and \cite[Lemma~4.8]{anosova2025recognition} proved that $S,Q$ have a common lattice with a unit cell $U$ such that $S=\La+(S\cap U)$ and $Q=\La+(Q\cap U)$.
Then $S,Q$ share a unit cell $U$ and have the same number $m=m(S)=m(Q)$ of points in $U$. 
The arguments below also work for any finite sets $S,Q$ in a large enough $U$. 
\smallskip

Expand $\PDDh$ of both $S,Q$ to the matrices with $m$ equally weighted rows.
Reorder all $m$ rows of 
$D(S,S\cap U;h,k)$ and $D(Q,Q\cap U;h,k)$ according to the bijection $g:S\cap U\to Q\cap U$.
Since any $p\in S$ is perturbed up to $\ep$, any distance $L_q(p,q)$ between $p,q\in S$ and hence any average sum $a$ from Definition~\ref{dfn:PDDh} changes by at most $2\ep$ due to the triangle inequality for the Minkowski metric $L_q$.
\smallskip
 
Some of the average sums from the original matrix $D(S, S\cap U;h,k)$ can increase up to $2\ep$ and will be outside the $k$ smallest average sums in the new matrix $D(S, S\cap U;h,k)$. 
In this case, for each row $i=1,\dots,m$, let $k_i\geq k$ be the maximum index such that the $k_i$-th smallest average sum (of pairwise distances between $h+1$ points including $p_i\in S$) for $S$ is at most $2\ep$ plus the largest average sum on $h+1$ points from the original matrix $D(S, S\cap U;h,k)$ in the $i$-th row.  
Set $k'=\max\limits_{i=1,\dots,m} k_i\geq k$.
Then the $i$-th row of $D(Q, Q\cap U;h,k')$ is obtained from the $i$-th row of $D(S, S\cap U;h,k')$ of $k'$ numbers by changing every value by at most $2\ep$, putting them in increasing order, and taking the first $k\leq k'$ smallest values.
\smallskip

For each $i=1,\dots,m$, Lemma~\ref{lem:perturb_list} implies that the corresponding components in the extended $i$-th rows of $k'$ numbers in $D(S, S\cap U;h,k')$ and $D(Q, Q\cap U;h,k')$ differ by at most $2\ep$. 
The same conclusion holds for the shorter $i$-th rows $R_{i}(S)$ and $R_{i}(Q)$ of $k$ values in the matrices $D(S, S\cap U;h,k)$ and $D(Q, Q\cap U;h,k)$, respectively.
Then $L_q(R_{i}(S),R_{i}(Q))\leq \sqrt[q]{k(2\ep)^q}=2\ep\sqrt[q]{k}$.
The Euclidean $L_2$ normalized with the factor $\frac{1}{\sqrt{k}}$ has the upper bound $\RMS(R_{i}(S),R_{i}(Q))\leq 2\ep$.
By Lemma~\ref{lem:EMD_upper_bound}, the distributions of rows $R_{i}(S)$ and $R_{i}(Q)$ have the same upper bound for their EMD metrics: $\EMD_q\big(\PDDh(S;k),\PDDh(Q;k)\big)\leq 2\ep\sqrt[q]{k}$ and $\EMD\leq 2\ep$. 
\qed
\smallskip

\begin{lem}[Lipschitz continuity of $\PSD$]
\label{lem:PSD_continuous}
For all finite or periodic sequences $S\subset\R$, for the ground metrics $\RMS$ and $L_q$, define the $\EMD$ and $\EMD_q$ respectively, on distributions $\PSD(S;k)$ for $k\geq 1$, see Definition~\ref{dfn:PSD}.
Let $Q\subset\R$ be a finite or periodic sequence obtained by perturbing every point of $S$ up to $\ep\in[0,r(S))$.
Then $\EMD_q\big(\PSD(S;k),\PSD(Q;k)\big)\leq 2\ep\sqrt[q]{k}$,
$\EMD\big(\PSD(S;k), \PSD(Q;k) \big)\leq 2\ep$. 
\end{lem}
\begin{proof}
Since $\ep$ is less than the packing radius $r(S)$, the given $\ep$-perturbation defines a bijection $g:S\to Q$, which changes inter-point distance by at most $2\ep$.
The bijection $g$ induces a 1-1 correspondence between rows $R_i(S)$ and $R_i(Q)$ of $\PSD(S;k)$ and $\PSD(Q;k)$, respectively with ground distances $L_q(R_{i}(S),R_{i}(Q))\leq 2\ep\sqrt[q]{k}$ and $\RMS(R_{i}(S),R_{i}(Q))\leq 2\ep$, which guarantee the required 
bounds 
by Lemma~\ref{lem:EMD_upper_bound}.
\end{proof}
\smallskip

Recall the brief notation of a maximum metric from Definition~\ref{dfn:EMD}(b): 
$$\begin{array}{l}
\EMDm_q[k](S,Q) = \max\limits_{i=1,\dots,h}\Big\{ \EMD_q \big(\, \PDD^{\{i\}}(S;k), \, \PDD^{\{i\}}(Q;k) \, \big) \Big\}.
\end{array}$$

\begin{lem}[lower bounds of $\EMD$]
\label{lem:EMD_PDDh_bounds}
Fix any real $q\in[1,+\infty]$ and integers $h,k\geq 1\leq l\leq n$.
Let $S,Q\subset\R^n$ be any finite or $l$-periodic point sets.
Then 
\medskip

\noindent
\textbf{(a)}
$\EMDm_q[k](S,Q)\geq \EMD_q\big(\, \PDD^{(g)}(S;k), \, \PDD^{(g)}(Q;k) \, \big)$
for $1\leq g\leq h$;
\medskip

\noindent
\textbf{(b)}
$\EMDh_q[k](S,Q)\geq \EMD_q\big(\, \PDDh(S;k'), \, \PDDh(Q;k') \, \big)$
for $1\leq k'\leq k$;
\medskip

\noindent
\textbf{(c)}
$\EMDh_q[k](S,Q)\geq 
L_q\big(\, \muo[\PDDh(S;k)], \, \muo[\PDDh(Q;k)] \, \big)$.
\medskip

\noindent
The same inequalities hold for the ground metric $\RMS$ instead of $L_q$.
\end{lem}
\begin{proof}
\textbf{(a)}
If the order $h$ drops to $g$, the maximum of a fewer number of distances cannot become larger by Definition~\ref{dfn:EMD}(b): $\EMDm_q[k](S,Q)\geq \EMD^{(g)}_q[k](S,Q)$.
\medskip

\noindent
\textbf{(b)}
Let $f_{ij}\in[0,1]$ be the parameters that minimize the EMD in Definition~\ref{dfn:EMD}(b):
$$\EMDh_q[k](S,Q)=\sum\limits_{i=1}^{m(S)}\sum\limits_{j=1}^{m(Q)} f_{ij} L_q(R_i(S),R_j(Q)),$$ where $R_i(S),R_j(Q)$ are rows in the distributions $\PDDh(S;k),\PDDh(Q;k)$, respectively. 
If $k$ drops to $k'$, the smaller distributions $\PDDh(S;k'),\PDDh(Q;k')$ are obtained from $\PDDh(S;k),\PDDh(Q;k)$ by removing the last $k-k'$ columns. 
The shortened rows $R'_i(S),R'_j(Q)$ of $k'\leq k$ components in the smaller distributions $\PDDh(S;k'),\PDDh(Q;k')$ satisfy $L_q(R_i(S),R_j(Q))\geq L_q(R'_i(S),R'_j(Q))$. 
Then
$$\begin{array}{l}
\EMDh_q[k](S,Q)\geq\sum\limits_{i=1}^{m(S)}\sum\limits_{j=1}^{m(Q)} f_{ij} L_q(R_i(S),R_j(S))\geq \\
\min\limits_{f_{ij}\in[0,1]} \sum\limits_{i=1}^{m(S)}\sum\limits_{j=1}^{m(Q)} f_{ij} L_q(R'_i(S),R'_j(Q))
=\EMDh_q[k'](S,Q).\end{array}$$

\noindent
\textbf{(c)}
Considering $\PDDh(S;k)$ as a weighted distribution of rows, $\muo[\PDDh(S;k)]$ is its centroid from \cite[section~3]{cohen1997earth}.
The argument below follows the proof for $q=+\infty$ of \cite[Theorem~1]{cohen1997earth}.
Below we use the inequality $||u||_q + ||v||_q|| \geq ||u+v||_q$ for the $q$-norm $||v||_q=\big(\sum\limits_{i=1} |v_i|^q \big)^{1/q}$ of the Minkowski metric $L_q$.
Let $f_{ij}\in[0,1]$ be the parameters that minimize the EMD in Definition~\ref{dfn:EMD}(b):
$$\begin{array}{l}
\EMD_q(\PDDh(S;k),\PDDh(Q;k))=
\sum\limits_{i=1}^{m(S)} \sum\limits_{j=1}^{m(Q)} f_{ij} L_q(R_i(S),R_j(Q))  =\\
\sum\limits_{i=1}^{m(S)} \sum\limits_{j=1}^{m(Q)} ||f_{ij} \big(R_i(S) -R_j(Q)\big)||_q \geq
||\sum\limits_{i=1}^{m(S)} \sum\limits_{j=1}^{m(Q)} f_{ij} (R_i(S) -R_j(Q)) ||_q = \\
|| \sum\limits_{i=1}^{m(S)} \big(\sum\limits_{j=1}^{m(Q)}
f_{ij} R_i(S) \big)-
\sum\limits_{j=1}^{m(Q)} \big(\sum\limits_{i=1}^{m(S)} f_{ij} R_j(Q) \big) ||_q = \\
|| \sum\limits_{i=1}^{m(S)} w_i(S) R_i(S) -
\sum\limits_{j=1}^{m(Q)} w_j(Q) R_j(Q) ||_q = \\
L_q\big( \, \muo[\PDDh(S;k)], \, \muo[\PDDh(Q;k)] \big).
\end{array}$$
All proofs are the same for the ground metric $\RMS=\dfrac{L_2}{\sqrt{k}}$ instead of $L_q$.
\qed
\end{proof}
\smallskip

Corollary~\ref{cor:AMD_continuous} extends the case $h=1$ from \cite[Theorem~9]{widdowson2022average}, where $\AMD(S;k)=\muo[\PDD(S;k)]$ is the vector of Average Minimum Distances, to any order $h\geq 1$.

\begin{cor}[Lipschitz continuity of {$\muo[\PDDh]$}]
\label{cor:AMD_continuous}
Fix integers $h,k\geq 1\leq l\leq n$.
Let $Q$ be a finite or an $l$-periodic set obtained from a finite or an $l$-periodic point set $S\subset\R^n$, respectively, by perturbing every point of $S$ up to a Euclidean distance $\ep\in[0,r(S))$.
Then 
$L_q\big( \, \muo[\PDDh(S;k)], \, \muo[\PDDh(Q;k)] \big)\leq 2\ep\sqrt[q]{k}$ and 
$\RMS\big( \, \muo[\PDDh(S;k)], \, \muo[\PDDh(Q;k)] \big)\leq 2\ep$. 
\end{cor}
\begin{proof}
The required bounds follow from Theorem~\ref{thm:PDDh_continuous} and Lemma~\ref{lem:EMD_PDDh_bounds}(c). 
\qed
\end{proof}

We conjecture that higher moments $\mut[\PDDh]$ for $t>1$ are continuous under perturbations of points, possibly in a weaker (than Lipschitz) sense.

\section{The asymptotic curves and computational complexity of $\PDDh$}
\label{sec:time}

To analyze the asymptotic of $\PDDh(S;k)$ as $k\to+\infty$, we choose a real
$b\geq h$ such that $\vect{b}{h}=\dfrac{b(b-1)\dots(b-h+1)}{h!}$ belongs to the interval $(k-1,k]$.
Then set $b(h,k)=b+1$ e.g. $b(1,k)=k+1$,  $b(2,k)=1.5+\sqrt{2k}$.
Let $V_n$ be the unit ball volume in $\R^n$, e.g. $V_2=\pi$.
Any periodic set $S\subset\R^n$ with a motif of $m$ points and unit cell of volume $\vol[U]$ has the \emph{point packing coefficient} $\PPC(S)=\sqrt[n]{\dfrac{\vol[U]}{mV_n}}$. 

\begin{thm}[asymptotic of {$\PDDh(S;k)$} as $k\to+\infty$] 
\label{thm:asymptotic}
Let a periodic point set $S\subset\R^n$ have a cell with a longest diagonal $d$. 
For any integers $h,k\geq 1$, let $a(h,k)$ be the average sum of the $k$-th column of $\PDDh(S;k)$ from Definition~\ref{dfn:PDDh}. 
$$\text{Then }
\dfrac{2}{h+1}\Big(\PPC(S)\sqrt[n]{b(h,k)}-d\Big)\leq a(h,k)\leq \dfrac{2h}{h+1}\Big(\PPC(S)\sqrt[n]{b(h,k)}+d\Big)$$ 
for any $k\geq 1$.
If $h=1$, then $\lim\limits_{k\to+\infty}\dfrac{a(1,k)}{\sqrt[n]{k}}=\PPC(S)$.
If $h=2$, then we have the bounds $\dfrac{2}{3}\PPC(S)\leq \dfrac{a(2,k)}{\sqrt[2n]{2k}}\leq\dfrac{4}{3}\PPC(S)$ for large enough $k$.
\bs
\end{thm}

Since any lattice $\La\subset\R^n$ has a single point in a motif, any Pointwise Distance Distribution $\PDDh(\La;k)$ is a single row of the $k$ numbers, which coincides with the vector $\muo[\PDDh(\La;k)]$ and can be visualized as a piecewise linear curve through $k$ points.
Fig.~\ref{fig:lattices2D} shows six 2D lattices illustrating the asymptotic behavior of $\PDDh$  for $h=2,3$ in Fig.~\ref{fig:lattices2Dasymptotic}.
Theorem~\ref{thm:asymptotic} supports the following conjecture. 

\begin{con}[$h$-order limit]
\label{con:h-limit}
In the notations of Theorem~\ref{thm:asymptotic}
for any periodic point set $S\subset\R^n$,  
$\lim\limits_{k\to+\infty}\dfrac{a(h,k)}{\sqrt[hn]{h!k}}$ exists for any $h\geq 2$.
If this limit differs from $\PPC(S)$, it can be called the $h$-order \emph{point packing coefficient} $\PPC(S;h)$.
\end{con}

\newcommand{\size}{15mm}
\begin{figure}[h!]
\centering
\includegraphics[height=\size]{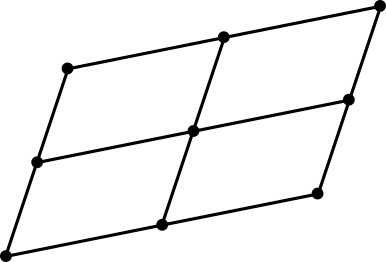}
\hspace*{0mm}
\includegraphics[height=\size]{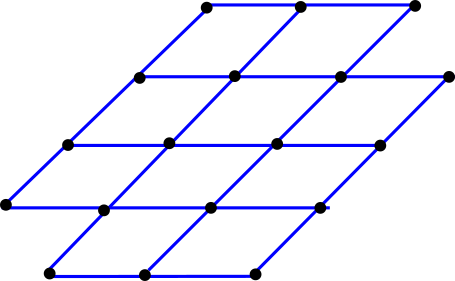}
\hspace*{0mm}
\includegraphics[height=\size]{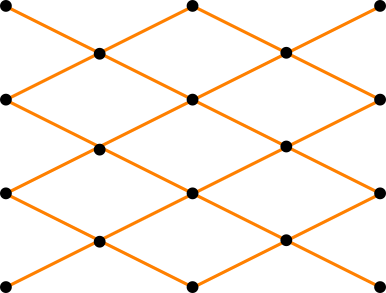}
\includegraphics[height=\size]{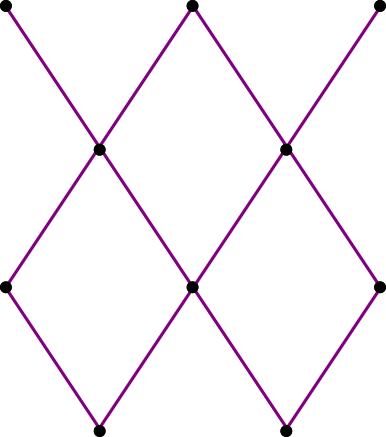}
\hspace*{0mm}
\includegraphics[height=\size]{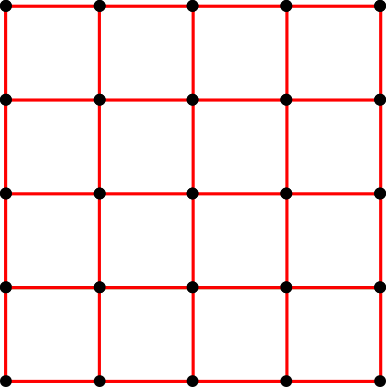}
\hspace*{0mm}
\includegraphics[height=\size]{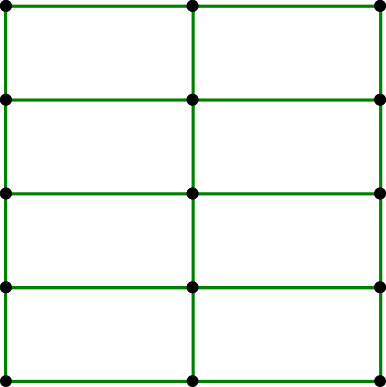}
\vspace*{-2mm}
\caption{The six 2-dimensional lattices whose invariants appeared in Fig.~\ref{fig:lattices2Dasymptotic}.
\textbf{1st}: a generic black lattice $\La_1$ with the basis $(1.25,0.25),(0.25,0.75)$ and $\PPC(\La_1)=\sqrt{\dfrac{7}{8\pi}}\approx 0.525$.
\textbf{2nd}: the blue hexagonal lattice $\La_2$ with the basis $(1,0),(1/2,\sqrt{3}/2)$ and $\PPC(\La_2)=\sqrt{\dfrac{\sqrt{3}}{2\pi}}\approx 0.528$.
\textbf{3rd}: the orange rhombic lattice $\La_3$ with the basis $(1,0.5),(1,-0.5)$ and $\PPC(\La_3)=\sqrt{\dfrac{1}{\pi}}\approx 0.564$.
\textbf{4th}: the purple rhombic lattice $\La_4$ with the basis $(1,1.5),(1,-1.5)$ and $\PPC(\La_4)=\sqrt{\dfrac{3}{\pi}}\approx 0.977$.
\textbf{5th}: the red square lattice $\La_5$ with the basis $(1,0),(0,1)$ and $\PPC(\La_5)=\sqrt{\dfrac{1}{\pi}}\approx 0.564$. 
\textbf{6th}: the green rectangular lattice $\La_6$ with the basis $(2,0),(0,1)$ and $\PPC(\La_6)=\sqrt{\dfrac{2}{\pi}}\approx 0.798$.
}
\label{fig:lattices2D}
\end{figure}

\begin{figure}[h!]
\centering
\includegraphics[width=\linewidth]{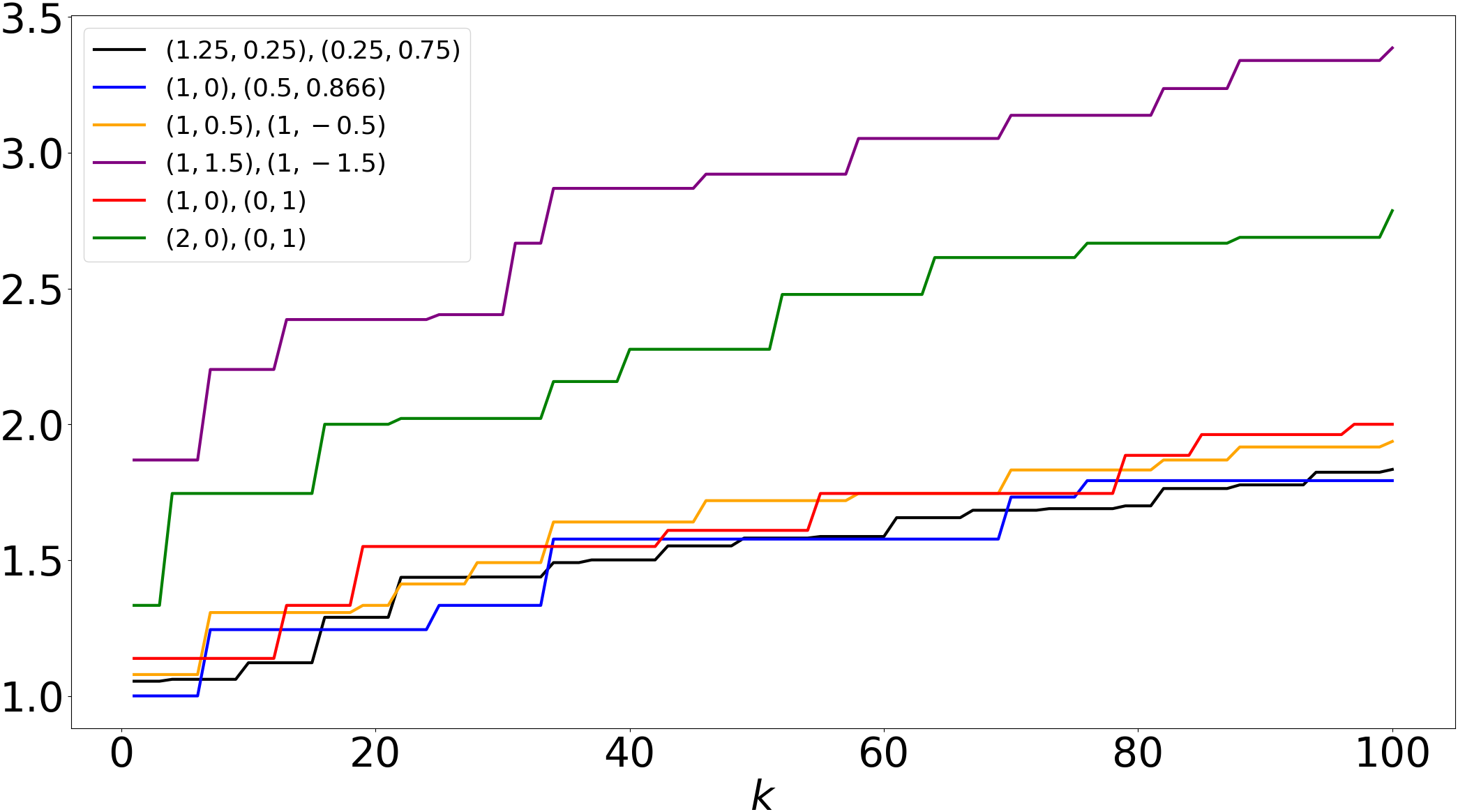}
\medskip

\includegraphics[width=\linewidth]{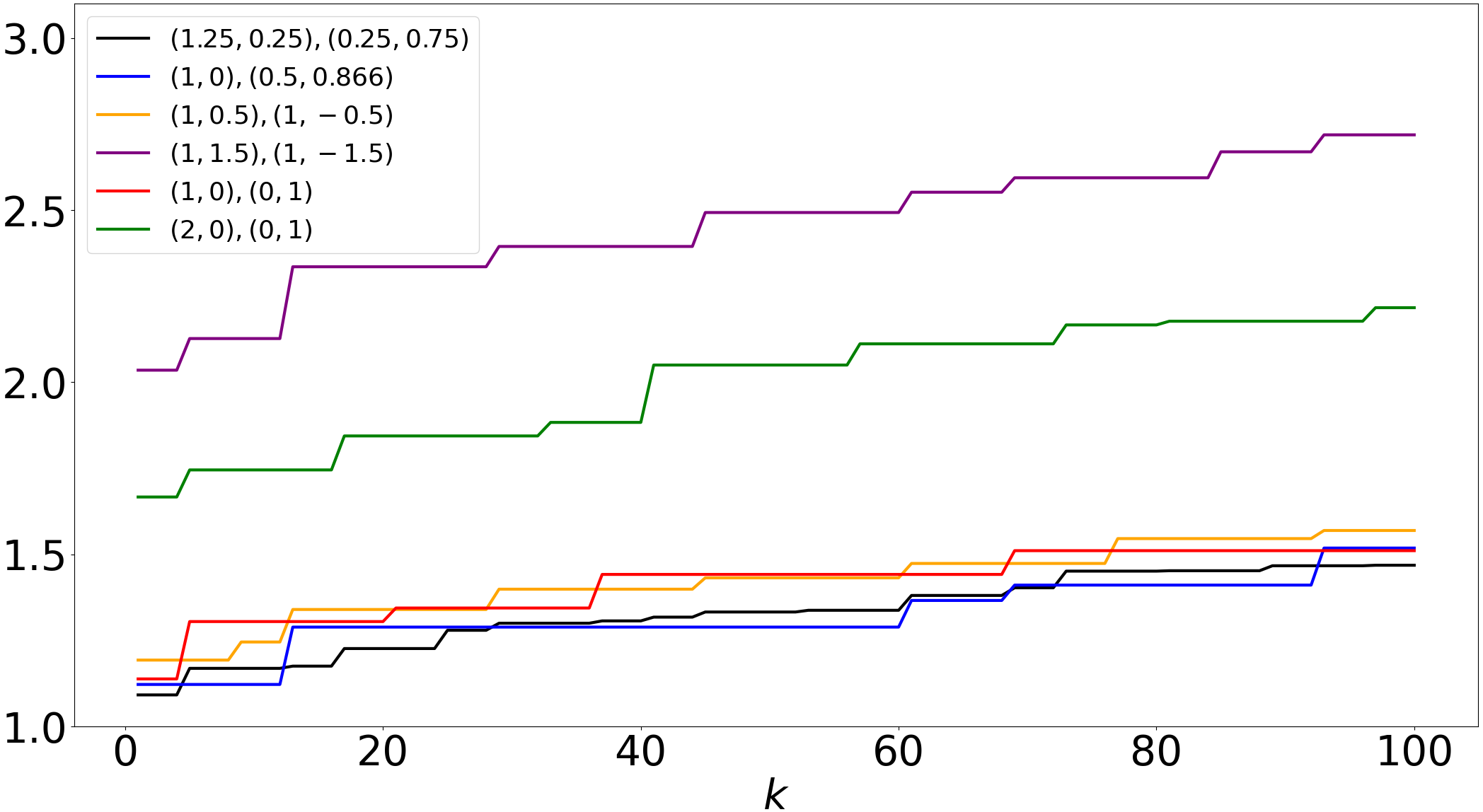}
\caption{The asymptotic behavior of the higher-order $\PDDt(\La;k)$ and $\PDD^{(3)}(\La;k)$ for the six lattices $\La\subset\R^2$ in Fig.~\ref{fig:lattices2D}, see their bases in the legends. 
\textbf{Top}: $h=2$. 
\textbf{Bottom}: $h=3$. 
}
\label{fig:lattices2Dasymptotic}
\end{figure}

Theorem~\ref{thm:asymptotic} 
justifies that there is no need to substantially increase the number $k$ of neighbors since $\PDDh(S;k)$ largely depends on $\PPC(S)$ when $k\to+\infty$. 
The practical advice is to choose $k$ depending on the size of a motif or constituent molecules so that all atoms have enough neighbors to capture the periodic 
connectivity.
We consider $k$ a degree of approximation similar to the number of decimal places on a calculator.
Theorem~\ref{thm:asymptotic} implies similar bounds for all $t$-moments and means that $\PDDh(S;k)$ and $\mut[\PDDh](S;k)$ are most discriminative for small values of $k$, so we used $k=100$, $t\leq 10$, and $h\leq 3$ in all experiments later.

\begin{lem}[distance bounds]
\label{lem:distance_bounds}
Let $S\subset\R^n$ be any periodic point set.
For any $h,k\geq 1$ and a point $p\in S$, let $a(h,k)$ be the $k$-th smallest average sum achieved for of all pairwise distances between $p$ and $h$ other points $p_1,\dots,p_h\in S$, see Definition~\ref{dfn:PDDh}.
Set $R=\max\limits_{i=1,\dots,h}|p_i-p|$.
Then $\dfrac{2R}{h+1}\leq a(h,k)\leq \dfrac{2hR}{h+1}$.  
\end{lem}
\begin{proof}
After translating $p\in S$ to the origin $0\in\R^n$, one can assume that $p=0$.
Let $p_1\in S$ be a point such that $R=|p_1|=\max\limits_{i=1,\dots,h}|p_i|$.
For any other point $p_i\neq p_1$, the triangle inequalities $|p_i|+|p_1-p_i|\geq |p_1|=R$ imply that
$$a(h,k)=\dfrac{2}{h(h+1)}\sum\limits_{0\leq i<j\leq h}|p_i-p_j|\geq$$ 
$$\geq\dfrac{2}{h(h+1)}\left(|p_1|+\sum\limits_{i=2}^h (|p_i|+|p_1-p_i|)\right)\geq\dfrac{2}{h(h+1)}\left(R+\sum\limits_{i=2}^h R\right)=\dfrac{2R}{h+1}.$$ 
For the upper bound of $a(h,k)$, we use $|p_i|\leq R$ and the triangle inequalities $|p_i-p_j|\leq|p_i|+|p_j|\leq 2R$ as follows:
$$a(h,k)=\dfrac{2}{h(h+1)}\left(\sum\limits_{i=1}^h |p_i|
+\sum\limits_{1\leq i<j\leq h} |p_i-p_j|\right)\leq$$
$$\leq \dfrac{2}{h(h+1)}\left(\sum\limits_{i=1}^h R + \sum\limits_{1\leq i<j\leq h} 2R\right)=\dfrac{2}{h(h+1)}\left( hR+ \dfrac{h(h-1)}{2}2R\right)=\dfrac{2hR}{h+1},$$ which finishes the proof of the upper bound.
\qed
\end{proof}

For $h=1$, the bounds of Lemma~\ref{lem:distance_bounds} give the exact equality $a(1,k)=R$.
Lemma~\ref{lem:ball_size} was proved in a slightly more general form in Lemma~11 from \cite{widdowson2022average}.

\begin{lem}[number of points in a ball]
\label{lem:ball_size}
Let $S\subset\R^n$ be any periodic point set with a unit cell $U$, which has $m$ points of $S$, generates a lattice $\La$, and has a longest diagonal of a length $d$.
For any point $p\in S\cap U$ and a radius $r$, 
consider 
$$U_-(p;r)=\bigcup\limits_{v\in\La} \{(U+v) \text{ such that } (U+v)\subset\bar B(p;r)\},$$ 
$$U_+(p;r)=\bigcup\limits_{v\in\La} \{(U+v) \text{ such that } (U+v)\cap\bar  B(p;r)\neq\emptyset\}.$$ Then
 the number of points of $S$ in the closed ball $\bar B(p;r)$ 
 has the bounds
$$\left(\dfrac{r-d}{\PPC(S)}\right)^n\leq
m\dfrac{\vol[U_-(p;r)]}{\vol[U]}\leq
|S\cap\bar B(p;r)|\leq 
m\dfrac{\vol[U_+(p;r)]}{\vol[U]}\leq
\left(\dfrac{r+d}{\PPC(S)}\right)^n.$$
\bs
\end{lem}
\smallskip

\begin{lem}[increasing binomial coefficient]
\label{lem:binomial}
For any fixed integer $h\geq 1$, the binomial coefficient $\vect{b}{h}=\dfrac{b(b-1)\dots(b-h+1)}{h!}$ is strictly increasing for any real $b\geq h$ so that if $h\leq b<c$ then $\vect{b}{h}<\vect{c}{h}$.
\end{lem}
\begin{proof}
The derivative $\dfrac{d}{d x}\vect{x}{h}>0$ for any $x\geq h$.
\end{proof}

\noindent
\textbf{Proof of Theorem}~\ref{thm:asymptotic}.
To prove the lower bound for the $k$-th smallest sum $a(h,k)$, set $r=\dfrac{h+1}{2}a(h,k)$.
For any point $p$ in a motif of $S$, consider the closed ball $\bar B(p;r)$ with the center $p$ and radius $r$.
By the lower bound of Lemma~\ref{lem:distance_bounds}, all points $p_1,\dots,p_h\in S$ that are used for computing $a(h,k)$ have the maximum distance $R=\max\limits_{i=1,\dots,h}|p_i-p|\leq \dfrac{h+1}{2}a(h,k)=r$ and hence belong to $\bar B(p;r)$.
\smallskip

By the upper bound of Lemma~\ref{lem:ball_size}, if this ball contains $c$ points of $S$ (excluding $p$), then $c+1\leq\left(\dfrac{r+d}{\PPC(S)}\right)^n$.
By using $p$ and any other $h$ distinct points $p_1,\dots,p_h$ among $c$ points in $S\cap\bar B(p;r)$, we can form $\vect{c}{h}=\dfrac{c(c-1)\dots(c-h+1)}{h!}$ tuples $p,p_1,\dots,p_h$ whose average sums of all pairwise distances 
should include all $k$ smallest values up to the $k$-th sum $a(h,k)$.
Hence $\vect{c}{h}\geq k$.
\smallskip

For $c\geq h=2$, the last inequality is $\dfrac{c(c-1)}{2}\geq k$, $c^2-c-2k\geq 0$, $c\geq\dfrac{1+\sqrt{1+8k}}{2}\geq 0.5+\sqrt{2k}$.
For any $h\geq 1$, let $b(h,k)=b+1$ satisfy $b\geq h$ and $\vect{b}{h}=\dfrac{b(b-1)\dots(b-h+1)}{h!}\in (k-1,k]$, e.g. one can set $b(2,k)=1.5+\sqrt{2k}$.
By Lemma~\ref{lem:binomial}, $\vect{c}{h}\geq k$ for $c\geq h$ implies that $c\geq b=b(h,k)-1$.
Then 
$$
\left(\dfrac{r+d}{\PPC(S)}\right)^n\geq c+1\geq b(h,k), \quad 
\dfrac{r+d}{\PPC(S)}\geq \sqrt[n]{b(h,k)},$$
$$\dfrac{h+1}{2}a(h,k)=r\geq \PPC(S)\sqrt[n]{b(h,k)}-d,\quad
a(h,k)\geq \dfrac{2}{h+1}\Big(\PPC(S)\sqrt[n]{b(h,k)}-d\Big).$$

To prove the upper bound for the $k$-th sum $a(h,k)$,
set $R=\dfrac{h+1}{2h}a(h,k)$ and consider any $r<R$.
By the upper bound of Lemma~\ref{lem:distance_bounds}, $p$ with any other $h$ points $p_1,\dots,p_h\in S\cap\bar B(p;r)$ have average sums that are at most $\dfrac{2hr}{h+1}<\dfrac{2hR}{h+1}=a(h,k)$, which is  less than the $k$-th smallest sum $a(h,k)$.
If the ball $\bar B(p;r)$ contains $c$ points of $S$ (excluding $p$), then these points can form at most $k-1$ tuples consisting of $p$ and $h$ (of $c$) other vertices, so $\vect{c}{h}\leq k-1$. 
Since $\vect{b}{h}=\dfrac{b(b-1)\dots(b-h+1)}{h!}\in (k-1,k]$ says that $\vect{b}{h}>k-1\geq \vect{c}{h}$, Lemma~\ref{lem:binomial} for $b=b(h,k)-1\geq h$ implies that $b>c$.
Lemma~\ref{lem:ball_size} gives
$$\left(\dfrac{r-d}{\PPC(S)}\right)^n\leq c+1<b+1=b(h,k), \quad
\dfrac{r-d}{\PPC(S)}<\sqrt[n]{b(h,k)}.$$
Since the resulting inequality $r<\PPC(S)\sqrt[n]{b(h,k)}+d$ holds for all
$r<R$, where $R=\dfrac{h+1}{2h}a(h,k)$ is fixed, we get
$\dfrac{h+1}{2h}a(h,k)=R\leq \PPC(S)\sqrt[n]{b(h,k)}+d$ and
$a(h,k)\leq \dfrac{2h}{h+1}\Big(\PPC(S)\sqrt[n]{b(h,k)}+d\Big).$
If $h=1$, both bounds have the same term: 
$$\PPC(S)\sqrt[n]{b(1,k)}-d\leq a(h,k)\leq \PPC(S)\sqrt[n]{b(1,k)}+d.$$
If we divide both sides of the last inequality by $\sqrt[n]{k}$, we get
$\lim\limits_{k\to+\infty}\dfrac{a(1,k)}{\sqrt[n]{k}}=\PPC(S)$.
We replaced $k+1$ with $k$ in $b(1,k)$, because $\lim\limits_{k\to+\infty}\dfrac{\sqrt[n]{k+1}}{\sqrt[n]{k}}=1$ for any fixed dimension $n$.
For similar reasons and $h=2$, the ratio $\dfrac{a(2,k)}{\sqrt[2n]{2k}}$ has the asymptotic bounds $\dfrac{2}{3}\PPC(S)$ and $\dfrac{4}{3}\PPC(S)$ as $k\to+\infty$.
\qed
\medskip

\begin{figure}[h!]
\centering
\includegraphics[width=\linewidth]{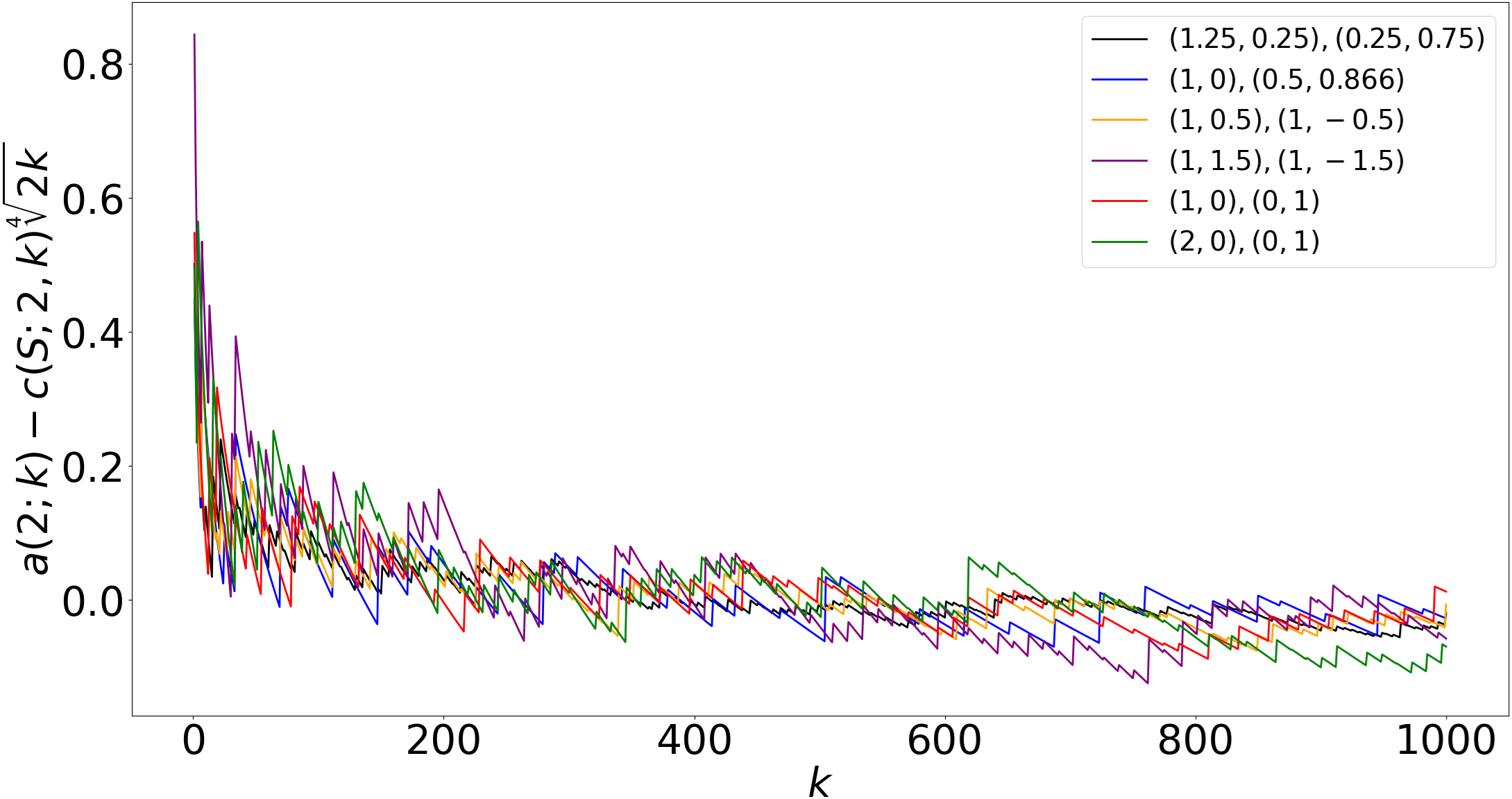}
\medskip

\includegraphics[width=\linewidth]{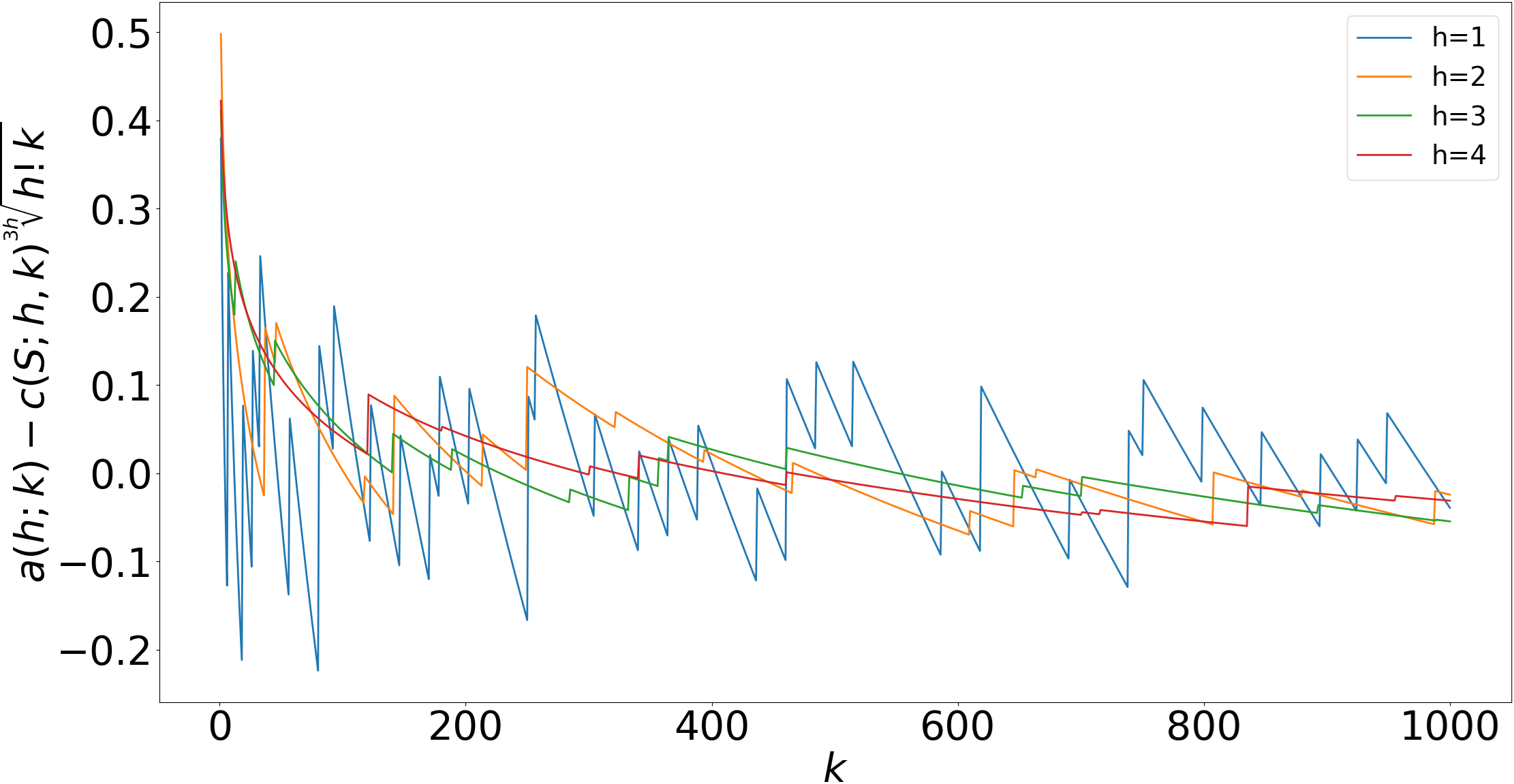}
\caption{
By Definition~\ref{dfn:PDAh}, any lattice $\La\subset\R^n$ has the vector $\ADAh(\La;k)$ consisting of the differences $a(h;k)-c(S;h,k)\sqrt[hn]{h!k}$, which converge to $0$ in the plots.
Here $a(h,k)$ is the $k$-th smallest average sum of pairwise distances from $0\in\La$ to $h$ other points in $\La$.
The coefficient $c(S;h,k)$ was experimentally fitted for $h\geq 2$ but should be independent of $k$ by Conjecture~\ref{con:h-limit}.
\textbf{Top}: six 2D lattices from Fig.~\ref{fig:lattices2D} and $h=2$. 
\textbf{Bottom}: cubic lattice $\Z^3$ and $h=1,2,3,4$. 
}
\label{fig:ADAh}
\end{figure}

Since the average sums $a(h,k)$ are increasing according to Theorem~\ref{thm:asymptotic} for $h=1,2$, comparing raw distances or sums for large $k$ is affected by deviating asymptotics.
To neutralize the effect of increasing deviations in $k$, \cite[Definition~3.7]{widdowson2021pointwise} adjusted $\PDD(S;k)$ by subtracting $\PPC(S)\sqrt[n]{k}$ from distances to $k$-th neighbors.
While Conjecture~\ref{con:h-limit} remains open for $h\geq 2$, supported by Fig.~\ref{fig:ADAh}, we find a coefficient $c$ minimizing the sum of squared deviations $E(c)=\sum\limits_{j=1}^k (a(h,k)- c\sqrt[hn]{h!k})^2$.
The polynomial $E(c)$ has the derivative $E'(c)=2\sum\limits_{j=1}^k \sqrt[hn]{h!k}(c\sqrt[hn]{h!k}-a(h,k))$ and a global minimum at $c=\dfrac{\sum_{j=1}^k a(h,j) \sqrt[hn]{h!j}}{\sum_{j=1}^k (\sqrt[hn]{h!j})^2}$.
Definition~\ref{dfn:PDAh} uses this $c$ to subtract from every row of $\PDDh$ the fitted sequence $C(j)=c\sqrt[hn]{h!j}$ for $j=1,\dots,k$. 

\begin{dfn}[Average/Pointwise Deviations from Asymptotic]
\label{dfn:PDAh}
\textbf{(a)}
Fix any integers $n,k\geq 1$ and $h\geq 2$.
For a finite or periodic point set $S\subset\R^n$ and a point $p\in S$, let $a(h,1)\leq\dots\leq a(h,k)$ be the $k$ column averages of the higher-order distribution $\PDDh(S;k)$, 
considered a matrix of $m$ unordered rows. 
Set $c(S;h,k)=\dfrac{\sum_{j=1}^k a(h,j) \sqrt[hn]{h!j}}{\sum_{j=1}^k (\sqrt[hn]{h!j})^2}$.
Let $A(S;h,k)$ denote the matrix of $m$ identical rows, each consisting of the $k$ ordered elements $c(S;h,k)\sqrt[hn]{h!j}$ for $j=1,\dots,k$.
\smallskip

\noindent
\textbf{(b)}
The \emph{Pointwise Deviation from Asymptotic} 
$\PDAh(S;k)=\PDDh(S;k)-A(S;h,k)$ is a distribution of unordered rows with the same weights as $\PDDh(S;k)$.
The $t$-moments from Definition~\ref{dfn:moments} give the $t\times k$ matrix $\mut[\PDAh(S;k)]$ of $m$ ordered rows, which can be flattened to a vector of $tk$ coordinates.
The \emph{Average Deviation from Asymptotic} is the vector $\ADAh(S;k)=\muo[\PDAh(S;k)]$ consisting of $k$ column averages (counted with weights) of the $m\times k$ matrix $\PDAh(S;k)$.
\bs
\end{dfn} 

Definition~\ref{dfn:PDAh} for $h=1$ uses the Point Packing Coefficient  $c(S;1,k)=\PPC(S)$, which depends only on $S$ (independent of $k$), so $\PDA^{\{1\}}(S;k)$ coincides with the previously defined $\PDA(S;k)$ in \cite[Definition~3.7]{widdowson2021pointwise}.
We adapt $\EMD$ from Definition~\ref{dfn:EMD}(a) to 
$\PDAh(S;k)$ with the ground metric $L_q$ on rows below. 

\begin{dfn}[EMD for $\PDAh$, $\PDA^{(h)}$ and Local Novelty Distances] 
\label{dfn:LND}
\textbf{(a)}
For any real $q\in[1,+\infty]$, any integers $n,h,k\geq 1$, and any periodic point sets $S,Q\subset\R^n$, Definition~\ref{dfn:EMD}(a) introduces the distances $\EMD,\,\EMD_q$ between $\PDAh(S;k)$ and $\PDAh(Q;k)$ with the ground metrics $\RMS,\,L_q$, respectively.
\smallskip

\noindent
\textbf{(b)}
The joint invariant $\PDA^{(h)}(S;k)$ is obtained by concatenating $\PDA^{\{i\}}(S;k)$ for $i=1,\dots,h$.
Define the max metric between $\PDA^{(h)}(S;k)$ and $\PDA^{(h)}(Q;k)$ as the maximum of all distances $\EMD_q\big(\PDA^{\{i\}}(S;k),\PDA^{\{i\}}(Q;k)\big)$ for $i=1,\dots,h$, similarly for the $\EMD$ based on the ground distance $\RMS$ instead of $L_q$.
\smallskip

\noindent
\textbf{(c)}
Fix an invariant distribution $I$ with a metric $d$, e.g. $I(S)=\PDAh(S;k)$ and $d=\EMD$ for all periodic point sets $S\subset\R^n$. 
Given a finite dataset $D$ of periodic sets, the $[I,d]$-based \emph{Local Novelty Distance} $\LND[I,d](S;D)=\min\limits_{Q\in D}d(I(S),I(Q))$ is the shortest distance from $S$ to some $Q\in D$ in the metric $d$ on values of $I$.
\end{dfn}

Lemma~\ref{lem:EMD_PDAh_bounds} justifies computations 
with smaller $h,k$ to filter out distant crystals. 

\begin{lem}[bounds for metrics on $\PDAh$]
\label{lem:EMD_PDAh_bounds}
Let $S,Q\subset\R^n$ be any periodic point sets.
Fix any real $q\in[1,+\infty]$ and any integers $1\leq g\leq h$, $1\leq k'\leq k$.
Then
$\EMD_q\big( \PDA^{(h)}(S;k), \PDA^{(h)}(Q;k)  \big) \geq 
\EMD_q\big( \PDA^{(g)}(S;k'), \PDA^{(g)}(Q;k') \big).$
The same inequality holds for $\EMD$ with the ground metric $\RMS$ instead of $L_q$.
\end{lem}
\begin{proof}
follows similar to 
Lemma~\ref{lem:EMD_PDDh_bounds}ab after replacing $\PDDh$ with $\PDAh$.
\qed
\end{proof}
\smallskip

\begin{cor}[Lipschitz continuity of $\PDAh$]
\label{cor:PDAh_continuous}
Fix any integers $n,k,h\geq 1$ and $q\in[1,+\infty]$.
Let $Q$ be a periodic point set obtained from a periodic point set $S\subset\R^n$ by perturbing every point of $S$ up to a Euclidean distance $\ep\in[0,r(S))$.
\smallskip

\noindent
\textbf{(a)}
We have that
$\big|c(S;h,k)-c(Q;h,k)\big|\leq 2\rho_k\ep$
for $\rho_k=\dfrac{\sum_{j=1}^k \sqrt[hn]{h!j}}{\sum_{j=1}^k (\sqrt[hn]{h!j})^2}$.
If $h=1$, then  $c(S;1,k)=\PPC(S)$ and $c(Q;1,k)=\PPC(Q)$ are equal, so we set $\rho_1=0$.
\medskip

\noindent
\textbf{(b)}
Let $h=1$.
Then the following identities hold: 
\smallskip

\noindent
$\EMD(\PDA(S;k), \PDA(Q;k))=\EMD(\PDD(S;k), \PDD(Q;k))$,
\smallskip

\noindent
$\EMD_q(\PDA(S;k), \PDA(Q;k))=\EMD_q(\PDD(S;k), \PDD(Q;k))$, 
\smallskip

\noindent
$\RMS(\ADA(S;k), \ADA(Q;k))=\RMS(\AMD(S;k), \AMD(Q;k))$,
\smallskip

\noindent
$L_q(\ADA(S;k), \ADA(Q;k))=L_q(\AMD(S;k), \AMD(Q;k))$.
\smallskip

\noindent
Under given $\ep$-perturbations, the Lipschitz constants of the metrics $\EMD$, $\EMD_q$, $\RMS$, $L_q$ above are $2,2\sqrt[q]{k},2,2\sqrt[q]{k}$, respectively, for any parameter $q\in[1,+\infty]$.
\medskip

\noindent
\textbf{(c)}
For $h\geq 2$, the upper bounds
$\EMD_q(\PDAh(S;k), \PDAh(Q;k))\leq 4\ep\sqrt[q]{k}$ and
$\EMD(\PDAh(S;k), \PDAh(Q;k))\leq 4\ep$ hold for the ground metric $\RMS$. 
\medskip

\noindent
\textbf{(d)}
To get a known crystal $Q\in D$ from a new crystal $S$, some atom of $S$ should be perturbed by at least $0.5\LND(S;D)$ for $\LND$ with the ground metric $\EMD_\infty$.  
\end{cor}
\begin{proof}
\textbf{(a)}
\cite[Lemma~4.1]{edelsbrunner2021density} proved that $S,Q$ have a common lattice with a unit cell $U$ such that $S=\La+(S\cap U)$ and $Q=\La+(Q\cap U)$.
Then $S,Q$ share a unit cell $U$ and have the same number $m=m(S)=m(Q)$ of points in $U$, so $\PPC(S)=\PPC(Q)$, which proves the case $h=1$.
For $h\geq 2$, by Definition~\ref{dfn:PDAh}(a), we estimate the difference
$\big|c(S;h,k)-c(Q;h,k)\big|\leq
\dfrac{\sum_{j=1}^k \big|a(S;h,j)-a(Q;h,j)\big|\sqrt[hn]{h!j}}{\sum_{j=1}^k (\sqrt[hn]{h!j})^2}$.
Since every point of $S$ is obtained as an $\ep$-perturbation of a point of $Q$, there is a bijection $S\to Q$ that shifts every point by at most $\ep$.
This bijection induces a 1-1 map between pairwise distances in $S,Q$, which changes every distance up to $2\ep$.
\smallskip

By Lemma~\ref{lem:perturb_list}, after writing the $k$ smallest $2\ep$-perturbed average sums in increasing order in every row of $\PDDh(S;k)$, the corresponding ordered sums still differ by at most $2\ep$, so $\big|\PDDh_{i,j}(S;k) - \PDDh_{i',j}(Q;k) \big|\leq 2\ep$.
Then the column averages $a(h,j)$ from Definition~\ref{dfn:PDAh}(a) also differ by at most $2\ep$.
\smallskip

Finally, $|a(S;h,j)-a(Q;h,j)|\leq 2\ep$ gives the required upper bound 
$$\big|c(S;h,k)-c(Q;h,k)\big|\leq\dfrac{\sum_{j=1}^k 2\ep \sqrt[hn]{h!j}}{\sum_{j=1}^k (\sqrt[hn]{h!j})^2}=2\rho_k\ep \text{ for } \rho_k=\dfrac{\sum_{j=1}^k \sqrt[hn]{h!j}}{\sum_{j=1}^k (\sqrt[hn]{h!j})^2}.$$

\noindent
\textbf{(b)}
For $h=1$, part (a) proved that $\PPC(S)=\PPC(Q)$.
By Definition~\ref{dfn:PDAh}, the matrices $\PDA(S;k),\PDA(Q;k)$ are obtained from $\PDD(S;k),\PDD(Q;k)$, respectively, by subtracting the same vector consisting of $\PPC(S)\sqrt[n]{j}$, $j=1,\dots,k$.
Then any $\RMS$ or $L_q$ distance between a row in $\PDD(S;k)$ and a row in $\PDD(Q;k)$ has the same value as between the corresponding rows in $\PDA(S;k)$ and $\PDA(Q;k)$.
\smallskip

The minimisation in Definition~\ref{dfn:EMD} gives the same values $\EMD$ and $\EMD_q$ when $\PDD$ is replaced with $\PDA$.
The same argument proves that $\RMS$ and $L_q$ remain the same when $\AMD$ is replaced with $\ADA$.
Hence the Lipschitz constants are the same as in Theorem~\ref{thm:PDDh_continuous} and Corollary~\ref{cor:AMD_continuous} restricted to order $h=1$.
\medskip

\noindent
\textbf{(c)}
By Definition~\ref{dfn:PDAh}, any element $\PDAh_{i,j}(S;k)$ in a row $i$, and a column $j$ of $\PDAh(S;k)$ equals $\PDDh_{i,j}(S;k)-c(S;h,k)\sqrt[hn]{h!j}$.
Estimate the difference of $i$-th elements from the same column $j$ in 
$\PDAh(S;k)$ and $\PDAh(Q;k)$.
$$\begin{array}{ll}
\De & = \big|\PDAh_{i,j}(S;k) - \PDAh_{i,j}(Q;k) \big| = \\
& = \big| \big(\PDDh_{i,j}(S;k)-c(S;h,k)\sqrt[hn]{h!j} \big) -
 \big(\PDDh_{i,j}(Q;k)-c(Q;h,k)\sqrt[hn]{h!j} \big) \big| \\
 & = \big| \big(\PDDh_{i,j}(S;k) - \PDDh_{i,j}(Q;k) \big)
 - \big( c(S;h,k)-c(Q;h,k)\big)\sqrt[hn]{h!j}\big| \leq \\
 & \leq \big|\PDDh_{i,j}(S;k) - \PDDh_{i,j}(Q;k) \big| +
  \big|c(S;h,k)-c(Q;h,k)\big|\sqrt[hn]{h!j} \leq \\
 & \leq 2\ep + 2\rho_k \ep \sqrt[hn]{h!j} = 2(1+\rho_k \sqrt[hn]{h!j})\ep\leq 4\ep,  
 \end{array}$$
where we used the upper bounds from part (a) and also 
 $\rho_k \sqrt[hn]{h!j}\leq 1$ for any $j=1,\dots,k$.
Let $R_i(S),R_i(Q)$ denote the $i$-th rows of $\PDAh(S;k)$, $\PDAh(Q;k)$, respectively. 
Then $L_q(R_i(S),R_i(Q))\leq \sqrt[q]{k(4\ep)^q}=4\ep\sqrt[q]{k}$.
The same proof for $\RMS=\dfrac{L_2}{\sqrt{k}}$ multiplies the Lipschitz constant by the factor $\dfrac{1}{\sqrt{k}}$.
Lemma~\ref{lem:EMD_upper_bound} guarantees the same upper bounds $4\ep\sqrt[q]{k}$ and $4\ep$ for $\EMD$ and $\EMD_q$, respectively.
If $h=1$, then $\rho_1=0$ by part  (a), so $4\ep$ can be replaced with $2\ep$.
\medskip

\noindent
\textbf{(d)}
Assume the contrary that $Q$ can be obtained from $S$ by perturbing every atom of $S$ by at most $\ep=0.5\LND(S;D)=\min\limits_{Q\in D}\EMD_\infty(\PDAh(S;k),\PDAh(Q;k))$.  
Part (c) for $q=+\infty$ implies that $\EMD_\infty(\PDAh(S;k),\PDAh(Q;k))\leq 2\ep=\LND(S;D)$, which contradicts the assumption and hence proves the lemma.
\qed
\end{proof}

\begin{thm}[time of $\PDDh$]
\label{thm:PDDh_time}
For any $h,k\geq 1$ and a periodic point set $S\subset\R^n$ with a motif of $m$ points and a unit cell $U$ with a longest diagonal $d$,
let 
$$a=\max\left\{h\Big(1+\dfrac{2.5d}{\PPC(S)}\Big),\sqrt[h]{16}\right\},
b=\log(2h!)+\log(\PPC(S)+d)-\log r(S),$$ 
where $r(S)$ is the packing radius of $S$.
Then $\PDDh(S;k)$ is computable in time
$$O\big(2^{8n} a^n\sqrt[h]{h!k}(b+\log k) +2^{12n}m(\log k)\log(h!k) + a^{hn}mk\log k\big).$$
\end{thm}
\begin{proof}
Fix the origin $0\in\R^n$ at the center of the unit cell $U$.
Then any point $p\in M=S\cap U$ is covered by the closed ball $\bar B(0,0.5d)$.
By Theorem~\ref{thm:asymptotic}, the distance $a(1,k)$ from any point $p\in M$ to its $k$-th nearest neighbor in $S$ has the upper bound $a(1,k)\leq \PPC(S)\sqrt[n]{k+1}+d$.
Then all $k$ neighbors of $p$ in $S$ are covered by the single ball $\bar B(0;r)$ of the radius $r=\PPC(S)\sqrt[n]{k+1}+1.5d$.
\smallskip

For a fixed point $p$ and any $h>1$, to find a similar ball including all points that are needed to compute the $k$ smallest average sums $a(h,1)\leq\dots\leq a(h,k)$, we start from the integer number $c=\lceil b(h,k)-1 \rceil$ of closest neighbors $p_1,\dots,p_c$ of $p$, where $b(h,k)$ is any real $b+1$ such that $b\geq h$ and $\vect{b}{h}\in(k-1,k]$.
Then $\vect{c}{h}\geq k$ by Lemma~\ref{lem:binomial}.
Since the $c+1$ points $p,p_1,\dots,p_c$ are covered by the ball $\bar B(p;R)$ of the radius $R=\max\limits_{i=1,\dots,c}|p_i-p|$, the lower bound of Lemma~\ref{lem:ball_size} gives 
$\left(\dfrac{R-d}{\PPC(S)}\right)^n\leq c+1\leq\ga$, where we set $\ga=b(h,k)+1$, so $R\leq \PPC(S)\sqrt[n]{\ga}+d$.
\smallskip

All $\vect{c}{h}\geq k$ average sums of pairwise distances between $p$ and any $h$ of $c$ points from $S\cap\bar B(p;R)$ have the upper bound $\dfrac{2hR}{h+1}$ by Lemma~\ref{lem:distance_bounds}.
If the $k$ smallest values of these sums are not greater than $\dfrac{2R}{h+1}$, which clearly holds for $h=1$, these $k$ smallest values form the required row $a(h,1)\leq\dots\leq a(h,k)$ of the point $p=p_0$ in $\PDDh(S;k)$.
Indeed, in this case for any $h$ points $p_1,\dots,p_h\in S$ with at least one distance (say) $|p_h-p_0|>R$, the lower bound of Lemma~\ref{lem:distance_bounds} implies that the average sum $\dfrac{2}{h(h+1)}\sum\limits_{0\leq i<j\leq h}|p_i-p_j|>\dfrac{2R}{h+1}$ cannot be among the sought after $k$ smallest values $a(h,1)\leq\dots\leq a(h,k)$. 
If we could not find $k$ smallest sums up to $\dfrac{2R}{h+1}$, we extend the radius $R$ to $hR$.
\smallskip

Similar to the above argument for the smaller radius $R$, the lower bound of Lemma~\ref{lem:distance_bounds} guarantees than any average sum involving at least one point at a distance $|p_h-p_0|>hR$ is greater than $\dfrac{2hR}{h+1}$ and hence cannot be among $k\leq \vect{c}{h}$ smallest sums that were already considered for the smaller ball $\bar B(p;R)$.
So the larger ball $\bar B(p;hR)$ is guaranteed to contain the required $k$ smallest sums.
\smallskip

To cover the necessary neighbors of all points $p$ from a motif $M=S\cap U$, we further increase the radius $hR$ by $0.5d$ and will use the earlier upper bound $R\leq \PPC(S)\sqrt[n]{b}+d$ for $\ga=b(h,k)+1\geq 1$.
Let the ball $\bar B(p;hR+0.5d)$ contain $\nu$ points of $S$, including its center $p$.
The upper bound $\nu\leq\left(\dfrac{hR+1.5d}{\PPC(S)}\right)^n$ from Lemma~\ref{lem:ball_size} and the earlier upper bound $R\leq \PPC(S)\sqrt[n]{\ga}+d$ imply that 
$$\begin{array}{l}
\nu\leq\left(\dfrac{hR+1.5d}{\PPC(S)}\right)^n
\leq\left( h\sqrt[n]{\ga}+\dfrac{(h+1.5)d}{\PPC(S)}\right)^n
=h^n\ga\left(1+\dfrac{(1+1.5/h)d}{\PPC(S)\sqrt[n]{\ga}}\right)^n\leq \\
\leq h^n \ga\left(1+\dfrac{2.5d}{\PPC(S)}\right)^n 
\leq a^n \ga \text{ for } a=\max\left\{h\Big(1+\dfrac{2.5d}{\PPC(S)}\Big),\sqrt[h]{16}\right\}.
\end{array}$$

To find $\nu$ nearest neighbors of all $m$ points $p$ from the motif $M=S\cap U$, we gradually extend the cell $U$ in spherical layers by adding shifted copies of $U$ until we get the upper union of shifted unit cells from Lemma~\ref{lem:ball_size}:
$$U_+=U_+(0;\PPC(S)h\sqrt[n]{\ga}+1.5d)\supset\bar B(0;hR+0.5d).$$

To estimate the neighbor search time \cite{elkin2022counterexamples}, we build a compressed cover tree on $\nu$ points of $U_+$ in time $O(\nu c_{\min}^{8}\log\frac{2R}{d_{\min}})$ by \cite[Theorem~3.7]{elkin2023new}, where $c_{\min}\leq 2^n$ is the minimized expansion constant of $T$, and $\dfrac{R}{r(S)}$ is the upper bound for the ratio of max/min inter-point distances.
Recall that $\ga=b(h,k)+1=b+2$ and $\vect{b}{h}\in(k-1,k]$.
If $h=1$, then $\ga=k+2=O(k)$.
For any $h\geq 2$, we have $\vect{b}{h}=\dfrac{O(b^h)}{h!}\leq k$. The rough upper bounds are $\ga\leq O(\sqrt[h]{h!k})$ and $\log\ga\leq O(\log(h!k))$ for any fixed $h$ and $k\to+\infty$.
Then $R\leq\PPC(S)\sqrt[n]{\ga}+d$ gives
$$\log (2R)\leq \log(\sqrt[n]{\ga}(2\PPC(S)+2d))=\log(2\PPC(S)+2d)+\log\ga.$$
Then
$O\left(\log\dfrac{R}{r(S)}\right)\leq b+\log k$ for 
$b=\log(2h!)+\log(\PPC(S)+d)-\log r(S)$.
\smallskip

Then the time for a compressed cover tree on $T$ is
$O\left(\nu c_{\min}^{8}\log\frac{R}{r(S)}\right)
\leq O\Big(\nu c_{\min}^{8} (b+\log k) \Big)$.
Below we use the upper bounds
$\nu\leq a^n\ga\leq a^n O(\sqrt[h]{h!k})$ and
$\log\nu\leq\log\ga + n\log a\leq O(\log(h!k))$, where the second term was absorbed by the first one.
Using 
\cite[Theorem~4.9]{elkin2023new}, we find $k$ neighbors of $m$ points among $\nu$ points of $T$ in time $O(m c^2 (\log k) (c_{\min}^{10}\log\nu+ ck))$, where $c_{\min}\leq c\leq 2^n$ are expansion constants of $T$. 
Then we can compute all distances from each of $m$ points from the motif $S\cap U$ to their $k$ nearest neighbors in $T$ in a time bounded as follows:
\begin{align*}
& O\big(\nu c_{\min}^{8}(b+\log k) \big) + 
O\big(m c^2 \log k (c_{\min}^{10}\log\nu+ ck)\big)\leq \\
& O\big(2^{8n} a^n  \sqrt[h]{h!k} (b+\log k)\big) + 
O\big(m 2^{2n} \log k (2^{10n}(\log(h!k)+2^{2n} k)\big)\leq \\
& O\big(2^{8n} a^n\sqrt[h]{h!k}(b+\log k) +2^{12n}m(\log k)\log(h!k) + 2^{4n}mk\log k \big). \qquad (*)
\end{align*}

By Definition~\ref{dfn:PDDh}, to compute the $k$ smallest average sums $a(h,1)\leq\dots\leq a(h,k)$, we consider all unordered $h$-tuples of points among the found $\nu$ neighbors.
Due to $\nu\leq a^n\ga\leq a^n O(\sqrt[h]{h!k})$, the number of these $h$-tuples is
$N=\vect{\nu}{h}\leq \dfrac{\nu^h}{h!}\leq\dfrac{a^{hn}}{h!}(O(\sqrt[h]{h!k}))^h=a^{hn}O(k)$. 
For each of $m$ points in the motif $S\cap U$, we sort $N$ average sums in time $O(N\log N)=a^{hn}O(k\log k)$ and select the $k$ smallest average sums in time $a^{hn}O(mk\log k)$.
When adding the latest time to the upper bound in $(*)$,
we use $a^{hn}\geq 2^{4n}$, $a^h\geq 16$, $a\geq\sqrt[h]{16}$ to get the expected total time: 
$$O\big(2^{8n} a^n\sqrt[h]{h!k}(b+\log k) +2^{12n}m(\log k)\log(h!k) + a^{hn}mk\log k\big).\qquad\qed$$  
\end{proof}

For small dimensions $n=2,3$ and orders $h=2,3$, the upper bound for the time of $\PDDh$ becomes $O(mk\log k+k\sqrt[h]{k})$, which is close to be linear in both key inputs sizes: the motif size $m$ and the number $k$ of smallest average sums.
\smallskip

For any $h\geq 1$ and a periodic set $S\subset\R^n$ of up to $m$ points in a unit cell, $\PDDh(S;k)$, the exact EMD can be found in time $O(m^3\log m)$ \cite{orlin1993faster}.
By \cite[Theorem~4.4]{widdowson2022resolving}, $\PDD(S;k)$ for a large enough $k$ (and hence the stronger $\PDDall$) together with a lattice of $S$ and the minimum number $m$ of points in a unit cell of $S$ can be inverted to any generic $S$ (outside a subspace of measure 0), uniquely under isometry.
Then by Lemma~\ref{lem:PDDh_invariant} and Theorems~\ref{thm:PDDh_continuous}, \ref{thm:PDDh_time}, $\PDDh$ satisfies almost all conditions of Problem~\ref{pro:invariants} with generic completeness instead of completeness.

\section{Big-scale experiments on high-profile databases of inorganic crystals}
\label{sec:experiments}

This section applies new invariants to quantify the novelty of materials reported by A-lab \cite{szymanski2023autonomous} and MatterGen \cite{zeni2025generative}, and then reports pairwise comparisons by the hierarchy of new invariants across three large databases of inorganic crystals:  
\smallskip

\noindent
\textbf{ICSD}: Inorganic Crystal Structure Database \cite{zagorac2019recent}, 170,206 entries \\
http://icsd.products.fiz-karlsruhe.de/en (version of February 25, 2025). 
\smallskip

\noindent
\textbf{MP}: Materials Project by the Berkeley lab \cite{jain2013commentary}, 153,235 entries \\
http://next-gen.materialsproject.org (version v2023.11.1). 
\smallskip

\noindent
\textbf{GNoME}: Graph Network Materials Exploration \cite{merchant2023scaling}, 384,938 entries, \\ 
http://github.com/google-deepmind/materials\_discovery (November 29, 2023).
\smallskip

Only experimentally measured non-disordered inorganic crystals from the ICSD were included. The Materials Project contains theoretical and experimental structures (including some sourced from the ICSD), but all entries undergo simulations which change their geometry before being added to the database. 
The GNoME dataset was generated by AI trained on crystals from the Materials Project. 
\smallskip

We start by comparing the recently released crystals by MatterGen \cite{zeni2025generative} with the already available structures in the ICSD and MP. 
Tables~\ref{tab:MatterGen-vs-ICSD},~\ref{tab:MatterGen-vs-MP} show several distances (based on the past $\PDD$ and new invariants $\PDDt$) from MatterGen crystals to their three nearest neighbors in the ICSD and MP respectively.
\smallskip

All distances are measured in Angstroms, where $1\angstrom$ is approximately the smallest inter-atomic distance. 
The physical meaning of all computed distances is justified by the Lipschitz continuity, which was proved for the past invariants $\ADA$ \cite[Theorem~9]{widdowson2022average}, $\PDA$ \cite[Theorem~6]{widdowson2025geographic}, and new invariants $\PDAh$, see Theorem~\ref{thm:PDDh_continuous}, as follows.
If every atom of a periodic crystal $S$ is perturbed up to $\ep=0.1\angstrom$, then all our distances between each invariant of $S$ and its perturbation is at most $2\ep=0.2\angstrom$.
Conversely, if a distance is $d=0.2\angstrom$, to match underlying crystals exactly, at least of their atoms should be shifted by at least $d/2=0.1\angstrom$.

\begin{table}[h!]
\centering
\setlength{\tabcolsep}{5pt}
\begin{tabular}{llllll}
MatterGen ID  & ICSD composition  & ICSD ID & $L_\infty[100]$ & $\EMD_{\infty}[100]$ & $\EMD^{(2)}_{\infty}[100]$ \\ \hline
\ce{Cr2MoO6} & \ce{LiMgFeF6} & 193630  & 0.022  & 0.037 & 0.074 \\
\ce{Cr2MoO6} & \ce{Cr2WO6}  & 24793   & 0.017  & 0.032 & 0.075  \\
\ce{Cr2MoO6} & \ce{V2WO6}   & 2576    & 0.045 & 0.064 & 0.098          \\ \hline
\ce{LaMoO4} & \ce{SmTaO4}  & 59218   & 0.065  & 0.117  & 0.146   \\
\ce{LaMoO4} & \ce{SmTaO4}  & 32996   & 0.056  & 0.125 & 0.188          \\
\ce{LaMoO4} & \ce{NdTaO4}   & 79498   & 0.063  & 0.104 & 0.195          \\ \hline
\ce{Mn3NiO6} & \ce{MgMnO3} & 690439  & 0.030  & 0.052 & 0.088          \\
\ce{Mn3NiO6} & \ce{MgGeO3}  & 171790  & 0.050  & 0.071 & 0.122          \\
\ce{Mn3NiO6} & \ce{MgGeO3}  & 171788  & 0.048  & 0.070 & 0.126          \\ \hline
\ce{Ta_{0.67}Cr_{1.33}O4} & \ce{MgF2}  & 9164  & 0.009  & 0.012 & 0.020  \\
\ce{Ta_{0.67}Cr_{1.33}O4} & \ce{MgF2}  & 117472  & 0.017 & 0.022 & 0.022          \\
\ce{Ta_{0.67}Cr_{1.33}O4} & \ce{MgF2}  & 8121    & 0.017  & 0.022 & 0.022  \\ \hline
\ce{TaCr2O6} & \ce{Ti Cr Sb O6} & 81932  & 0.014  & 0.025 & 0.047          \\
\ce{TaCr2O6} & \ce{MgF2} & 8121 & 0.021 & 0.031 & 0.050 \\
\ce{TaCr2O6} & \ce{MgF2} & 117472 & 0.021 & 0.031 & 0.050 
\end{tabular}
\caption{\textbf{Column 1}: IDs of 5 MatterGen crystals \cite{zeni2025generative} in the folder `experimental' \cite{MatterGen}.
\textbf{Columns 2-3}: compositions and IDs of three nearest neighbors in the ICSD, found by the new invariants, see column 6.
\textbf{Column 4}: distance $L_\infty$ on vector invariants $\ADA(S;100)$.
\textbf{Column 5}: distance $\EMD_\infty$ on matrix invariants $\PDA(S;100)$.
\textbf{Column 6}: max distance $\EMD_\infty^{(2)}$ on new invariants $\PDAh(S;100)$ for orders $h=1,2$.
All distances are in Angstroms.} 
\label{tab:MatterGen-vs-ICSD}
\end{table}

\cite{juelshol2025continued} suggested that the MatterGen crystal
\ce{TaCr2O6} is ``identical'' to ICSD entry 9516, which was reported in 1972 \cite{astrov1972atomic}.
However, these crystals have $L_\infty[100]=0.089\angstrom$, $\EMD_{\infty}[100]=0.098\angstrom$, and $\EMD^{(2)}_{\infty}[100]=0.196\angstrom$, which are larger than the distances 
in the last three rows of Table~\ref{tab:MatterGen-vs-ICSD}.
In fact, entry 9516 is outside the first 1000 neighbors of \ce{TaCr2O6} by $\EMD^{(2)}_{\infty}$ in the ICSD.
Tables \ref{tab:Alab-vs-ICSD} and~\ref{tab:Alab-vs-MP} show the first neighbors of 43 A-lab crystals \cite{szymanski2023autonomous} in the ICSD and MP, respectively. 
\smallskip

\begin{table}[h]
\centering
\setlength{\tabcolsep}{5pt}
\begin{tabular}{llllll}
MatterGen ID   & MP composition & MP ID & $L_\infty[100]$ & $\EMD_{\infty}[100]$ & $\EMD^{(2)}_{\infty}[100]$ \\ \hline
\ce{Cr2MoO6} & \ce{Cr2VO6} & mp-1101261 & 0.010 & 0.018 & 0.032    \\
\ce{Cr2MoO6} & \ce{Cr2WO6} & mp-19894  & 0.033  & 0.042 & 0.058     \\
\ce{Cr2MoO6} & \ce{Ga2WO6} & mp-770737  & 0.018 & 0.028 & 0.059  \\ \hline
\ce{LaMoO4} & \ce{NdTaO4} & mp-4718 & 0.065  & 0.108  & 0.207   \\
\ce{LaMoO4} & \ce{NbCeO4} & mp-7550  & 0.083  & 0.133 & 0.233     \\
\ce{LaMoO4} & \ce{SmTaO4} & mp-3756  & 0.077  & 0.131 & 0.269 \\ \hline
\ce{Mn3NiO6} & \ce{MnMgO3}  & mp-770618  & 0.028 & 0.055 & 0.099 \\
\ce{Mn3NiO6} & \ce{MnCoO3}  & mp-20641   & 0.032 & 0.062 & 0.110             \\
\ce{Mn3NiO6} & \ce{FeMgO3} & mp-754508  & 0.059 & 0.093 & 0.127    \\ \hline
\ce{Ta_{0.67}Cr_{1.33}O4} & \ce{MgF2} & mp-1249  & 0.029 & 0.037 & 0.038 \\
\ce{Ta_{0.67}Cr_{1.33}O4} & \ce{NiF2} & mp-559798  & 0.050  & 0.063 & 0.063 \\
\ce{Ta_{0.67}Cr_{1.33}O4} & \ce{TiO2} & mp-2657 & 0.051 & 0.067 & 0.067  \\ \hline
\ce{TaCr2O6} & \ce{LiNiRhF6} & mp-1222366 & 0.027  & 0.048 & 0.051     \\
\ce{TaCr2O6} & \ce{MgF2} & mp-1249  & 0.033 & 0.046 & 0.058             \\
\ce{TaCr2O6} & \ce{TiVO4}  & mp-690490  & 0.019  & 0.029 & 0.061
\end{tabular}
\caption{\textbf{Column 1}: IDs of 5 MatterGen crystals \cite{zeni2025generative} in the folder `experimental' \cite{MatterGen}.
\textbf{Columns 2-3}: compositions and IDs of three nearest neighbors in the MP, found by the new invariants, see column 6.
\textbf{Column 4}: distance $L_\infty$ on vector invariants $\ADA(S;100)$.
\textbf{Column 5}: distance $\EMD_\infty$ on matrix invariants $\PDA(S;100)$.
\textbf{Column 6}: max distance $\EMD_\infty^{(2)}$ on new invariants $\PDAh(S;100)$ for orders $h=1,2$.
All distances are in Angstroms.} 
\label{tab:MatterGen-vs-MP}
\end{table}

The distance $\EMD_\infty^{(2)}[k]$ between crystals $S,Q$ is defined as the maximum of $\EMD_\infty(\PDAh(S;k),\PDAh(Q;k))$ for two orders $h=1,2$. 
In most cases, the maximum distance is achieved for order $h=2$, because the 2nd order invariant $\PDA^{\{2\}}$ collects geometric data for triples of atoms instead of pairs (inter-atomic distances).
However, very symmetric crystals can have many equal triangles, so the same number $k$ of smallest inter-atomic distances can be more separating than $k$ smallest perimeters of triangles.  
For example, the A-lab crystal \ce{KMn3O6} has the nearest neighbors $\ce{K_{1.39}Mn3O6}$ (ICSD id 261406) and \ce{KMn2O4} (mp-2765485 in the Materials Project) with $\EMD_\infty[100]$ distances $0.103\angstrom$ and $0.51\angstrom$, which are larger than the $\EMD_\infty^{\{2\}}[100]$ distances $0.19\angstrom$ and $0.444\angstrom$, respectively. 
\smallskip

\begin{table}[h]
\centering
\begin{tabular}{lllll}
A-lab ID                & ICSD composition & ICSD ID  & $\EMD_\infty$ & $\EMD_\infty^{(2)}$   \\ \hline
\ce{Ba2ZrSnO6}          & \ce{B2Ho2Pd6} & 44417            & $<0.001$ & $<0.001$ \\
\ce{Ba6Na2Ta2V2O17}     & \ce{Ba6Na2Ru2V2O17} & 97524            & 0.092    & 0.132    \\
\ce{Ba6Na2Sb2V2O17}     & \ce{Ba6Na2Ru2V2O17} & 97524            & 0.081    & 0.147    \\
\ce{Ba9Ca3La4(Fe4O15)2} & \ce{Ba10Ca2Pr4(Fe4O_{15})2} & 405911           & 0.187    & 0.212    \\
\ce{CaCo(PO3)4}         & \ce{Cd_{0.5}Co_{1.5}(PO3)4} & 81574            & 0.144    & 0.236    \\
\ce{CaFe2P2O9}          & \ce{CaV2P2O9} & 79735            & 0.073    & 0.093    \\
\ce{CaGd2Zr(GaO3)4}     & \ce{Fe5Tb3O_{12}} & 80550            & 0.108    & 0.168    \\
\ce{CaMn(PO3)4}         & \ce{Cd2(PO3)4} & 260975           & 0.168    & 0.220    \\
\ce{CaNi(PO3)4}         & \ce{Cd_{0.5}Co_{1.5}(PO_{3})4} & 81574            & 0.157    & 0.235    \\
\ce{FeSb3Pb4O13}        & \ce{Ni_{0.666}Sb_{3.33}Pb4O_{13}} & 88959            & 0.048    & 0.116    \\
\ce{Hf2Sb2Pb4O13}       & \ce{Ru4Pb4O_{13}} & 49531            & 0.095    & 0.183    \\
\ce{InSb3(PO4)6}        & \ce{Sc4(SeO4)6} & 1729             & 0.201    & 0.240    \\
\ce{InSb3Pb4O13}        & \ce{Ru4Pb4O_{13}} & 49531            & 0.149    & 0.284    \\
\ce{K2TiCr(PO4)3}       & \ce{K_{1.928}Ti_{1.515}Fe_{0.485}(PO4)3} & 418185           & 0.037    & 0.055    \\
\ce{K4MgFe3(PO4)5}      & \ce{K4MgFe3(PO4)5} & 161484           & 0.075    & 0.109    \\
\ce{K4TiSn3(PO5)4}      & \ce{K4Ti_{1.88}Sn_{2.12}(PO5)4} & 250087           & 0.091    & 0.163    \\
\ce{KBaGdWO6}           & \ce{K2NaF4NbO2} & 183827           & 0.004    & 0.010    \\
\ce{KBaPrWO6}           & \ce{H8F6N2NaV} & 246824           & 0.004    & 0.009    \\
\ce{KMn3O6}             & \ce{K_{1.39}Mn3O6} & 261406           & 0.103 & 0.103 \\
\ce{KNa2Ga3(SiO4)3}     & \ce{Na3Ga3(SiO4)3} & 46861            & 0.110    & 0.146    \\
\ce{KNaP6(PbO3)8}       & \ce{KNaP6(PbO3)8} & 182501           & 0.005    & 0.006    \\
\ce{KNaTi2(PO5)2}       & \ce{K_{1.04}Na_{0.96}Ti2(PO5)2} & 71239            & 0.062    & 0.105    \\
\ce{KPr9(Si3O13)2}      & \ce{Sr_{1.91}Nd_{8.09}(Si3O_{13})2} & 238283           & 0.144    & 0.172    \\
\ce{Mg3MnNi3O8}         & \ce{Mg_{1.2}MnNi_{4.8}O8} & 80303            & 0.020    & 0.031    \\
\ce{Mg3NiO4}            & \ce{Mg4O4} & 690939           & 0.000    & 0.000    \\
\ce{MgCuP2O7}           & \ce{Mg_{1.08}Co_{0.92}P2O7} & 69576            & 0.218    & 0.227    \\
\ce{MgNi(PO3)4}         & \ce{Mg2(PO_{3})4} & 4280             & 0.082    & 0.097    \\
\ce{MgTi2NiO6}          & \ce{Mn_{0.64}Ti2Ni_{1.36}O6} & 238957           & 0.045    & 0.056    \\
\ce{MgTi4(PO4)6}        & \ce{FeTi4(PO_{12})6} & 290966           & 0.132    & 0.152    \\
\ce{MgV4Cu3O14}         & \ce{V4Cu4O14} & 164189           & 0.146    & 0.193    \\
\ce{Mn2VPO7}            & \ce{Mn2V_{0.91}P_{1.09}O7} & 250126           & 0.219    & 0.333    \\
\ce{Mn4Zn3(NiO6)2}      & \ce{Mg6Ti3O12} & 65793            & 0.128    & 0.186    \\
\ce{Mn7(P2O7)4}         & \ce{Fe7(P2O7)4} & 67514            & 0.126    & 0.155    \\
\ce{MnAgO2}             & \ce{MnAgO2} & 670065           & 0.097    & 0.142    \\
\ce{Na3Ca18Fe(PO4)14}   & \ce{K2Sr18Mg2(PO4)14} & 127462           & 0.173    & 0.252    \\
\ce{Na7Mg7Fe5(PO4)12}   & \ce{Na8Ni8Fe4(PO4)12} & 169444           & 0.157    & 0.157 \\ 
\ce{NaCaMgFe(SiO3)4}    & \makecell[tl]{\ce{V_{0.28}Cr_{0.49}Mn_{0.004}Ti_{0.002}\\Na_{0.792}Ca_{1.208}Mg_{1.17}\\Fe_{0.016}Si_{3.98}O12}} & 117172           & 0.066    & 0.096    \\
\ce{NaMnFe(PO4)2}       & \makecell[tl]{\ce{Na_{1.17}Mg_{0.19}Mn_{0.46}Fe_{1.35}\\(PO4)2}} & 168037           & 0.232   & 0.232 \\  
\ce{Sn2Sb2Pb4O13}       & \ce{Ru4Pb4O_{13}} & 49531            & 0.088    & 0.188    \\
\ce{Y3In2Ga3O12}        & \ce{Y_{2.74}Sc_{2.19}Ga_{3.01}O_{12}} & 39834            & 0.018    & 0.041    \\
\ce{Zn2Cr3FeO8}         & \ce{Mg2Cr4O8} & 160954           & 0.022    & 0.035    \\
\ce{Zn3Ni4(SbO6)2}      & \ce{CoLi2Ti_{2.8}O8} & 19999            & 0.173    & 0.211    \\
\ce{Zr2Sb2Pb4O13}       & \ce{Ru4Pb4O_{13}} & 49531            & 0.106    & 0.218 
\end{tabular}
\caption{\textbf{Column 1}: IDs of 43 A-lab crystals reported in \cite{szymanski2023autonomous}.
\textbf{Columns 2-3}: compositions and IDs of the nearest neighbor in the ICSD, found by the new invariants, see column 5.
\textbf{Column 4}: distance $\EMD_\infty$ on matrix invariants $\PDA(S;100)$.
\textbf{Column 5}: max distance $\EMD_\infty^{(2)}$ on new invariants $\PDAh(S;100)$ for orders $h=1,2$.
All distances are in Angstroms.} 
\label{tab:Alab-vs-ICSD}
\end{table}

\begin{table}[h]
\centering
\begin{tabular}{lllll}
A-lab ID                & MP composition & MP ID       & $\EMD_\infty$   & $\EMD_\infty^{(2)}$  \\ \hline
\ce{Ba2ZrSnO6}          & \ce{Hf2KPrO6} & mp-1522216  & $<0.001$        & $<0.001$ \\
\ce{Ba6Na2Ta2V2O17}     & \ce{Ba6Na2Ta2V2O17} & mp-1214664  & 0.029           & 0.051 \\
\ce{Ba6Na2Sb2V2O17}     & \ce{Ba6Na2Sb2V2O17} & mp-1214658  & 0.021           & 0.030 \\
\ce{Ba9Ca3La4(Fe4O15)2} & \ce{Ba9Ca3La4(Fe4O_{15})2} & mp-1228537  & 0.136           & 0.141 \\
\ce{CaCo(PO3)4}         & \ce{CaCo(PO3)4} & mp-1045787  & 0.090  & 0.090 \\ 
\ce{CaFe2P2O9}          & \ce{CaV2P2O9} & mp-21541    & 0.061           & 0.088 \\
\ce{CaGd2Zr(GaO3)4}     & \ce{CaGd2Zr(GaO_{3})4} & mp-686296   & 0.069           & 0.072 \\
\ce{CaMn(PO3)4}         & \ce{CaTi(PO3)4} & mp-1045626  & 0.071           & 0.097 \\
\ce{CaNi(PO3)4}         & \ce{CaCo(PO3)4} & mp-1045787  & 0.105           & 0.121 \\
\ce{FeSb3Pb4O13}        & \ce{FeSb3Pb4O_{13}} & mp-1224890  & 0.027           & 0.034 \\
\ce{Hf2Sb2Pb4O13}       & \ce{Hf2Sb2Pb4O_{13}} & mp-1224490  & 0.012           & 0.022 \\
\ce{InSb3(PO4)6}        & \ce{InSb3(PO_{4})6} & mp-1224667  & 0.011           & 0.018 \\
\ce{InSb3Pb4O13}        & \ce{InSb3Pb4O_{13}} & mp-1223746  & 0.029           & 0.036 \\
\ce{K2TiCr(PO4)3}       & \ce{K2TiCr(PO_{4})3} & mp-1224541  & 0.009           & 0.019 \\
\ce{K4MgFe3(PO4)5}      & \ce{K4MgFe3(PO_{4})5} & mp-532755   & 0.076           & 0.088 \\
\ce{K4TiSn3(PO5)4}      & \ce{K4TiSn3(PO_{5})4} & mp-1224290  & 0.014           & 0.025 \\
\ce{KBaGdWO6}           & \ce{NaSmEuWO6} & mp-1523299  & 0.001           & 0.003 \\
\ce{KBaPrWO6}           & \ce{NaNiRb2F6} & mp-556353   & 0.003           & 0.007 \\
\ce{KMn3O6}             & \ce{KMn2O4} & mp-2765485  & 0.510           & 0.510 \\ 
\ce{KNa2Ga3(SiO4)3}     & \ce{KNa2Ga3(SiO_{4})3} & mp-1211711  & 0.022           & 0.032 \\
\ce{KNaP6(PbO3)8}       & \ce{Na2P6(PbO_{3})8} & mp-690977   & 0.090           & 0.121 \\
\ce{KNaTi2(PO5)2}       & \ce{KNaTi2(PO_{5})2} & mp-1211611  & 0.012           & 0.016 \\
\ce{KPr9(Si3O13)2}      & \ce{KPr9(Si3O_{13})2} & mp-1223421  & 0.009           & 0.021 \\
\ce{Mg3MnNi3O8}         & \ce{Mg3MnNi3O8} & mp-1222170  & 0.029           & 0.032 \\
\ce{Mg3NiO4}            & \ce{Mg3CuO4} & mp-1099249  & 0.001           & 0.002 \\
\ce{MgCuP2O7}           & \ce{MgCuP2O7} & mp-1041741  & 0.093           & 0.088 \\
\ce{MgNi(PO3)4}         & \ce{MgNi(PO_{3})4} & mp-1045786  & 0.018           & 0.024 \\
\ce{MgTi2NiO6}          & \ce{MgTi2NiO6} & mp-1221952  & 0.009           & 0.023 \\
\ce{MgTi4(PO4)6}        & \ce{MgTi4(PO_{4})6} & mp-1222070  & 0.075           & 0.076 \\
\ce{MgV4Cu3O14}         & \ce{MgV4Cu3O_{14}} & mp-1222158  & 0.060           & 0.070 \\
\ce{Mn2VPO7}            & \ce{Mn2VPO7} & mp-1210613  & 0.125           & 0.153 \\
\ce{Mn4Zn3(NiO6)2}      & \ce{Mn4Zn3(NiO_{6})2} & mp-1222033  & 0.054           & 0.063 \\
\ce{Mn7(P2O7)4}         & \ce{Mn7(P2O_{7})4} & mp-778008   & 0.123           & 0.132 \\
\ce{MnAgO2}             & \ce{MnAgO2} & mp-996995   & 0.098           & 0.112 \\
\ce{Na3Ca18Fe(PO4)14}   & \ce{Na3Ca_{18}Fe(PO_{4})14} & mp-725491   & 0.031           & 0.038 \\
\ce{Na7Mg7Fe5(PO4)12}   & \ce{Na7Mg7Fe5(PO_{4})12} & mp-1173791  & 0.028           & 0.036 \\
\ce{NaCaMgFe(SiO3)4}    & \ce{NaCaMgFe(SiO_{3})4} & mp-1221075  & 0.026           & 0.032 \\
\ce{NaMnFe(PO4)2}       & \ce{NaMnFe(PO4)2} & mp-1173592  & 0.032           & 0.034 \\
\ce{Sn2Sb2Pb4O13}       & \ce{Sn2Sb2Pb4O_{13}} & mp-1219056  & 0.025           & 0.038 \\
\ce{Y3In2Ga3O12}        & \ce{Y3In2Ga3O_{12}} & mp-1207946  & 0.008           & 0.028 \\
\ce{Zn2Cr3FeO8}         & \ce{Mg2Ga4O8} & mp-4590     & 0.022           & 0.040 \\
\ce{Zn3Ni4(SbO6)2}      & \ce{Zn3Ni4(SbO_{6})2} & mp-1216023  & 0.092           & 0.108 \\
\ce{Zr2Sb2Pb4O13}       & \ce{Zr2Sb2Pb4O_{13}} & mp-1215826  & 0.025           & 0.042 \\
\end{tabular}
\caption{\textbf{Column 1}: IDs of 43 A-lab crystals reported in \cite{szymanski2023autonomous}.
\textbf{Columns 2-3}: compositions and IDs of the nearest neighbor in the MP, found by the new invariants, see column 5.
\textbf{Column 4}: distance $\EMD_\infty$ on matrix invariants $\PDA(S;100)$.
\textbf{Column 5}: max distance $\EMD_\infty^{(2)}$ on new invariants $\PDAh(S;100)$ for orders $h=1,2$.
All distances are in Angstroms.} 
\label{tab:Alab-vs-MP}
\end{table}
    
One reason that it was previously impossible to detect geometric duplicates in each of these databases and find substantial overlaps between different databases is their huge size and the slow speed of traditional comparisons.
Our experiments were on a typical desktop (AMD Ryzen 5 5600X 6-core, 32GB RAM).
\smallskip

Another drawback of any distance is very limited information (a single number) per pair of crystals, while invariants such as $\PDDh$ include many more numerical values per crystal.
Detecting near-duplicates by invariants is much faster than by distances due to the hierarchy starting with vectors $\ADAh(S;100)$, which quickly filter out distant crystals with $L_\infty>0.01\angstrom$.
The stronger invariants $\PDDh(S;100)$ cannot have smaller distances due to Lemma~\ref{lem:EMD_PDDh_bounds}(c). 
\smallskip

The invariants $\PDA^{(h)}$ obtained by concatenating $\PDA,\PDA^{\{2\}},\dots,\PDA^{\{h\}}$ form a natural hierarchy so that increasing the order $h=1,2,\dots$ adds more invariant information to better distinguish given crystals under isometry.
\smallskip

In addition to $L_\infty$-based distances in Tables~\ref{tab:MatterGen-vs-ICSD}-~\ref{tab:Alab-vs-MP}, below we also use metrics based on $\RMS$ (Root Mean Square) between vectors or rows of $\PDA$ matrices, so the resulting $\EMD$ on $\PDAh$ is written without a subscript for simplicity.
The $\RMS$-based metrics have Lipschitz constant 2 (or $4$ for $h>1$) by Corollary~\ref{cor:PDAh_continuous}.
\smallskip

Since any computations accumulate arithmetic errors, we start by finding geometric near-duplicates (under isometry including reflections) with the threshold $10^{-10}\angstrom=10^{-18}m$ 
for all distances. 
Then we gradually increase the threshold to $0.01\angstrom$, which is about 1\% of the smallest interatomic distance and considered experimental noise.
Tables~\ref{tab:EMD_PDA2_leq10^-6A}-\ref{tab:db_pairs-vs-thresholds} summarize all-vs-all comparisons across the databases ICSD, MP, and GNoME by using two distances on the new $\PDA^{(2)}$.

\begin{table}[h]
	\centering
	\setlength{\tabcolsep}{2pt}
	\begin{center}
		\caption{Count and percentage of pure periodic crystals in each database (left) found to have a near-duplicate in other databases (top) by $\EMD$ and $\EMD_\infty$ under $10^{-6}\angstrom$ on $\PDA^{(2)}(S;100)$.}
		\label{tab:EMD_PDA2_leq10^-6A}
		\begin{tabular}{l|cc|cc|cc|cc|cc|cc|}
			Data  &                    \multicolumn{4}{c|}{ICSD}                    &                    \multicolumn{4}{c|}{MP}                     &                   \multicolumn{4}{c|}{GNoME}                    \\
			      & \multicolumn{2}{c|}{$L_\infty$} &   \multicolumn{2}{c|}{RMS}    & \multicolumn{2}{c|}{$L_\infty$} &   \multicolumn{2}{c|}{RMS}   & \multicolumn{2}{c|}{$L_\infty$} &   \multicolumn{2}{c|}{RMS}    \\
			      &     count     &       \%        &     count     &      \%       &    count     &        \%        &    count     &      \%       &     count     &       \%        &     count     &      \%       \\ \hline
			  ICSD  & \textbf{9454} & \textbf{8.05} & \textbf{9462} & \textbf{8.05} & 53 & 0.05 & 154 & 0.13 & 1 & 0.00 & 8 & 0.01\\
              MP    & 26 & 0.02 & 87 & 0.06 & \textbf{80} & \textbf{0.05} & \textbf{293} & \textbf{0.19} & 10 & 0.01 & 21 & 0.01\\
              GNoME & 1 & 0.00 & 8 & 0.00 & 10 & 0.00 & 20 & 0.01 & \textbf{4351} & \textbf{1.13} & \textbf{4392} & \textbf{1.14}
		\end{tabular}
	\end{center}
\end{table}

\begin{table}[h]
	\centering
	\setlength{\tabcolsep}{2pt}
	\begin{center}
		\caption{Count and percentage of pure periodic crystals in each database (left) found to have a near-duplicate in other databases (top) by $\EMD$ and $\EMD_\infty$ under $10^{-5}\angstrom$ on $\PDA^{(2)}(S;100)$.}
		\label{tab:EMD_PDA2_leq10^-5A}
		\begin{tabular}{l|cc|cc|cc|cc|cc|cc|}
			Data  &                    \multicolumn{4}{c|}{ICSD}                    &                    \multicolumn{4}{c|}{MP}                     &                   \multicolumn{4}{c|}{GNoME}                    \\
			      & \multicolumn{2}{c|}{$L_\infty$} &   \multicolumn{2}{c|}{RMS}    & \multicolumn{2}{c|}{$L_\infty$} &   \multicolumn{2}{c|}{RMS}   & \multicolumn{2}{c|}{$L_\infty$} &   \multicolumn{2}{c|}{RMS}    \\
			      &     count     &       \%        &     count     &      \%       &    count     &        \%        &    count     &      \%       &     count     &       \%        &     count     &      \%       \\ \hline
			  ICSD  & \textbf{9509} & \textbf{8.09} & \textbf{9779} & \textbf{8.32} & 273 & 0.23 & 1021 & 0.87 & 18 & 0.02 & 84 & 0.07\\
              MP    & 176 & 0.11 & 764 & 0.50 & \textbf{545} & \textbf{0.36} & \textbf{2067} & \textbf{1.35} & 41 & 0.03 & 161 & 0.11\\
              GNoME & 14 & 0.00 & 55 & 0.01 & 38 & 0.01 & 138 & 0.04 & \textbf{4432} & \textbf{1.15} & \textbf{4590} & \textbf{1.19}
		\end{tabular}
	\end{center}
\end{table}

\begin{table}[h]
	\centering
	\setlength{\tabcolsep}{2pt}
	\begin{center}
		\caption{Count and percentage of pure periodic crystals in each database (left) found to have a near-duplicate in other databases (top) by $\EMD$ and $\EMD_\infty$ under $10^{-4}\angstrom$ on $\PDA^{(2)}(S;100)$.}
		\label{tab:EMD_PDA2_leq10^-4A}
		\begin{tabular}{l|cc|cc|cc|cc|cc|cc|}
			Data  &                    \multicolumn{4}{c|}{ICSD}                    &                    \multicolumn{4}{c|}{MP}                     &                   \multicolumn{4}{c|}{GNoME}                    \\
			      & \multicolumn{2}{c|}{$L_\infty$} &   \multicolumn{2}{c|}{RMS}    & \multicolumn{2}{c|}{$L_\infty$} &   \multicolumn{2}{c|}{RMS}   & \multicolumn{2}{c|}{$L_\infty$} &   \multicolumn{2}{c|}{RMS}    \\
			      &     count     &       \%        &     count     &      \%       &    count     &        \%        &    count     &      \%       &     count     &       \%        &     count     &      \%       \\ \hline
			  ICSD  & \textbf{10411} & \textbf{8.86} & \textbf{12845} & \textbf{10.93} & 1910 & 1.63 & 4708 & 4.01 & 170 & 0.14 & 636 & 0.54\\
              MP    & 1595 & 1.04 & 5182 & 3.38 & \textbf{3709} & \textbf{2.42} & \textbf{7018} & \textbf{4.58} & 343 & 0.22 & 1289 & 0.84\\
              GNoME & 122 & 0.03 & 393 & 0.10 & 268 & 0.07 & 507 & 0.13 & \textbf{4808} & \textbf{1.25} & \textbf{5070} & \textbf{1.32}
		\end{tabular}
	\end{center}
\end{table}

\begin{table}[h]
	\centering
	\setlength{\tabcolsep}{2pt}
	\begin{center}
		\caption{Count and percentage of pure periodic crystals in each database (left) found to have a near-duplicate in other databases (top) by $\EMD$ and $\EMD_\infty$ under $10^{-3}\angstrom$ on $\PDA^{(2)}(S;100)$.}
		\label{tab:EMD_PDA2_leq10^-3A}
		\begin{tabular}{l|cc|cc|cc|cc|cc|cc|}
			Data  &                    \multicolumn{4}{c|}{ICSD}                    &                    \multicolumn{4}{c|}{MP}                     &                   \multicolumn{4}{c|}{GNoME}                    \\
			      & \multicolumn{2}{c|}{$L_\infty$} &   \multicolumn{2}{c|}{RMS}    & \multicolumn{2}{c|}{$L_\infty$} &   \multicolumn{2}{c|}{RMS}   & \multicolumn{2}{c|}{$L_\infty$} &   \multicolumn{2}{c|}{RMS}    \\
			      &     count     &       \%        &     count     &      \%       &    count     &        \%        &    count     &      \%       &     count     &       \%        &     count     &      \%       \\ \hline
			ICSD  & \textbf{16052} & \textbf{13.66} & \textbf{21073} & \textbf{17.94} & 6722 & 5.72 & 9975 & 8.49 & 1263 & 1.08 & 3637 & 3.10\\
            MP    & 7228 & 4.72 & 9275 & 6.05 & \textbf{8301} & \textbf{5.42} & \textbf{10511} & \textbf{6.86} & 2460 & 1.61 & 5821 & 3.80\\
            GNoME & 589 & 0.15 & 793 & 0.21 & 625 & 0.16 & 906 & 0.24 & \textbf{5581} & \textbf{1.45} & \textbf{8049} & \textbf{2.09}
		\end{tabular}
	\end{center}
\end{table}

\begin{table}[h]
	\centering
	\setlength{\tabcolsep}{2pt}
	\begin{center}
		\caption{Count and percentage of pure periodic crystals in each database (left) found to have a near-duplicate in other databases (top) by $\EMD$ and $\EMD_\infty$ under $0.01\angstrom$ on $\PDA^{(2)}(S;100)$.}
		\label{tab:EMD_PDA2_leq10^-2A}
		\begin{tabular}{l|cc|cc|cc|cc|cc|cc|}
			Data  &                    \multicolumn{4}{c|}{ICSD}                     &                     \multicolumn{4}{c|}{MP}                      &                    \multicolumn{4}{c|}{GNoME}                    \\
			      & \multicolumn{2}{c|}{$L_\infty$} &    \multicolumn{2}{c|}{RMS}    & \multicolumn{2}{c|}{$L_\infty$} &    \multicolumn{2}{c|}{RMS}    & \multicolumn{2}{c|}{$L_\infty$} &    \multicolumn{2}{c|}{RMS}    \\
			      &     count      &       \%       &     count      &      \%       &     count      &       \%       &     count      &      \%       &     count     &       \%        &     count      &      \%       \\ \hline
			ICSD  & \textbf{30855} & \textbf{25.9}  & \textbf{27898} & \textbf{23.4} &     12540      &      10.5      &     12004      &     10.1      &     6120     &      5.14       &      5853      &     4.92      \\
			MP    &     12607      &      5.99      &     12588      &     5.98      & \textbf{18466} & \textbf{8.77}  & \textbf{18047} & \textbf{8.57} &     11283     &      5.35       &     11296      &     5.36      \\
			GNoME &      1379      &      0.36      &      1230      &     0.32      &      4645      &      1.21      &      4998      &     1.30      & \textbf{35314} &  \textbf{9.17}  & \textbf{49403} & \textbf{12.8}
		\end{tabular}
	\end{center}
\end{table}

\begin{figure}[h!]
\centering
\includegraphics[height=30mm]{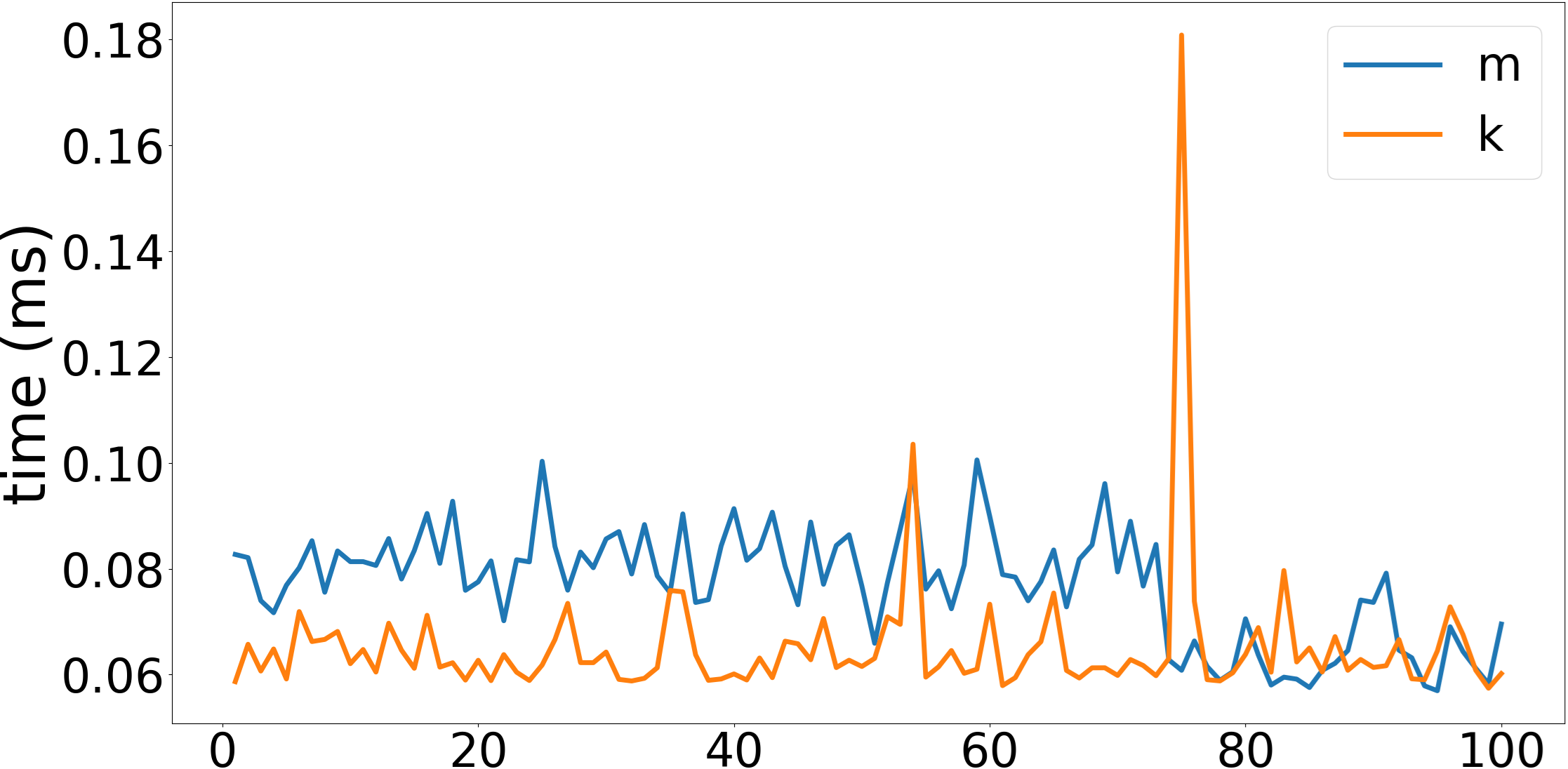}
\includegraphics[height=30mm]{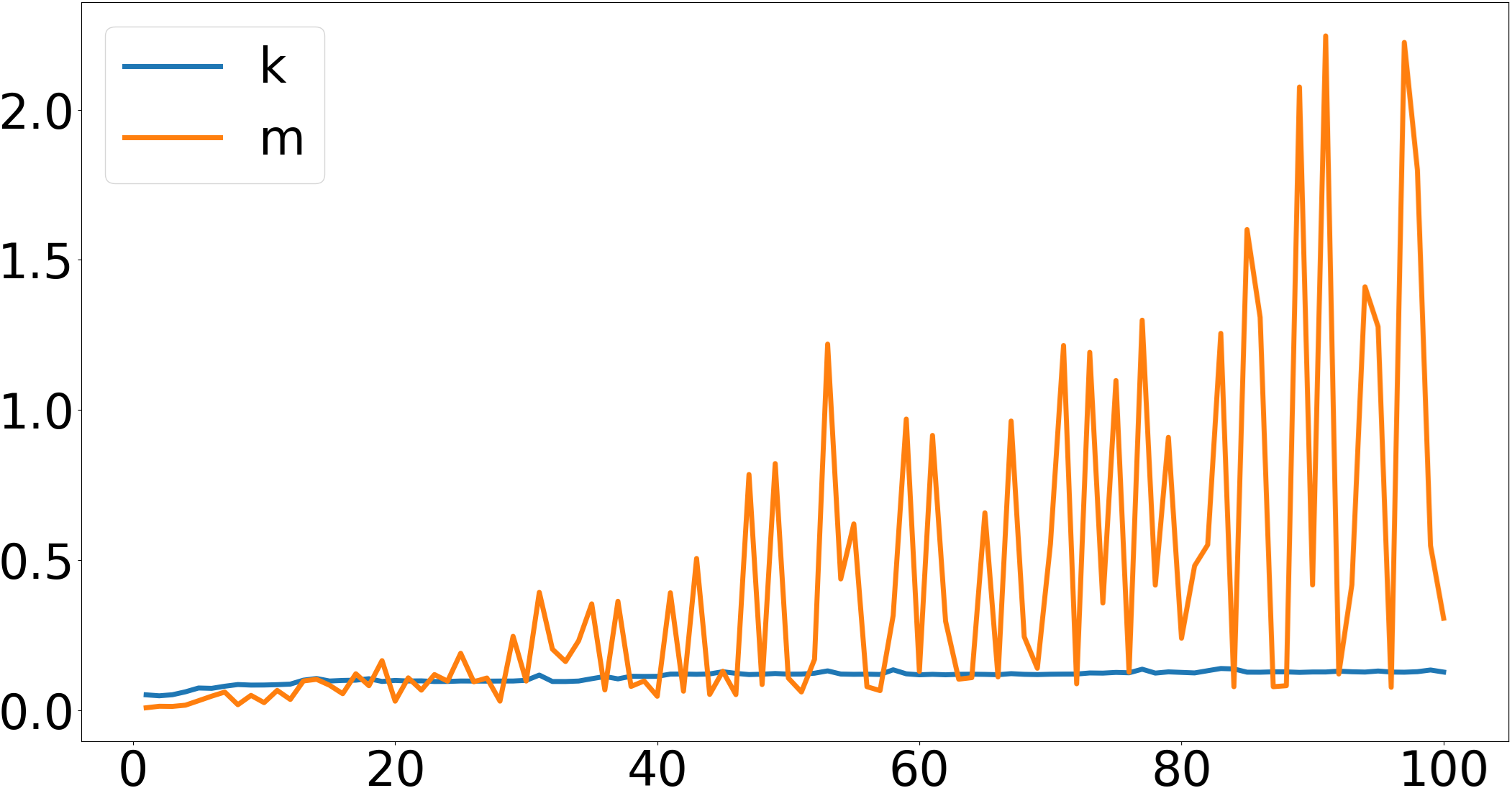}
\caption{Time in milliseconds to compare invariants $\ADA$ (left) and $\PDA$ (right).
Blue: average over pairwise comparisons of 10 random crystals from 3 databases for $k=100$ and a fixed size $m$ of an asymmetric unit.
Orange: average per atom over 500 random crystals for $k=1,\dots,100$.}
\label{fig:ADA_PDA_times_asym_unit}
\end{figure}

Tables~\ref{tab:EMD_PDA2_leq10^-6A}-\ref{tab:EMD_PDA2_leq10^-2A}  count near-duplicates (under isometry not distinguishing mirror images) within each database, which can be filtered out for any analysis or training, else the data becomes skewed.
The ultra-fast $\ADA(S;100)$ finds nearest neighbors within and between all databases using KD-trees \cite{gieseke2014buffer}.
All pairs within a given threshold by $\ADA(S;100)$ were re-compared by the stronger $\ADA^{(2)}(S;100)$, followed by $\PDA(S; 100)$ and finally $\PDA^{(2)}(S;100)$, each time keeping only the pairs with distances within the threshold. 
To avoid repeated calculations, all invariants were computed separately before making comparisons, see Fig.~\ref{fig:ADA_PDA_times_asym_unit} and Table~\ref{tab:times_PDA100}.  
\smallskip

\begin{table}[!h]
\centering
\setlength{\tabcolsep}{4pt}
\caption{
Each database has thousands of (near-)duplicates (by $L_\infty$) whose all atomic positions can be matched by tiny perturbations.
Duplication with different compositions is unexpected for very low thresholds as replacing an atom with a different one should affect geometry.
  }
    \label{tab:db_pairs-vs-thresholds}
    \begin{tabular}{c|c|ccccccc}
           near-duplicates & database & $10^{-10}\angstrom$  & $10^{-6}\angstrom$ & $10^{-5}\angstrom$ & $10^{-4}\angstrom$ & $10^{-3}\angstrom$ & $10^{-2}\angstrom$ \\ \hline
\multirowcell{3}{pairs of entries \\ within a threshold \\ by $\EMD$ on $\PDA^{(2)}$ }
    & ICSD      & 8994 & 8995 & 9036 & 10353 & 33314 & 259169\\
    & MP        & 5 & 40 & 283 & 2718 & 26703 & 278739\\
    & GNoME     & 1852 & 2482 & 2524 & 2719 & 3284 & 39487\\ \hline
\multirowcell{3}{percentage of all entries \\ in close pairs vs \\ the full database}
    & ICSD      & 8.05 & 8.05 & 8.09 & 8.86 & 13.66 & 26.24\\
    & MP        & 0.01 & 0.05 & 0.36 & 2.42 & 5.42 & 9.42\\
    & GNoME     & 0.84 & 1.13 & 1.15 & 1.25 & 1.45 & 9.17\\ \hline
\multirowcell{3}{\noindent percentage of close \\ pairs with different \\ chemical compositions}
    & ICSD      & 46.91 & 46.90 & 46.70 & 45.36 & 51.96 & 71.19\\
    & MP        & 60.00 & 85.00 & 97.53 & 99.01 & 99.88 & 99.93\\
    & GNoME     & 33.86 & 47.38 & 46.71 & 44.28 & 47.05 & 90.96 \\ \hline
\end{tabular}
\end{table}

\begin{table}[h!]
	\centering
	\setlength{\tabcolsep}{5pt}
	\begin{center}
		\caption{Times (hours:minutes:seconds) to calculate $\PDA$ and $\PDA^{\{2\}}$ for each database, and times to compare each pair of databases by the metric $\EMD_\infty$ to produce Table~\ref{tab:EMD_PDA2_leq10^-2A}. 
		The vectors $\ADA$ and $\ADA^{\{2\}}$ are near-instantly computable from $\PDA$ and $\PDA^{\{2\}}$, respectively.}
		\label{tab:times_PDA100}
		\begin{tabular}{l|cc|ccc|c}
			data  & \multicolumn{2}{c|}{Invariants} & \multicolumn{3}{c|}{Comparisons} &  Sum of  \\
			      & $\PDA$  &    $\PDA^{\{2\}}$     &  ICSD   &   MP    &    GNoME     &  times   \\ \hline
			ICSD  & 0:01:07 &        5:57:26        & 0:04:02 & 0:04:47 &   0:00:50    & 6:08:12  \\
			MP    & 0:04:35 &       25:44:33        & 0:04:10 & 0:05:32 &   0:01:44    & 26:00:34 \\
			GNoME & 0:03:47 &        9:54:48        & 0:00:43 & 0:01:51 &   0:14:53    & 10:16:02
		\end{tabular}
	\end{center}
\end{table}

Some experimental materials of different compositions may have very close geometries because their structures were determined under different conditions, such as temperature and pressure, which will be discussed in future work.
\smallskip

Table~\ref{tab:MP-duplicates} shows five pairs that were found in the Materials Project within $10^{-10}\angstrom$ for $L_\infty$ on $\ADA(S;100)$.
Three pairs with different compositions have identical numerical data and likely need updating because changing chemical elements should perturb geometry. 
Two pairs with identical compositions have identical cells and atomic coordinates that can be matched by reflection, see the appendix.

\begin{table}[!h]
\centering
\setlength{\tabcolsep}{5pt}
\caption{Geometrically identical entries in MP, three of which have different compositions.}
\label{tab:MP-duplicates}
\begin{tabular}{lllll}
MP id 1    & MP id 2    & composition 1 & composition 2 & compositional distance \cite{hargreaves2020earth} \\ \hline
mp-1100417 & mp-631388  & \ce{VSbRh} & \ce{CdIrRu}  & 8             \\
mp-1013559 & mp-1013733 & \ce{Sr3As2}   & \ce{Ca3BiSb}     & 1.2           \\
mp-1013536 & mp-1013552 & \ce{Sr3PN}      & \ce{Sr3P2}         & 0.2           \\
mp-771976 & mp-1345479  & \ce{Rb2Be2O3}      & \ce{Rb2Be2O3}  & 0 \\
mp-29783 & mp-1338697   & \ce{B5H9}          & \ce{B5H9}  & 0           
\end{tabular}
\end{table}

\section{Discussion of limitations, scientific integrity, and growing significance} 
\label{sec:discussion}

Diffraction patterns helped predict cell-based representations of crystals for 100+ years.
Recently, \cite{shen2022general} showed how to convert any crystal into many different homometric structures that have identical diffraction.
Fig.~\ref{fig:hexagonal_lattice}~(right) illustrated how any known crystal can be easily disguised by changing or expanding its cell, perturbing atoms to make the new cell primitive, and changing chemical elements. 
\smallskip

As a result, artificially generated structures threaten the integrity of experimental databases \cite{chawla2024crystallography}, which are already skewed by previously undetectable near-duplicates in other databases \cite{anosova2025complete}.
These practical challenges motivated us to formalize the fundamental questions \emph{Same or different, and by how much?} \cite{sacchi2020same}.
Problem~\ref{pro:invariants} asked for a complete, Lipschitz continuous, and polynomial-time invariant of all periodic point sets with up to $m$ points in a unit cell, and is being addressed for other real objects in the emerging area of Geometric Data Science \cite{kurlin2025complete}. 
\smallskip

While diffraction patterns and $\PDD$s cannot distinguish infinitely many homometric crystals, $\PDDt$ distinguished all known (infinitely many) counter-examples to the completeness of the $\PDD$ under isometry in Examples~\ref{exa:6-point_pairs} and~\ref{exa:Pauling_homometric}.
For practical dimensions and orders $n,h\leq 3$, the time of $\PDDh$ is near-linear in both key input sizes $k,m$ by Theorem~\ref{thm:PDDh_time}.
The new hierarchy of $\ADA^{\{h\}}$ and $\PDA^{\{h\}}$ for $h\geq 1$ allows us to use the stronger invariants $\PDA^{\{2\}}$ only in rare cases to confirm exact duplicates after much faster filtering by $\ADA,\PDA$.
\smallskip

The limitations are Conjectures~\ref{con:PDDh_complete} (completeness of $\PDD^{(h)}$ in $\R^h$) and~\ref{con:h-limit} (exact asymptotic of $\PDD^{\{h\}}$ for $h>1$), which we plan to tackle in future work.
\smallskip

Before Theorem~\ref{thm:PSD}, there was no complete, continuous, and polynomial-time invariant of periodic sets even in dimension $n=1$.
The developed hierarchy quickly detects near-duplicates of any newly claimed materials in existing databases and hence becomes an efficient barrier for noisy disguises of known crystals.
\smallskip

This research was supported by the Royal Society APEF fellowship ``New geometric methods for mapping the space of periodic crystals' '(APX/R1/231152) and EPSRC New Horizons grant `` Inverse design of periodic crystals'' (EP/X018474/1). 

\vspace*{-5mm}

\bibliographystyle{spmpsci}      
\bibliography{PDDh2025arxiv2}   


\renewcommand{\thesection}{\Alph{section}}
\setcounter{section}{0}
\section{Appendix: details of hierarchical comparisons across three databases} 
\label{sec:exp_details}

\begin{figure}[h!]
\centering
\includegraphics[width=\linewidth]{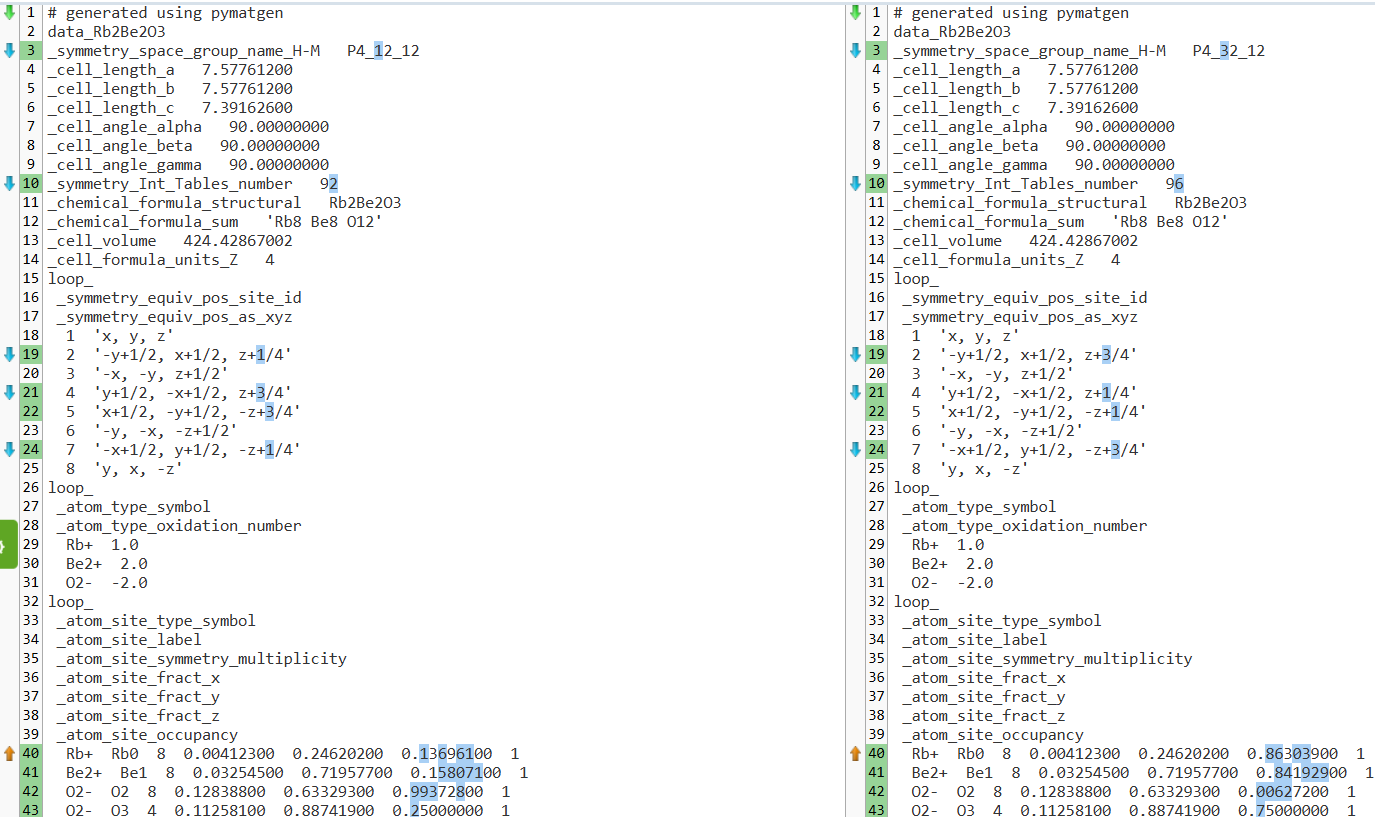}
\caption{The CIFs of the MP entries mp-771976 (left) and mp-1345479 (right) have identical cells and different coordinates, which can be matched under reflection $(x,y,z)\mapsto(x,y,1-z)$.}
\label{fig:mp-771976-vs-mp-1345479_Rb2Be2O3}
\end{figure}

\begin{figure}[h!]
\centering
\includegraphics[width=\linewidth]{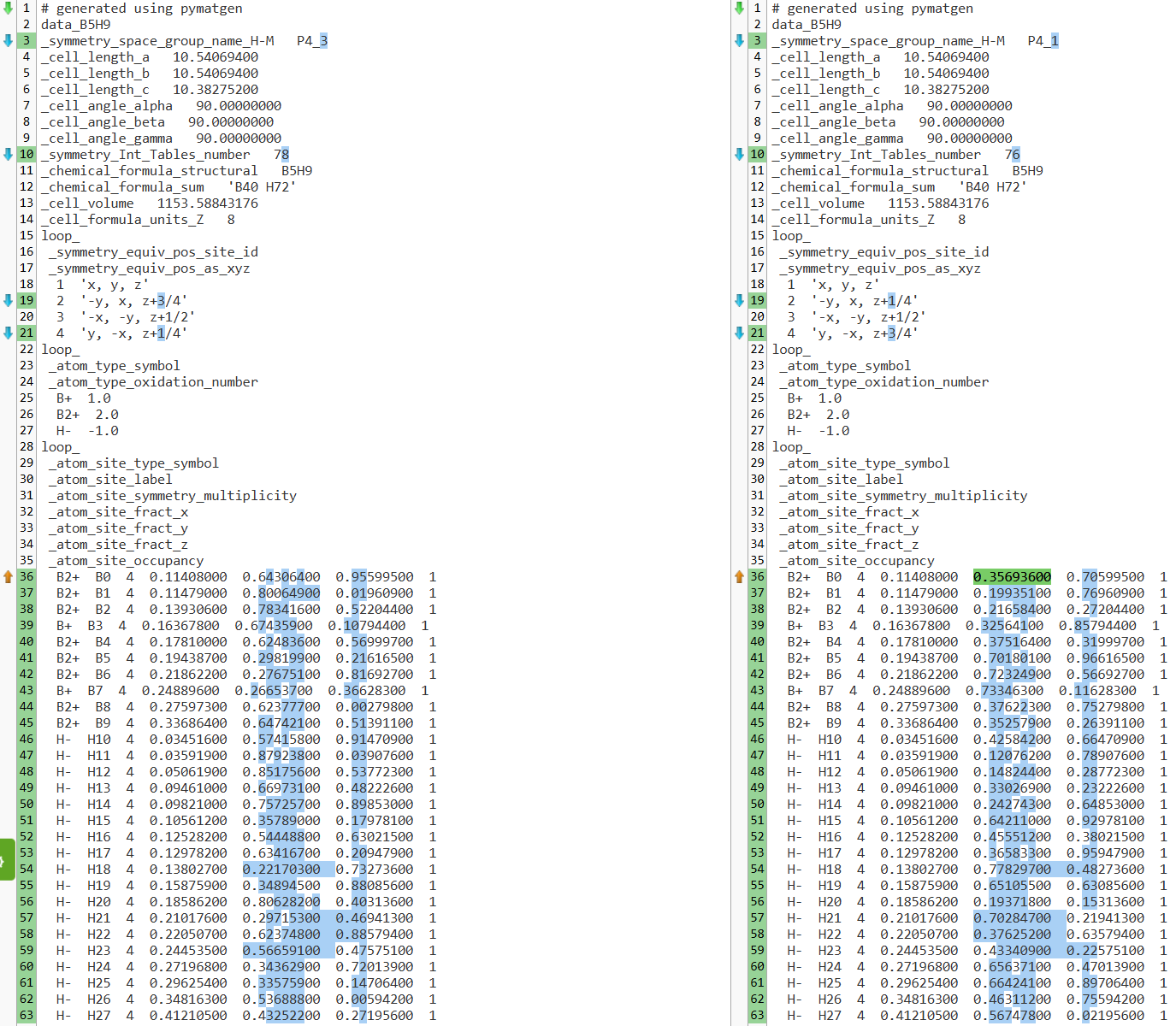}
\caption{The CIFs of the MP entries mp-29783 (left) and mp-1338697 (right) have identical compositions and unit cells, but different fractional coordinates of atoms, which can be exactly matched under reflection $(x,y,z)\mapsto(x,1-y,z-0.25)$, where $z-0.25$ is taken modulo 1.}
\label{fig:mp-29783-vs-mp-1338697_B5H9}
\end{figure}

Fig.~\ref{fig:mp-771976-vs-mp-1345479_Rb2Be2O3}-\ref{fig:mp-29783-vs-mp-1338697_B5H9} include screenshots (from https://text-compare.com) of different CIFs for the pairs from the last two rows of Table~\ref{tab:MP-duplicates}. 
Though distance-based invariants, such as $\PDAh$, cannot distinguish mirror images, our slower metric on isosets \cite{anosova2025recognition}, which are complete under rigid motion, has approximate values $0.468\angstrom$ and $0.33552\angstrom$, so these mirror images are not related by translations and rotations. 

Tables~\ref{tab:EMD_PDA100_leq10^-10A}--\ref{tab:EMD_PDA100_leq10^-2A} include the running times and numbers of compared pairs and resulting unique entries for two versions of ($L_\infty$ and $\RMS$-based) distances
between the invariants $\ADA(S;100)$, $\PDA(S;100)$, $\ADA^{(2)}(S;100)$, $\PDA^{(2)}(S;100)$. 
\smallskip

The smallest threshold $10^{-10}\angstrom$ in Table~\ref{tab:EMD_PDA100_leq10^-10A} is considered a floating-point error, and the resulting pairs of geometric duplicates are available by request. 
At the higher threshold $10^{-6}\angstrom$ in Table~\ref{tab:EMD_PDA100_leq10^-6A} only a few extra duplicates appear.
However, the further tables show that the numbers of near-duplicates substantially grow for larger thresholds up to $0.01\angstrom$, which is still considered experimental noise.
\smallskip

\begin{table}[h!]
	\centering
	\setlength{\tabcolsep}{2pt}
	\begin{center}
		\caption{Number of pairs, unique entries (and as a percentage of the database size), and running time (in seconds) taken at each stage of the duplicate finding process, using the threshold $10^{-10}\angstrom$ and $k=100$ atomic neighbors with sequentially stronger invariants.}
		\label{tab:EMD_PDA100_leq10^-10A}
		\begin{tabular}{l|c|cccc|cccc|}
			                                       &           &         \multicolumn{4}{c|}{$L_\infty$}         &             \multicolumn{4}{c|}{RMS}             \\
			                                       &           & $\ADA$  & $\PDA$  & $\ADA^{(2)}$ & $\PDA^{(2)}$ &  $\ADA$  & $\PDA$  & $\ADA^{(2)}$ & $\PDA^{(2)}$ \\ \hline
			\multirowcell{4}{ICSD \\ vs \\ ICSD}   &   Pairs   & 8994 & 8994 & 8994 & 8994 & 8994 & 8994 & 8994 & 8994\\
			                                       &  Entries  & 9452 & 9452 & 9452 & 9452 & 9452 & 9452 & 9452 & 9452\\
			                                       &    \%     & 8.05 & 8.05 & 8.05 & 8.05 & 8.05 & 8.05 & 8.05 & 8.05\\
			                                       & Time (s)  & 3.1 & 2.6 & 0.0 & 2.2 & 6.5 & 1.8 & 0.0 & 1.9\\ \hline
			\multirowcell{4}{ICSD \\ vs \\ MP}     &   Pairs   & 33 & 33 & 33 & 33 & 33 & 33 & 33 & 33\\
			                                       &  Entries  & 33 & 33 & 33 & 33 & 33 & 33 & 33 & 33\\
			                                       &    \%     & 0.03 & 0.03 & 0.03 & 0.03 & 0.03 & 0.03 & 0.03 & 0.03\\
			                                       & Time (s)  & 4.3 & 0.0 & 0.0 & 0.0 & 4.9 & 0.0 & 0.0 & 0.0\\ \hline
			\multirowcell{4}{ICSD \\ vs \\ GNoME}  &   Pairs   & 0 & 0 & 0 & 0 & 0 & 0 & 0 & 0\\
			                                       &  Entries  & 0 & 0 & 0 & 0 & 0 & 0 & 0 & 0\\
			                                       &    \%     & 0.00 & 0.00 & 0.00 & 0.00 & 0.00 & 0.00 & 0.00 & 0.00\\
			                                       & Time (s)  & 11.2 & 0.0 & 0.0 & 0.0 & 10.5 & 0.0 & 0.0 & 0.0\\ \hline
			\multirowcell{4}{MP \\ vs \\ ICSD}     &   Pairs   & 33 & 33 & 33 & 33 & 33 & 33 & 33 & 33\\
			                                       &  Entries  & 15 & 15 & 15 & 15 & 15 & 15 & 15 & 15\\
			                                       &    \%     & 0.01 & 0.01 & 0.01 & 0.01 & 0.01 & 0.01 & 0.01 & 0.01\\
			                                       & Time (s)  & 3.6 & 0.0 & 0.0 & 0.0 & 4.7 & 0.0 & 0.0 & 0.0\\ \hline
			\multirowcell{4}{MP \\ vs \\ MP}       &   Pairs   & 5 & 5 & 5 & 5 & 5 & 5 & 5 & 5\\
			                                       &  Entries  & 10 & 10 & 10 & 10 & 10 & 10 & 10 & 10\\
			                                       &    \%     & 0.01 & 0.01 & 0.01 & 0.01 & 0.01 & 0.01 & 0.01 & 0.01\\
			                                       & Time  (s) & 5.1 & 0.0 & 0.0 & 0.0 & 7.6 & 0.0 & 0.0 & 0.0\\ \hline
			\multirowcell{4}{MP \\ vs \\ GNoME}    &   Pairs   & 4 & 4 & 4 & 4 & 4 & 4 & 4 & 4\\
			                                       &  Entries  & 4 & 4 & 4 & 4 & 4 & 4 & 4 & 4\\
			                                       &    \%     & 0.00 & 0.00 & 0.00 & 0.00 & 0.00 & 0.00 & 0.00 & 0.00\\
			                                       & Time (s)  & 8.7 & 0.0 & 0.0 & 0.0 & 11.2 & 0.0 & 0.0 & 0.0\\ \hline
			\multirowcell{4}{GNoME \\ vs \\ ICSD}  &   Pairs   & 0 & 0 & 0 & 0 & 0 & 0 & 0 & 0\\
			                                       &  Entries  & 0 & 0 & 0 & 0 & 0 & 0 & 0 & 0\\
			                                       &    \%     & 0.00 & 0.00 & 0.00 & 0.00 & 0.00 & 0.00 & 0.00 & 0.00\\
			                                       & Time (s)  & 4.3 & 0.0 & 0.0 & 0.0 & 8.5 & 0.0 & 0.0 & 0.0\\ \hline
			\multirowcell{4}{GNoME \\ vs \\ MP}    &   Pairs   & 4 & 4 & 4 & 4 & 4 & 4 & 4 & 4\\
			                                       &  Entries  & 4 & 4 & 4 & 4 & 4 & 4 & 4 & 4\\
			                                       &    \%     & 0.00 & 0.00 & 0.00 & 0.00 & 0.00 & 0.00 & 0.00 & 0.00\\
			                                       & Time  (s) & 5.2 & 0.0 & 0.0 & 0.0 & 9.7 & 0.0 & 0.0 & 0.0\\ \hline
			\multirowcell{4}{GNoME \\ vs \\ GNoME} &   Pairs   & 1852 & 1852 & 1852 & 1852 & 1852 & 1852 & 1852 & 1852\\
			                                       &  Entries  & 3248 & 3248 & 3248 & 3248 & 3248 & 3248 & 3248 & 3248\\
			                                       &    \%     & 0.84 & 0.84 & 0.84 & 0.84 & 0.84 & 0.84 & 0.84 & 0.84\\
			                                       & Time (s)  & 14.5 & 0.5 & 0.0 & 0.5 & 21.4 & 0.5 & 0.0 & 0.5
		\end{tabular}
	\end{center}
\end{table}

In the bottom section of Table~\ref{tab:EMD_PDA100_leq10^-10A}, the number 3248 of geometric duplicates in the GNoME was previously found in \cite[Table~1]{anosova2024importance} by comparisons of CIFs by numerical data (unit cell parameters and atomic coordinates) than by invariants.

\begin{table}[h!]
	\centering
	\setlength{\tabcolsep}{2pt}
	\begin{center}
		\caption{Number of pairs, unique entries (and as a percentage of the database size), and running time (in seconds) taken at each stage of the duplicate finding process, using the threshold $10^{-6}\angstrom$ and $k=100$ atomic neighbors with sequentially stronger invariants.}
		\label{tab:EMD_PDA100_leq10^-6A}
		\begin{tabular}{l|c|cccc|cccc|}
			                                       &           &         \multicolumn{4}{c|}{$L_\infty$}         &             \multicolumn{4}{c|}{RMS}             \\
			                                       &           & $\ADA$  & $\PDA$  & $\ADA^{(2)}$ & $\PDA^{(2)}$ &  $\ADA$  & $\PDA$  & $\ADA^{(2)}$ & $\PDA^{(2)}$ \\ \hline
			\multirowcell{4}{ICSD \\ vs \\ ICSD}   &   Pairs   & 8999 & 8998 & 8996 & 8995 & 9036 & 9027 & 9003 & 8999\\
			                                       &  Entries  & 9462 & 9460 & 9456 & 9454 & 9510 & 9493 & 9470 & 9462\\
			                                       &    \%     & 8.05 & 8.05 & 8.05 & 8.05 & 8.09 & 8.08 & 8.06 & 8.05\\
			                                       & Time (s)  & 3.1 & 1.9 & 0.0 & 1.9 & 8.3 & 5.2 & 0.1 & 4.7\\ \hline
			\multirowcell{4}{ICSD \\ vs \\ MP}     &   Pairs   & 102 & 101 & 53 & 53 & 310 & 283 & 168 & 163\\
			                                       &  Entries  & 102 & 101 & 53 & 53 & 292 & 265 & 159 & 154\\
			                                       &    \%     & 0.09 & 0.09 & 0.05 & 0.05 & 0.25 & 0.23 & 0.14 & 0.13\\
			                                       & Time (s)  & 3.3 & 0.1 & 0.0 & 0.0 & 6.2 & 0.2 & 0.0 & 0.1\\ \hline
			\multirowcell{4}{ICSD \\ vs \\ GNoME}  &   Pairs   & 3 & 3 & 1 & 1 & 15 & 15 & 8 & 8\\
			                                       &  Entries  & 3 & 3 & 1 & 1 & 15 & 15 & 8 & 8\\
			                                       &    \%     & 0.00 & 0.00 & 0.00 & 0.00 & 0.01 & 0.01 & 0.01 & 0.01\\
			                                       & Time (s)  & 8.8 & 0.0 & 0.0 & 0.0 & 13.5 & 0.0 & 0.0 & 0.0\\ \hline
			\multirowcell{4}{MP \\ vs \\ ICSD}     &   Pairs   & 102 & 101 & 53 & 53 & 310 & 283 & 168 & 163\\
			                                       &  Entries  & 57 & 56 & 26 & 26 & 186 & 169 & 91 & 87\\
			                                       &    \%     & 0.04 & 0.04 & 0.02 & 0.02 & 0.12 & 0.11 & 0.06 & 0.06\\
			                                       & Time (s)  & 2.8 & 0.0 & 0.0 & 0.0 & 4.9 & 0.1 & 0.0 & 0.0\\ \hline
			\multirowcell{4}{MP \\ vs \\ MP}       &   Pairs   & 91 & 90 & 40 & 40 & 290 & 279 & 148 & 148\\
			                                       &  Entries  & 182 & 180 & 80 & 80 & 558 & 537 & 293 & 293\\
			                                       &    \%     & 0.12 & 0.12 & 0.05 & 0.05 & 0.36 & 0.35 & 0.19 & 0.19\\
			                                       & Time  (s) & 7.1 & 0.1 & 0.0 & 0.0 & 8.7 & 0.3 & 0.0 & 0.1\\ \hline
			\multirowcell{4}{MP \\ vs \\ GNoME}    &   Pairs   & 12 & 12 & 10 & 10 & 44 & 42 & 22 & 22\\
			                                       &  Entries  & 12 & 12 & 10 & 10 & 43 & 41 & 21 & 21\\
			                                       &    \%     & 0.01 & 0.01 & 0.01 & 0.01 & 0.03 & 0.03 & 0.01 & 0.01\\
			                                       & Time (s)  & 12.3 & 0.0 & 0.0 & 0.0 & 19.1 & 0.2 & 0.0 & 0.1\\ \hline
			\multirowcell{4}{GNoME \\ vs \\ ICSD}  &   Pairs   & 3 & 3 & 1 & 1 & 15 & 15 & 8 & 8\\
			                                       &  Entries  & 3 & 3 & 1 & 1 & 13 & 13 & 8 & 8\\
			                                       &    \%     & 0.00 & 0.00 & 0.00 & 0.00 & 0.00 & 0.00 & 0.00 & 0.00\\
			                                       & Time (s)  & 4.7 & 0.0 & 0.0 & 0.0 & 13.1 & 0.1 & 0.0 & 0.0\\ \hline
			\multirowcell{4}{GNoME \\ vs \\ MP}    &   Pairs   & 12 & 12 & 10 & 10 & 44 & 42 & 22 & 22\\
			                                       &  Entries  & 12 & 12 & 10 & 10 & 40 & 38 & 20 & 20\\
			                                       &    \%     & 0.00 & 0.00 & 0.00 & 0.00 & 0.01 & 0.01 & 0.01 & 0.01\\
			                                       & Time  (s) & 8.0 & 0.0 & 0.0 & 0.0 & 10.5 & 0.1 & 0.0 & 0.0\\ \hline
			\multirowcell{4}{GNoME \\ vs \\ GNoME} &   Pairs   & 2511 & 2490 & 2489 & 2482 & 2547 & 2516 & 2513 & 2504\\
			                                       &  Entries  & 4406 & 4367 & 4365 & 4351 & 4477 & 4416 & 4410 & 4392\\
			                                       &    \%     & 1.14 & 1.13 & 1.13 & 1.13 & 1.16 & 1.15 & 1.15 & 1.14\\
			                                       & Time (s)  & 18.4 & 3.3 & 0.0 & 2.5 & 30.2 & 3.5 & 0.0 & 2.6
		\end{tabular}
	\end{center}
\end{table}

\begin{table}[h!]
	\centering
	\setlength{\tabcolsep}{2pt}
	\begin{center}
		\caption{Number of pairs, unique entries (and as a percentage of the database size), and running time (in seconds) taken at each stage of the duplicate finding process, using the threshold $10^{-5}\angstrom$ and $k=100$ atomic neighbors with sequentially stronger invariants.}
		\label{tab:EMD_PDA100_leq10^-5A}
		\begin{tabular}{l|c|cccc|cccc|}
			                                       &           &         \multicolumn{4}{c|}{$L_\infty$}         &             \multicolumn{4}{c|}{RMS}             \\
			                                       &           & $\ADA$  & $\PDA$  & $\ADA^{(2)}$ & $\PDA^{(2)}$ &  $\ADA$  & $\PDA$  & $\ADA^{(2)}$ & $\PDA^{(2)}$ \\ \hline
			\multirowcell{4}{ICSD \\ vs \\ ICSD}   &   Pairs   & 9173 & 9143 & 9041 & 9036 & 10400 & 10190 & 9397 & 9339\\
			                                       &  Entries  & 9661 & 9620 & 9511 & 9509 & 10465 & 10294 & 9843 & 9779\\
			                                       &    \%     & 8.22 & 8.19 & 8.10 & 8.09 & 8.91 & 8.76 & 8.38 & 8.32\\
			                                       & Time (s)  & 4.9 & 5.4 & 0.1 & 4.7 & 9.5 & 5.7 & 0.1 & 5.0\\ \hline
			\multirowcell{4}{ICSD \\ vs \\ MP}     &   Pairs   & 788 & 774 & 291 & 291 & 2641 & 2486 & 1245 & 1199\\
			                                       &  Entries  & 702 & 690 & 273 & 273 & 1946 & 1848 & 1066 & 1021\\
			                                       &    \%     & 0.60 & 0.59 & 0.23 & 0.23 & 1.66 & 1.57 & 0.91 & 0.87\\
			                                       & Time (s)  & 5.0 & 0.6 & 0.0 & 0.2 & 7.3 & 1.6 & 0.0 & 0.7\\ \hline
			\multirowcell{4}{ICSD \\ vs \\ GNoME}  &   Pairs   & 67 & 61 & 18 & 18 & 191 & 171 & 89 & 85\\
			                                       &  Entries  & 67 & 61 & 18 & 18 & 173 & 159 & 88 & 84\\
			                                       &    \%     & 0.06 & 0.05 & 0.02 & 0.02 & 0.15 & 0.14 & 0.07 & 0.07\\
			                                       & Time (s)  & 14.5 & 0.1 & 0.0 & 0.0 & 16.6 & 0.1 & 0.0 & 0.1\\ \hline
			\multirowcell{4}{MP \\ vs \\ ICSD}     &   Pairs   & 788 & 774 & 291 & 291 & 2641 & 2486 & 1245 & 1199\\
			                                       &  Entries  & 490 & 477 & 176 & 176 & 1628 & 1537 & 788 & 764\\
			                                       &    \%     & 0.32 & 0.31 & 0.11 & 0.11 & 1.06 & 1.00 & 0.51 & 0.50\\
			                                       & Time (s)  & 4.4 & 0.3 & 0.0 & 0.1 & 7.0 & 0.8 & 0.0 & 0.4\\ \hline
			\multirowcell{4}{MP \\ vs \\ MP}       &   Pairs   & 821 & 792 & 289 & 283 & 2746 & 2659 & 1309 & 1285\\
			                                       &  Entries  & 1430 & 1378 & 557 & 545 & 3740 & 3620 & 2104 & 2067\\
			                                       &    \%     & 0.93 & 0.90 & 0.36 & 0.36 & 2.44 & 2.36 & 1.37 & 1.35\\
			                                       & Time  (s) & 6.3 & 0.7 & 0.0 & 0.2 & 9.8 & 1.5 & 0.0 & 0.7\\ \hline
			\multirowcell{4}{MP \\ vs \\ GNoME}    &   Pairs   & 116 & 111 & 44 & 42 & 381 & 368 & 170 & 169\\
			                                       &  Entries  & 110 & 106 & 43 & 41 & 346 & 333 & 162 & 161\\
			                                       &    \%     & 0.07 & 0.07 & 0.03 & 0.03 & 0.23 & 0.22 & 0.11 & 0.11\\
			                                       & Time (s)  & 14.9 & 0.1 & 0.0 & 0.0 & 17.6 & 0.3 & 0.0 & 0.1\\ \hline
			\multirowcell{4}{GNoME \\ vs \\ ICSD}  &   Pairs   & 67 & 61 & 18 & 18 & 191 & 171 & 89 & 85\\
			                                       &  Entries  & 41 & 36 & 14 & 14 & 123 & 115 & 57 & 55\\
			                                       &    \%     & 0.01 & 0.01 & 0.00 & 0.00 & 0.03 & 0.03 & 0.01 & 0.01\\
			                                       & Time (s)  & 6.6 & 0.1 & 0.0 & 0.0 & 12.7 & 0.2 & 0.0 & 0.1\\ \hline
			\multirowcell{4}{GNoME \\ vs \\ MP}    &   Pairs   & 116 & 111 & 44 & 42 & 381 & 368 & 170 & 169\\
			                                       &  Entries  & 99 & 94 & 40 & 38 & 271 & 262 & 139 & 138\\
			                                       &    \%     & 0.03 & 0.02 & 0.01 & 0.01 & 0.07 & 0.07 & 0.04 & 0.04\\
			                                       & Time  (s) & 8.1 & 0.1 & 0.0 & 0.0 & 12.7 & 0.3 & 0.0 & 0.1\\ \hline
			\multirowcell{4}{GNoME \\ vs \\ GNoME} &   Pairs   & 2640 & 2564 & 2549 & 2524 & 2721 & 2678 & 2648 & 2606\\
			                                       &  Entries  & 4658 & 4506 & 4479 & 4432 & 4809 & 4726 & 4674 & 4590\\
			                                       &    \%     & 1.21 & 1.17 & 1.16 & 1.15 & 1.25 & 1.23 & 1.21 & 1.19\\
			                                       & Time (s)  & 22.1 & 3.7 & 0.0 & 2.7 & 28.5 & 3.4 & 0.0 & 2.6
		\end{tabular}
	\end{center}
\end{table}

\begin{table}[h!]
	\centering
	\setlength{\tabcolsep}{2pt}
	\begin{center}
		\caption{Number of pairs, unique entries (and as a percentage of the database size), and running time (in seconds) taken at each stage of the duplicate finding process, using the threshold $10^{-4}\angstrom$ and $k=100$ atomic neighbors with sequentially stronger invariants.}
		\label{tab:EMD_PDA100_leq10^-4A}
		\begin{tabular}{l|c|cccc|cccc|}
			                                       &           &         \multicolumn{4}{c|}{$L_\infty$}         &             \multicolumn{4}{c|}{RMS}             \\
			                                       &           & $\ADA$  & $\PDA$  & $\ADA^{(2)}$ & $\PDA^{(2)}$ &  $\ADA$  & $\PDA$  & $\ADA^{(2)}$ & $\PDA^{(2)}$ \\ \hline
			\multirowcell{4}{ICSD \\ vs \\ ICSD}   &   Pairs   & 14009 & 13669 & 10360 & 10353 & 33936 & 31231 & 18669 & 18021\\
			                                       &  Entries  & 12045 & 11857 & 10425 & 10411 & 15732 & 15004 & 13061 & 12845\\
			                                       &    \%     & 10.25 & 10.09 & 8.87 & 8.86 & 13.39 & 12.77 & 11.12 & 10.93\\
			                                       & Time (s)  & 4.5 & 6.7 & 0.1 & 5.0 & 5.5 & 10.8 & 0.1 & 6.7\\ \hline
			\multirowcell{4}{ICSD \\ vs \\ MP}     &   Pairs   & 7304 & 7087 & 2586 & 2586 & 26146 & 24318 & 11831 & 11481\\
			                                       &  Entries  & 3795 & 3729 & 1910 & 1910 & 6852 & 6645 & 4833 & 4708\\
			                                       &    \%     & 3.23 & 3.17 & 1.63 & 1.63 & 5.83 & 5.66 & 4.11 & 4.01\\
			                                       & Time (s)  & 4.6 & 3.5 & 0.0 & 1.3 & 9.3 & 14.2 & 0.2 & 7.3\\ \hline
			\multirowcell{4}{ICSD \\ vs \\ GNoME}  &   Pairs   & 504 & 481 & 184 & 184 & 1892 & 1725 & 844 & 820\\
			                                       &  Entries  & 434 & 416 & 170 & 170 & 1285 & 1173 & 656 & 636\\
			                                       &    \%     & 0.37 & 0.35 & 0.14 & 0.14 & 1.09 & 1.00 & 0.56 & 0.54\\
			                                       & Time (s)  & 16.0 & 0.9 & 0.0 & 0.3 & 18.0 & 2.3 & 0.5 & 1.0\\ \hline
			\multirowcell{4}{MP \\ vs \\ ICSD}     &   Pairs   & 7304 & 7087 & 2586 & 2586 & 26146 & 24318 & 11831 & 11481\\
			                                       &  Entries  & 3881 & 3828 & 1595 & 1595 & 7275 & 7115 & 5266 & 5182\\
			                                       &    \%     & 2.53 & 2.50 & 1.04 & 1.04 & 4.75 & 4.64 & 3.44 & 3.38\\
			                                       & Time (s)  & 3.1 & 4.2 & 0.0 & 1.7 & 6.3 & 11.1 & 0.1 & 4.9\\ \hline
			\multirowcell{4}{MP \\ vs \\ MP}       &   Pairs   & 7843 & 7710 & 2718 & 2718 & 27091 & 26019 & 12634 & 12426\\
			                                       &  Entries  & 6196 & 6121 & 3709 & 3709 & 8406 & 8176 & 7107 & 7018\\
			                                       &    \%     & 4.04 & 3.99 & 2.42 & 2.42 & 5.49 & 5.34 & 4.64 & 4.58\\
			                                       & Time  (s) & 4.2 & 3.5 & 0.0 & 1.4 & 8.9 & 9.4 & 0.1 & 5.3\\ \hline
			\multirowcell{4}{MP \\ vs \\ GNoME}    &   Pairs   & 1083 & 1067 & 378 & 377 & 3855 & 3689 & 1743 & 1713\\
			                                       &  Entries  & 883 & 873 & 344 & 343 & 2470 & 2368 & 1312 & 1289\\
			                                       &    \%     & 0.58 & 0.57 & 0.22 & 0.22 & 1.61 & 1.55 & 0.86 & 0.84\\
			                                       & Time (s)  & 12.7 & 1.6 & 0.0 & 0.6 & 13.5 & 2.6 & 0.0 & 1.3\\ \hline
			\multirowcell{4}{GNoME \\ vs \\ ICSD}  &   Pairs   & 504 & 481 & 184 & 184 & 1892 & 1725 & 844 & 820\\ 
			                                       &  Entries  & 292 & 280 & 122 & 122 & 589 & 565 & 403 & 393\\
			                                       &    \%     & 0.08 & 0.07 & 0.03 & 0.03 & 0.15 & 0.15 & 0.10 & 0.10\\
			                                       & Time (s)  & 6.1 & 0.7 & 0.0 & 0.3 & 10.9 & 1.5 & 0.0 & 0.7\\ \hline
			\multirowcell{4}{GNoME \\ vs \\ MP}    &   Pairs   & 1083 & 1067 & 378 & 377 & 3855 & 3689 & 1743 & 1713\\
			                                       &  Entries  & 456 & 453 & 269 & 268 & 629 & 601 & 516 & 507\\
			                                       &    \%     & 0.12 & 0.12 & 0.07 & 0.07 & 0.16 & 0.16 & 0.13 & 0.13\\
			                                       & Time  (s) & 7.1 & 1.4 & 0.0 & 0.6 & 12.6 & 2.5 & 0.0 & 1.0\\ \hline
			\multirowcell{4}{GNoME \\ vs \\ GNoME} &   Pairs   & 2859 & 2804 & 2741 & 2719 & 3461 & 3144 & 2955 & 2901\\
			                                       &  Entries  & 5038 & 4941 & 4845 & 4808 & 5830 & 5366 & 5159 & 5070\\
			                                       &    \%     & 1.31 & 1.28 & 1.26 & 1.25 & 1.51 & 1.39 & 1.34 & 1.32\\
			                                       & Time (s)  & 14.6 & 4.7 & 0.1 & 3.6 & 28.5 & 4.1 & 0.0 & 3.8
		\end{tabular}
	\end{center}
\end{table}

\begin{table}[h!]
	\centering
	\setlength{\tabcolsep}{2pt}
	\begin{center}
		\caption{Number of pairs, unique entries (and as a percentage of the database size), and running time (in seconds) taken at each stage of the duplicate finding process, using the threshold $10^{-3}\angstrom$ and $k=100$ atomic neighbours with sequentially stronger invariants.}
		\label{tab:EMD_PDA100_leq10^-3A}
		\begin{tabular}{l|c|cccc|cccc|}
			                                       &           &         \multicolumn{4}{c|}{$L_\infty$}         &             \multicolumn{4}{c|}{RMS}             \\
			                                       &           & $\ADA$  & $\PDA$  & $\ADA^{(2)}$ & $\PDA^{(2)}$ &  $\ADA$  & $\PDA$  & $\ADA^{(2)}$ & $\PDA^{(2)}$ \\ \hline
			\multirowcell{4}{ICSD \\ vs \\ ICSD}   &   Pairs   & 82033 & 78770 & 33672 & 33314 & 258044 & 227421 & 119734 & 115102\\
			                                       &  Entries  & 20066 & 18893 & 16352 & 16052 & 27894 & 24479 & 22104 & 21073\\
			                                       &    \%     & 17.08 & 16.08 & 13.92 & 13.66 & 23.74 & 20.84 & 18.81 & 17.94\\
			                                       & Time (s)  & 4.7 & 27.0 & 0.5 & 14.2 & 9.8 & 75.9 & 1.0 & 36.7\\ \hline
			\multirowcell{4}{ICSD \\ vs \\ MP}     &   Pairs   & 75607 & 73226 & 25476 & 25416 & 275005 & 251022 & 121371 & 117451\\
			                                       &  Entries  & 9455 & 9320 & 6759 & 6722 & 11921 & 11406 & 10130 & 9975\\
			                                       &    \%     & 8.05 & 7.93 & 5.75 & 5.72 & 10.15 & 9.71 & 8.62 & 8.49\\
			                                       & Time (s)  & 4.0 & 23.6 & 0.4 & 9.6 & 9.0 & 88.7 & 3.3 & 38.4\\ \hline
			\multirowcell{4}{ICSD \\ vs \\ GNoME}  &   Pairs   & 6255 & 5751 & 1881 & 1853 & 27803 & 23438 & 10315 & 9543\\
			                                       &  Entries  & 3036 & 2755 & 1279 & 1263 & 5814 & 5359 & 3952 & 3637\\
			                                       &    \%     & 2.58 & 2.35 & 1.09 & 1.08 & 4.95 & 4.56 & 3.36 & 3.10\\
			                                       & Time (s)  & 12.5 & 3.4 & 0.0 & 1.3 & 15.7 & 10.5 & 0.8 & 5.1\\ \hline
			\multirowcell{4}{MP \\ vs \\ ICSD}     &   Pairs   & 75607 & 73226 & 25476 & 25416 & 275005 & 251022 & 121371 & 117451\\
			                                       &  Entries  & 9014 & 8884 & 7240 & 7228 & 11124 & 10415 & 9414 & 9275\\
			                                       &    \%     & 5.88 & 5.80 & 4.72 & 4.72 & 7.26 & 6.80 & 6.14 & 6.05\\
			                                       & Time (s)  & 4.4 & 25.0 & 0.4 & 10.2 & 10.3 & 75.1 & 2.0 & 36.1\\ \hline
			\multirowcell{4}{MP \\ vs \\ MP}       &   Pairs   & 79298 & 77364 & 26760 & 26703 & 284625 & 267243 & 127150 & 124642\\
			                                       &  Entries  & 10482 & 9973 & 8369 & 8301 & 14386 & 12798 & 10962 & 10511\\
			                                       &    \%     & 6.84 & 6.51 & 5.46 & 5.42 & 9.39 & 8.35 & 7.15 & 6.86\\ 
			                                       & Time  (s) & 6.0 & 25.1 & 0.4 & 9.5 & 13.2 & 82.5 & 1.3 & 37.6\\ \hline
			\multirowcell{4}{MP \\ vs \\ GNoME}    &   Pairs   & 12057 & 11622 & 3890 & 3864 & 44774 & 41267 & 19263 & 18717\\
			                                       &  Entries  & 5011 & 4823 & 2475 & 2460 & 7959 & 7345 & 6016 & 5821\\
			                                       &    \%     & 3.27 & 3.15 & 1.62 & 1.61 & 5.19 & 4.79 & 3.93 & 3.80\\
			                                       & Time (s)  & 13.8 & 5.7 & 0.1 & 2.5 & 21.6 & 16.0 & 0.3 & 8.2\\ \hline
			\multirowcell{4}{GNoME \\ vs \\ ICSD}  &   Pairs   & 6255 & 5751 & 1881 & 1853 & 27803 & 23438 & 10315 & 9543\\
			                                       &  Entries  & 802 & 760 & 603 & 589 & 1224 & 1059 & 838 & 793\\
			                                       &    \%     & 0.21 & 0.20 & 0.16 & 0.15 & 0.32 & 0.28 & 0.22 & 0.21\\
			                                       & Time (s)  & 6.1 & 3.9 & 0.0 & 1.5 & 16.5 & 10.6 & 0.1 & 4.1\\ \hline
			\multirowcell{4}{GNoME \\ vs \\ MP}    &   Pairs   & 12057 & 11622 & 3890 & 3864 & 44774 & 41267 & 19263 & 18717\\
			                                       &  Entries  & 930 & 848 & 638 & 625 & 1655 & 1317 & 981 & 906\\
			                                       &    \%     & 0.24 & 0.22 & 0.17 & 0.16 & 0.43 & 0.34 & 0.25 & 0.24\\
			                                       & Time  (s) & 7.3 & 5.8 & 0.1 & 2.4 & 18.9 & 15.3 & 0.2 & 6.5\\ \hline
			\multirowcell{4}{GNoME \\ vs \\ GNoME} &   Pairs   & 9932 & 4542 & 3595 & 3284 & 74039 & 14086 & 9894 & 5781\\
			                                       &  Entries  & 13720 & 6889 & 6016 & 5581 & 49640 & 14993 & 12992 & 8049\\
			                                       &    \%     & 3.56 & 1.79 & 1.56 & 1.45 & 12.90 & 3.89 & 3.38 & 2.09\\
			                                       & Time (s)  & 19.9 & 11.1 & 0.0 & 4.5 & 44.7 & 51.0 & 0.1 & 8.7
		\end{tabular}
	\end{center}
\end{table}

\begin{table}[h!]
	\centering
	\setlength{\tabcolsep}{2pt}
	\begin{center}
		\caption{Number of pairs, unique entries (and as a percentage of the database size), and running time (in seconds) taken at each stage of the duplicate finding process, using the threshold $10^{-2}\angstrom$ and $k=100$ atomic neighbors with sequentially stronger invariants.}
		\label{tab:EMD_PDA100_leq10^-2A}
		\begin{tabular}{l|c|cccc|cccc|}
			                                       &           &         \multicolumn{4}{c|}{$L_\infty$}         &             \multicolumn{4}{c|}{RMS}             \\
			                                       &           & $\ADA$  & $\PDA$  & $\ADA^{(2)}$ & $\PDA^{(2)}$ &  $\ADA$  & $\PDA$  & $\ADA^{(2)}$ & $\PDA^{(2)}$ \\ \hline
			\multirowcell{4}{ICSD \\ vs \\ ICSD}   &   Pairs   & 687635 & 631318 & 268187 & 259169 & 2342691 & 1976021 & 1047622 & 966117\\
			                                       &  Entries  & 46723 & 37356 & 33746 & 30831 & 68312 & 53203 & 50481 & 43508\\
			                                       &    \%     & 39.77 & 31.80 & 28.72 & 26.24 & 58.15 & 45.29 & 42.97 & 37.03\\
			                                       & Time (s)  & 9.1 & 193.4 & 3.5 & 81.6 & 54.5 & 581.1 & 6.7 & 251.7\\ \hline
			\multirowcell{4}{ICSD \\ vs \\ MP}     &   Pairs   & 851607 & 792015 & 273619 & 269722 & 3467416 & 2888095 & 1359712 & 1274212\\
			                                       &  Entries  & 19875 & 15834 & 12968 & 12398 & 47467 & 28057 & 23532 & 18785\\
			                                       &    \%     & 16.92 & 13.48 & 11.04 & 10.55 & 40.40 & 23.88 & 20.03 & 15.99\\
			                                       & Time (s)  & 10.0 & 209.3 & 4.8 & 60.6 & 60.6 & 819.0 & 10.7 & 316.2\\ \hline
			\multirowcell{4}{ICSD \\ vs \\ GNoME}  &   Pairs   & 164605 & 125416 & 35516 & 30837 & 1307587 & 779813 & 299214 & 223812\\
			                                       &  Entries  & 10637 & 9094 & 6460 & 6079 & 25425 & 16358 & 12792 & 10833\\
			                                       &    \%     & 9.05 & 7.74 & 5.50 & 5.17 & 21.64 & 13.92 & 10.89 & 9.22\\
			                                       & Time (s)  & 15.3 & 49.8 & 1.4 & 12.8 & 49.6 & 323.2 & 2.6 & 71.3\\ \hline
			\multirowcell{4}{MP \\ vs \\ ICSD}     &   Pairs   & 851607 & 792015 & 273619 & 269722 & 3467416 & 2888095 & 1359712 & 1274212\\
			                                       &  Entries  & 17575 & 14364 & 11759 & 11156 & 38866 & 23146 & 19735 & 16263\\
			                                       &    \%     & 11.47 & 9.37 & 7.67 & 7.28 & 25.36 & 15.10 & 12.88 & 10.61\\
			                                       & Time (s)  & 10.0 & 227.8 & 6.1 & 76.4 & 76.0 & 794.3 & 10.3 & 310.4\\ \hline
			\multirowcell{4}{MP \\ vs \\ MP}       &   Pairs   & 903434 & 828727 & 285041 & 278739 & 3906101 & 3071761 & 1404598 & 1324840\\
			                                       &  Entries  & 28806 & 19177 & 16067 & 14430 & 66908 & 34277 & 30425 & 23452\\
			                                       &    \%     & 18.80 & 12.51 & 10.49 & 9.42 & 43.66 & 22.37 & 19.86 & 15.30\\
			                                       & Time  (s) & 13.4 & 259.7 & 5.1 & 84.1 & 110.1 & 1417.2 & 11.5 & 335.9\\ \hline
			\multirowcell{4}{MP \\ vs \\ GNoME}    &   Pairs   & 202503 & 156999 & 51103 & 47040 & 1646545 & 928066 & 364659 & 291505\\
			                                       &  Entries  & 13362 & 10681 & 8411 & 7894 & 29364 & 18680 & 15365 & 12422\\
			                                       &    \%     & 8.72 & 6.97 & 5.49 & 5.15 & 19.16 & 12.19 & 10.03 & 8.11\\
			                                       & Time (s)  & 16.7 & 62.7 & 1.0 & 14.4 & 61.9 & 441.8 & 3.4 & 87.8\\ \hline
			\multirowcell{4}{GNoME \\ vs \\ ICSD}  &   Pairs   & 164605 & 125416 & 35516 & 30837 & 1307587 & 779813 & 299214 & 223812\\
			                                       &  Entries  & 4702 & 2515 & 1624 & 1374 & 60377 & 11310 & 6923 & 3631\\
			                                       &    \%     & 1.22 & 0.65 & 0.42 & 0.36 & 15.68 & 2.94 & 1.80 & 0.94\\
			                                       & Time (s)  & 8.6 & 47.6 & 0.8 & 13.1 & 103.2 & 326.8 & 2.7 & 70.2\\ \hline
			\multirowcell{4}{GNoME \\ vs \\ MP}    &   Pairs   & 202503 & 156999 & 51103 & 47040 & 1646545 & 928066 & 364659 & 291505\\
			                                       &  Entries  & 11124 & 3401 & 2282 & 1733 & 97553 & 19661 & 14347 & 5792\\
			                                       &    \%     & 2.89 & 0.88 & 0.59 & 0.45 & 25.34 & 5.11 & 3.73 & 1.50\\
			                                       & Time  (s) & 9.8 & 61.5 & 0.8 & 14.7 & 116.5 & 439.3 & 3.6 & 91.3\\ \hline
			\multirowcell{4}{GNoME \\ vs \\ GNoME} &   Pairs   & 1815980 & 174478 & 123171 & 39487 & 30732727 & 2059788 & 1726547 & 421833\\
			                                       &  Entries  & 197340 & 82859 & 73733 & 35315 & 326550 & 216030 & 208265 & 127820\\
			                                       &    \%     & 51.27 & 21.53 & 19.15 & 9.17 & 84.83 & 56.12 & 54.10 & 33.21\\
			                                       & Time (s)  & 33.5 & 880.8 & 0.9 & 67.8 & 549.5 & 21659.9 & 12.7 & 1061.5
		\end{tabular}
	\end{center}
\end{table}

\end{document}